\documentclass[aps,prb,twocolumn,superscriptaddress,longbibliography,floatfix]{revtex4-2}
\usepackage{amsmath,amssymb,amsthm,graphicx,bm,xcolor,dsfont,booktabs,multirow,array,longtable}
\usepackage[colorlinks=true,linkcolor=blue,citecolor=blue,urlcolor=blue]{hyperref}
\newcommand{\Tr}{\mathrm{Tr}}
\newcommand{\ii}{\mathrm{i}}
\newcommand{\one}{\mathds{1}}
\newcommand{\Z}{\mathbb{Z}}
\newcommand{\T}{\Theta}

\newcommand{\Jt}{J_t}
\newcommand{\Je}{J_e}
\newcommand{\Kt}{K_t}
\newcommand{\Ke}{K_e}
\newcommand{\YK}{ZC011110}
\newtheorem{theorem}{Theorem}
\newtheorem{corollary}{Corollary}

\makeatletter\newcommand{\SMcite}[1]{\begingroup\@fileswfalse\cite{#1}\endgroup}\makeatother
\begin{document}

\newcommand{\papertitle}{Chirality, Klein duality, and the stability of Kitaev spin liquids on the star lattice}
\title{\papertitle}

\author{Harsh Raj}
\affiliation{Department of Physics, Indian Institute of Technology Madras, Chennai 600036, India}
\author{Yasir Iqbal}
\email{yiqbal@physics.iitm.ac.in}
\affiliation{Department of Physics, Indian Institute of Technology Madras, Chennai 600036, India}

\begin{abstract}
The star lattice hosts the Yao--Kivelson chiral spin liquid, an exactly solvable Kitaev spin liquid that breaks time reversal spontaneously and is non-Abelian or Abelian depending on the ratio of its two couplings. A density-matrix renormalization group study found that an added Heisenberg exchange leaves it a wide stability window for ferromagnetic but only a narrow one for antiferromagnetic Kitaev coupling. Using projective symmetry groups (PSGs) of fermionic partons, we show that the chirality is not a peculiarity of the solvable point: whenever the $\pi$ spin rotations act on the partons as in the exact solution (spin-gauge locking), a time-reversal-symmetric mean-field state has no amplitude on the triangles, at any range, so a Kitaev-type spin liquid on this lattice must break time reversal. The Abelian and non-Abelian liquids of both signs share a single chiral PSG, within which their transition is a band inversion. The Klein duality, which reverses the Heisenberg coupling, permutes the PSGs and leaves exactly the locked ones unchanged; a $\Z_2$ spin-liquid phase mapped onto itself is therefore locked and, if it spans a connected range of couplings, contains a Kitaev point. Unprojected mean-field theory in Kitaev's Majorana representation treats both signs alike. Exact second-order perturbation theory about the Kitaev points locates the asymmetry: the triangles respond identically for both signs, whereas on the bonds between triangles the two off-diagonal Heisenberg terms create the same four visons and interfere constructively for antiferromagnetic and almost destructively for ferromagnetic coupling, admixing a low-energy four-vison state $82$ times more strongly on the antiferromagnetic side. The PSG results rest on a classification of $352$ $\Z_2$ PSGs with spin--orbit-coupled amplitudes, which also identifies the fermionic partners of the known Schwinger-boson spin liquids.
\end{abstract}

\maketitle

\tableofcontents

\section{Introduction}
\label{sec:intro}

Quantum spin liquids are phases of magnets in which quantum fluctuations prevent conventional magnetic long-range order down to zero temperature and the elementary excitations carry fractional quantum numbers~\cite{SavaryBalents2017,ZhouKanodaNg2017,Broholm2020}. Kitaev's honeycomb model~\cite{Kitaev2006} is an exactly solvable model of a quantum spin liquid~\cite{TrebstHickey2022,MatsudaShibauchiKee2025}, in which the spins fractionalize into Majorana fermions coupled to a static $\Z_2$ gauge field. It remains solvable on other tricoordinated lattices, in which every site has three bonds, and there the lattice geometry determines the ground-state pattern of $\Z_2$ fluxes through the plaquettes and the type of spin liquid. The star lattice, obtained by replacing every site of the honeycomb lattice by a triangle, is a periodic tricoordinated lattice with plaquettes of odd length on which the Kitaev model remains solvable. Kitaev showed that odd loops break time reversal spontaneously, because the flux through an odd loop takes two values that are exchanged by time reversal, and noted that replacing every site of the honeycomb lattice by a triangle should give a gapped chiral phase without a magnetic field~\cite{Kitaev2006}. Yao and Kivelson solved this model and showed that its ground state is a chiral spin liquid with non-Abelian anyons, of Ising type~\cite{Kitaev2006}, when the ratio of the inter- to the intra-triangle coupling is below $\sqrt3$, a range that includes the isotropic point, and an Abelian one above it~\cite{YaoKivelson2007}. For isotropic triangles, Dusuel \emph{et al.}\ derived effective models about the limits of decoupled triangles and decoupled inter-triangle bonds and obtained the Majorana gap at weak inter-triangle coupling~\cite{Dusuel2008}, and Kells \emph{et al.}\ obtained the phase boundary for general triangle couplings and a uniform inter-triangle coupling, a sphere in the space of these couplings (the critical sphere)~\cite{Kells2010}. The thermodynamics and tensor-network representations of the model were studied in Refs.~\cite{NasuMotome2015,LeeKanekoOkuboKawashima2020,ChenHuangKao2021}, and a spin-orbital variant with three Majorana flavors, introduced on the star lattice~\cite{YaoLee2011}, has been studied further on the honeycomb lattice~\cite{NatoriKnolle2020,Akram2023}. The same odd-loop mechanism produces chiral spin liquids on the pentaheptite lattice~\cite{Peri2020} and in amorphous Kitaev models~\cite{Cassella2023,GrushinRepellin2023}.

d'Ornellas and Knolle studied the Kitaev--Heisenberg (KH) model on the star lattice~\cite{dOrnellasKnolle2024}. With infinite density-matrix renormalization group (iDMRG) calculations they found the Kitaev spin liquid in a window around the ferromagnetic (FM) Kitaev point that is several times wider than the window around the antiferromagnetic (AFM) one, flanked by FM and stripy order on one side and by singlet and triplet valence-bond solids (VBSs), in which the spins pair into a periodic pattern of bond singlets or bond triplets, on the other, and they showed that a spin rotation that repeats with a period of two unit cells along each lattice direction maps the model at $(K,J)$ onto the model at $(K,-J)$. They traced the asymmetry to the competing phases, as Chaloupka, Jackeli, and Khaliullin had done for the honeycomb lattice~\cite{ChaloupkaJackeliKhaliullin2013}: the two VBSs lie lower in energy than their FM counterparts, whereas the two Kitaev spin liquids have the same energy scale, so that the VBSs take over quickly on the AFM side. A Majorana mean-field theory in Kitaev's representation of each spin by four Majorana fermions reproduced the sequence of phases but gave windows of nearly equal width on both sides~\cite{dOrnellasKnolle2024}. Jahromi and Or\'us studied the spin-$1/2$ Heisenberg model with couplings $\Jt$ on the triangles and $\Je$ between them using infinite projected entangled-pair states (iPEPS) and found two VBS phases: a phase of singlets on the $\Je$ bonds, which contains the isotropic point, and a resonating phase of triangle dimers at large $\Jt/\Je$~\cite{JahromiOrus2018}; a recent iPEPS study instead finds at large $\Jt$ the $\sqrt3\times\sqrt3$ valence-bond crystal (VBC), a VBS with a threefold enlarged unit cell, proposed earlier~\cite{MisguichSindzingre2007,YangParamekantiKim2010,Ghosh2025}. Choy and Kim classified the mean-field states of bosonic spinons, the fractional spin-$1/2$ quasiparticles of a spin liquid, for this Heisenberg model by their projective symmetry groups (PSGs), which record how each symmetry acts on the spinons up to a gauge transformation, and found two $\Z_2$ spin liquids, that is, spin liquids with an emergent $\Z_2$ gauge field, with all nearest-neighbor amplitudes, a zero-flux and a $\pi$-flux state. In the large-$N$ limit they found a spin-liquid window between the two dimerized phases at finite boson density $\kappa$ that shrinks to the point $\Jt=\Je$ as the density goes to zero, and assigned it to the zero-flux state~\cite{ChoyKim2009}. Among all four nearest-neighbor bosonic $\Z_2$ states (Sec.~\ref{sec:heisenberg}), the lowest at the sampled points inside this window for $0.1\le\kappa\le0.5$ is a state in which time reversal acts on the bosons with opposite phases on the two triangles of the unit cell, which we call staggered time reversal. Yang, Paramekanti, and Kim compared the six fermionic flux states with nearest-neighbor hopping, which differ in the fluxes of the hopping phases through the triangles and dodecagons, after projecting them onto states with one fermion per site~\cite{YangParamekantiKim2010}, and the Heisenberg model was studied by exact diagonalization~\cite{Richter2004,MisguichSindzingre2007}, at finite temperature~\cite{Reingruber2024}, and with ferromagnetic inter-triangle coupling~\cite{Ran2018}.

These results leave open whether the chirality of the Kitaev spin liquid is a consequence of the solvable point or holds for every spin liquid of the same kind away from it, and whether the Abelian and the non-Abelian liquids, which differ in topological order, also differ in the symmetry fractionalization of the fermionic spinon, that is, in the projective way the symmetries act on a single spinon. It is also not known what the Klein duality, the spin rotation of d'Ornellas and Knolle that maps $J$ onto $-J$, does to spin liquids and which of them can be self-dual, that is, mapped onto themselves, what makes the AFM Kitaev spin liquid itself more sensitive to the Heisenberg coupling than the FM one when mean-field theory treats both alike, and how the bosonic $\Z_2$ states of Choy and Kim and the VBS phases of the Heisenberg model are related to the fermionic spin liquids of the same lattice.

We address these questions with the PSG of fermionic partons~\cite{Wen2002}, the fermions into which each spin is split in a parton description of a spin liquid, extended to the symmetry group of the KH model, in which every lattice operation is accompanied by a spin rotation that permutes the bond labels $x$, $y$, $z$ (colors) of the Kitaev couplings, and to spin-independent (singlet) and spin-dependent (triplet) hopping and pairing amplitudes, as they arise with spin--orbit coupling~\cite{LiuHalaszBalents2021}. The group is $D_2\rtimes p6mm$, where $D_2$ consists of the global $\pi$ rotations of the spins about the three cubic axes. On a spin-$1/2$ spinon the $\pi$ rotations about different axes anticommute, so $D_2$ acts only projectively in spin space; what distinguishes PSGs is whether the gauge transformations that accompany the $\pi$ rotations commute or anticommute. We call PSGs with anticommuting gauge parts spin-gauge locked, or locked for short, after the locking of spin and gauge rotations that You, Kimchi, and Vishwanath noted for the Kitaev spin liquid~\cite{YouKimchiVishwanath2012}, because the gauge sector then reproduces the quaternion algebra of the spin rotations, as for the exact Kitaev states. Parton mean fields of the honeycomb KH model in the spin frame obtained by the four-sublattice rotation~\cite{Khaliullin2005,ChaloupkaJackeliKhaliullin2010}, the honeycomb counterpart of this duality, including triplet amplitudes, were studied by Schaffer, Bhattacharjee, and Kim, who determined the gauge groups that leave the resulting states invariant and left a PSG classification with triplet channels open~\cite{SchafferBhattacharjeeKim2012}.

In every time-reversal-symmetric $\Z_2$ PSG with locked $D_2$, time reversal combined with an element $d\in D_2$ acts on the spinons as a gauge transformation that reduces to a sign $\theta(i)=\pm1$ on each site, a site sign (Sec.~\ref{sec:classification}). Every mean-field ansatz in such a class, that is, every set of hopping and pairing amplitudes compatible with it, of any range and including the Lagrange multipliers, then obeys a selection rule for its amplitude channels, which we call the channel rule: on-site terms vanish, singlet amplitudes and triplet amplitudes along $d$ join sites of opposite sign, and the other two triplet channels join sites of the same sign. When a symmetry permutes the three bond colors cyclically, as on the star lattice, $d=1$, every amplitude joins sites of opposite sign, and the bonds that carry amplitude form a bipartite graph. No odd loop can then carry amplitude on all of its bonds. On the star lattice the triangle edges carry none, while inter-triangle and second-neighbor amplitudes remain allowed in some locked classes, so a Kitaev-type spin liquid, which needs amplitude on the triangles, must break time reversal. This is a PSG-level form of the odd-loop arguments known for the exact solutions~\cite{Kitaev2006,YaoKivelson2007,Peri2020,Cassella2023}, for three-dimensional Kitaev models~\cite{OBrienHermannsTrebst2016}, and for Majorana mean fields~\cite{KimchiVishwanath2014}, and of the observation of You, Kimchi, and Vishwanath that time reversal forbids second-neighbor amplitudes in the Kitaev PSG of the honeycomb lattice~\cite{YouKimchiVishwanath2012}; it applies to every ansatz in the class and needs no conserved flux.

The exact Kitaev states of both coupling signs and of all nonzero finite ratios of the intra- to inter-triangle couplings, non-Abelian, critical, and Abelian, have one and the same chiral PSG, labeled \YK\ below by the signs of its symmetry relations, in which the mirrors are combined with time reversal (Sec.~\ref{sec:kitaev}). The gap closes at a single Majorana cone, a Dirac point of the Majorana bands, at $\Ke/\Kt=\sqrt3$, where the Chern number of these bands changes~\cite{YaoKivelson2007,Kells2010}; the PSG shows that this Abelian--non-Abelian transition is a band inversion within one PSG, permitted at a single cone by the projective sign $n_{C_6}=-1$ of this PSG, the sign that six successive sixfold rotations acquire in the gauge sector, while the change of topological order follows from the change of the Chern number through Kitaev's sixteenfold way~\cite{Kitaev2006}, which fixes the anyon types by the Chern number modulo $16$. For the nearest-neighbor honeycomb Kitaev model at zero field, the gapless phase has the PSG (I)(B) of You, Kimchi, and Vishwanath~\cite{YouKimchiVishwanath2012}, whereas the gapped phases require anisotropic couplings that lower the lattice symmetry. On the star lattice, by contrast, the Abelian and the non-Abelian phases share one PSG at full symmetry; this PSG is chiral, that is, it breaks time reversal but keeps its products with some lattice operations, it belongs to the only one of four chiral classes that admits a Chern number, and the transition changes the Chern number within it. The PSG is the $\Z_2$ descendant of the chiral state with gauge group $SU(2)$ and the same fluxes, the candidate for the Kalmeyer--Laughlin chiral spin liquid on the star lattice~\cite{KalmeyerLaughlin1987,YangParamekantiKim2010}: adding triplet amplitudes whose spin direction is tied to the bond color (color-locked triplets) reduces the gauge group to $\Z_2$.

The rotation of d'Ornellas and Knolle is a Klein duality in the sense of Kimchi and Vishwanath~\cite{KimchiVishwanath2014}, and we show that it acts on the PSGs through an outer automorphism of the symmetry group, a relabeling of the group elements that is not a conjugation by any element (Sec.~\ref{sec:duality}). Every self-dual PSG is spin-gauge locked, and on the star lattice the $32$ locked classes are exactly the self-dual ones; all others come in dual pairs, in which a singlet amplitude is exchanged for a color-locked triplet amplitude. The couplings $(K,J)$ are parametrized by an angle on a circle (Sec.~\ref{sec:lattice}); the duality reverses the orientation of this circle, and its only self-dual points are the Kitaev points~\cite{KimchiVishwanath2014}. A phase on a connected interval of the circle that is mapped onto itself therefore contains a Kitaev point, as the Kitaev spin-liquid windows do, and by the PSG result it lies in a locked class if it is a $\Z_2$ spin liquid.

Sections~\ref{sec:meanfield} to~\ref{sec:pt} turn to the asymmetry between the two Kitaev spin liquids. Whereas the explanation of Ref.~\cite{dOrnellasKnolle2024} concerns the competing VBS phases, we study the spin-liquid side. Any mean field in the Kitaev representation without projection onto the physical states (unprojected) maps every state without correlations between the itinerant Majorana fermions $c$ and the other three Majorana fermions $b^a$ at the coupling angle $\phi$ onto a state with exactly the same energy at $\phi+\pi$, which reverses both $K$ and $J$. The magnetically ordered product states carry such correlations, but in this mean-field energy each has the energy of a dimer covering at $\phi+\pi$, the FM state that of the covering by singlet dimers and the stripy state that of the covering by triplet dimers, so the Kitaev spin-liquid windows come out symmetric when these states are the lowest competitors; the asymmetry found by iDMRG lies beyond that comparison. We project the nearest-neighbor states of the chiral PSG and of its counterpart with alternating triangle fluxes onto exactly one fermion per site (Gutzwiller projection) and evaluate them by variational Monte Carlo (VMC); their second-order energy responses to $J$, estimated by finite differences, are about three times stronger on the AFM side. On $96$ sites, with the variational parameters optimized on $24$ sites, the energies follow the same trend; size-converged curvatures require optimization on the larger clusters. Compared with exact diagonalization (ED) on a $24$-site torus, where the flux sector with alternating triangle fluxes is lowest, the projected state is accurate over a much wider range on the FM side than on the AFM side, although most of this asymmetry of the accuracy ranges comes from the exact spectrum of the cluster. The exact second-order energy about the two Kitaev points (Sec.~\ref{sec:pt}) shows where the asymmetric response of the two spin liquids to $J$ resides. On a bond of color $a$, the two Heisenberg terms $\sigma^b_i\sigma^b_j$ with $b\neq a$ (off-color terms) create the same four visons, the flux excitations of the $\Z_2$ gauge field, and their relative phase is set by the sign of the Kitaev coupling, an interference identified by Zhang, Hal\'asz, Zhu, and Batista for the hopping of flux pairs on the honeycomb lattice~\cite{ZhangHalaszZhuBatista2021}. We evaluate the second-order energy exactly, including the creation of four visons from the vacuum, and find that on the star lattice the interference depends on the bond type. The triangles are exactly blind to the sign of the coupling, whereas at the AFM point the inter-triangle bonds admix a low-energy four-vison state at $0.131|K|$, with a second-order energy $37$ times and a fidelity susceptibility $82$ times larger than at the FM point (Table~\ref{tab:ratios} collects these and the other ratios quoted in this paper). The fidelity susceptibility measures how strongly the perturbation changes the ground state, and this admixture limits perturbation theory on the AFM side to a much smaller range of Heisenberg couplings. That the inter-triangle bonds are those that dimerize in the singlet and triplet VBSs bordering the AFM spin liquid suggests a link to the VBS side of the competition, which we do not establish.

For the Heisenberg model (Sec.~\ref{sec:heisenberg}), the arguments of Lu, Cho, and Vishwanath for the symmetry action on the vison (vison PSG)~\cite{LuChoVishwanath2017}, derived for the kagome lattice, imply that on the star lattice the $\pi$-flux and zero-flux states of Choy and Kim correspond to the fermionic PSGs Z000000(1) and Z100001(1), whose digits list the projective signs of the symmetry relations (Sec.~\ref{sec:parton}); we call these PSGs their fermionic partners, since they describe the same $\Z_2$ spin-liquid phases with fermionic spinons. Because the $C_6$ axes and the mirrors $\sigma_d$ contain no sites, time reversal can also act on the Schwinger bosons, the bosonic spinons of this construction, with opposite phases $\pm\ii$ on the two triangles of the unit cell (staggered time reversal); this doubles the bosonic $\Z_2$ PSGs to $32$, among them two further states with all nearest-neighbor amplitudes. Of the $64$ realizable fermionic $\Z_2$ PSGs, those that allow a nonzero mean-field amplitude, $32$ have no counterpart among the Schwinger-boson mean-field (SBMF) states: the $16$ in which the spinon is not a Kramers doublet (time reversal squares to $+1$ on it) and the $16$ that satisfy the criterion of Lu, Cho, and Vishwanath for edge modes protected by the mirrors that contain sites ($\sigma_v$). Before projection, the symmetric fermionic mean-field states, the Kalmeyer--Laughlin candidate among them, lie above decoupled cluster states at the sampled couplings. Projection overturns this ordering: the projected SL$[0,0,\pi]$ state of Yang, Paramekanti, and Kim, a special case with gauge group $U(1)$ of the zero-flux partner Z100001(1), lies below the product of valence bonds at $\Jt=2\Je$~\cite{YangParamekantiKim2010}, and our projected states of both partners lie about $1\%$ above the ED energy of the $24$-site torus at the isotropic point. We also classify the PSGs of the symmetry groups of the competing VBS and VBC states, and discuss the large-$\Jt$ regime of the copper compound DiMACuS~\cite{Sorolla2020,Ishikawa2024} in Appendix~\ref{app:dimacus}.

The classification underlying these statements contains $352$ algebraic $\Z_2$ PSGs with spin--orbit-coupled amplitudes (solutions of the symmetry relations, irrespective of whether they allow a nonzero ansatz) for the time-reversal-symmetric case and $80$ for each of the four chiral classes (Sec.~\ref{sec:classification}). Sections~\ref{sec:lattice} and~\ref{sec:parton} set up the lattice, the symmetry group, and the parton theory. Section~\ref{sec:conclusion} summarizes the results and their limitations. The appendices give the canonical forms of the PSGs, the locked classes, the locking theorem and the Klein duality on the honeycomb, square-octagon, and pentaheptite lattices, the analysis of DiMACuS, and remarks on published values that enter our comparisons (Appendix~\ref{app:remarks}); the Supplemental Material~\cite{SM} lists the allowed amplitudes of all classes.
\section{Lattice, Models, and the Kitaev--Heisenberg Symmetry Group}
\label{sec:lattice}

\subsection{Geometry and bonds}

The star lattice is the Archimedean tiling $3.12^2$, in which every vertex is shared by one triangle and two dodecagons~\cite{Richter2004,RichterLNP2004}. Its space group is $p6mm$; in the standard labels of the symmetry-distinct points of this group (Wyckoff positions), the dodecagon centers are at $1a$, the triangle centers at $2b$, and the midpoints of the inter-triangle bonds at $3c$. The six spins of a unit cell sit at the position $6e$, whose site-symmetry group contains one mirror. We use the ideal geometry with all nearest-neighbor bonds of unit length. The Bravais lattice is spanned by
\begin{equation}
\bm{T}_{1,2}=A\left(\tfrac{\sqrt3}{2},\mp\tfrac12\right),\qquad A=2+\sqrt3,
\label{eq:T}
\end{equation}
using the same orientation convention as on the square-hexagon-dodecagon (SHD) lattice~\cite{RajIqbalSHD}, whose lattice constant is instead $3+\sqrt3$, and a site $i=(x,y,\mu)$ sits at $\bm{r}_i=x\bm{T}_1+y\bm{T}_2+(1+\sqrt3)(\cos\frac{\pi(\mu-1)}{3},\sin\frac{\pi(\mu-1)}{3})$ with $\mu=1,\dots,6$. Each site of the reference cell is the vertex of its triangle that points away from the dodecagon center at the origin. The sites with odd and even $\mu$ belong to the two triangles of a unit cell, centered at $\pm(\bm{T}_1+\bm{T}_2)/3$ up to lattice vectors. The lattice is shown in Fig.~\ref{fig:lattice}(a).

\begin{table}
\caption{Bond orbits, i.e., classes of symmetry-equivalent bonds, of the star lattice up to the fourth distance shell (fourth bond length). The reference bond $i\to j$ starts in the unit cell $(0,0)$; lengths are in units of the nearest-neighbor bond. $N$ is the number of bonds per unit cell, and the last column gives the element of $p6mm$ that maps the reference bond onto itself with its endpoints exchanged. The orbits $\Jt$ and $J_{3b}$ contain odd cycles of their own bonds.}
\label{tab:bonds}
\begin{ruledtabular}
\begin{tabular}{lcccc}
bond & $i\to j$ & length & $N$ & reversal\\
\hline
$\Jt$ & $(0,0,1)\to(0,1,5)$ & $1$ & $6$ & $T_2C_6^4S$\\
$\Je$ & $(0,0,1)\to(1,1,4)$ & $1$ & $3$ & $T_1T_2C_6^3$\\
$J_2$ & $(0,0,1)\to(-1,1,6)$ & $2\cos15^\circ$ & $12$ & --\\
$J_{3a}$ & $(0,0,1)\to(0,0,2)$ & $1+\sqrt3$ & $6$ & $C_6S$\\
$J_{3b}$ & $(0,0,1)\to(0,2,5)$ & $1+\sqrt3$ & $6$ & $T_2^2C_6^4S$\\
$J_4$ & $(0,0,1)\to(0,1,4)$ & $2.9093$ & $6$ & $T_2C_6^3$
\end{tabular}
\end{ruledtabular}
\end{table}

The nearest-neighbor bonds form two orbits: the triangle edges ($\Jt$, six per cell) and the bonds between triangles ($\Je$, three per cell). Table~\ref{tab:bonds} lists these and the further orbits used below. The bond $J_2$ closes a triangle with one $\Jt$ and one $\Je$ bond and is mapped onto itself by no symmetry other than the identity; the two orbits at distance $1+\sqrt3$, $J_{3a}$ and $J_{3b}$, are inequivalent. The triangles and the $J_{3b}$ bonds are the odd cycles that appear within single bond orbits.

\subsection{Models and units}

\begin{figure*}
\centering
\includegraphics[width=\textwidth]{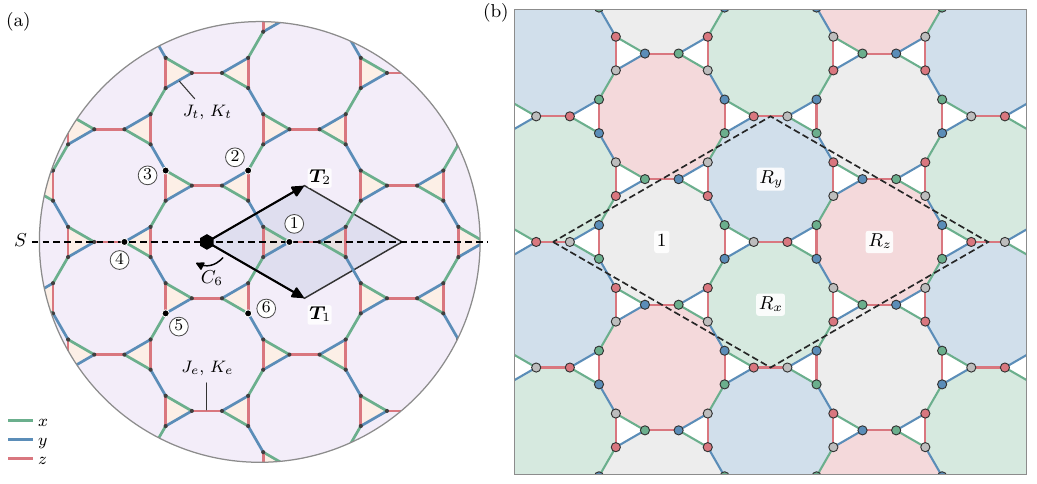}
\caption{(a) Star lattice with the Kitaev coloring of Ref.~\cite{dOrnellasKnolle2024}: every site has one $x$, one $y$, and one $z$ bond, and every inter-triangle bond carries the color of the opposite edge of each of its triangles. Triangle edges carry the couplings $\Jt$, $\Kt$ and inter-triangle bonds $\Je$, $\Ke$. The shaded unit cell is spanned by $\bm{T}_1$ and $\bm{T}_2$ from the sixfold rotation center $C_6$ at the dodecagon center; the circled numbers are the sublattice labels $\mu=1,\dots,6$ of the reference cell, and the dashed line is the mirror $S:y\to-y$. (b) The four-unit-cell Klein pattern of the duality of Sec.~\ref{sec:duality}: all spins of the cell $(x,y)$ are rotated by $\one$, $R_x$, $R_y$, or $R_z$ for $(x,y)\bmod2=(0,0)$, $(1,0)$, $(0,1)$, or $(1,1)$. The six sites of a cell are the outer vertices of the triangles around its dodecagon, which is tinted and labeled; the site colors give the rotation, and across every bond of color $a$ the two rotations differ by $R_a$. The dashed rhombus is the $2\times2$ supercell.}
\label{fig:lattice}
\end{figure*}

The Kitaev--Heisenberg (KH) model studied in Ref.~\cite{dOrnellasKnolle2024} reads
\begin{equation}
H=-\sum_{a=x,y,z}\sum_{\langle ij\rangle_a}\Big[K\sigma^a_i\sigma^a_j+J\sum_{b\neq a}\sigma^b_i\sigma^b_j\Big],
\label{eq:HKH}
\end{equation}
where $\langle ij\rangle_a$ runs over the nearest-neighbor bonds of color (Kitaev bond label) $a$ [Fig.~\ref{fig:lattice}(a)] and $\sigma^a$ are Pauli matrices, with
\begin{equation}
K=2\sin\phi+\cos\phi,\qquad J=\cos\phi.
\label{eq:phi}
\end{equation}
Equation~\eqref{eq:HKH} equals $-\sum[\cos\phi\,\bm{\sigma}_i\cdot\bm{\sigma}_j+2\sin\phi\,\sigma^a_i\sigma^a_j]$, so that $\phi=0$ and $\pi$ are the ferromagnetic (FM) and antiferromagnetic (AFM) Heisenberg points and $\phi=\pi/2$ and $3\pi/2$ the FM and AFM Kitaev points. Because of the overall minus sign, this $\phi$ is the angle of the honeycomb-lattice literature~\cite{ChaloupkaJackeliKhaliullin2013} shifted by $\pi$. In the Kitaev limit we allow different couplings $\Kt$ on the triangles and $\Ke$ between them, $H_K=-\sum\Kt\sigma^a_i\sigma^a_j-\sum\Ke\sigma^a_i\sigma^a_j$, with $K>0$ ferromagnetic. For the Heisenberg model studied in Refs.~\cite{Richter2004,JahromiOrus2018,Ghosh2025} we use spin operators,
\begin{equation}
H_{\rm H}=\Jt\sum_{\langle ij\rangle_t}\bm{S}_i\cdot\bm{S}_j+\Je\sum_{\langle ij\rangle_e}\bm{S}_i\cdot\bm{S}_j,
\label{eq:HH}
\end{equation}
with $\bm{S}=\bm{\sigma}/2$; the inter-triangle coupling $\Je$ is called $J_D$ in Ref.~\cite{Richter2004} and $J_d$ in Refs.~\cite{ChoyKim2009,Ghosh2025}. Energies of the KH model are quoted per site in the units of Eq.~\eqref{eq:HKH}, and energies of the Heisenberg model per site in the units of Eq.~\eqref{eq:HH}; an energy in the units of Eq.~\eqref{eq:HKH} is four times the same energy expressed with spin operators.

\subsection{Coloring and symmetry group}

In the coloring of Ref.~\cite{dOrnellasKnolle2024}, every site has one bond of each color, and each inter-triangle bond is perpendicular to, and has the color of, the opposite edge of each of its two triangles [Fig.~\ref{fig:lattice}(a)]. The coloring is invariant under translations, so the magnetic unit cell is the unit cell. Lattice operations permute the colors: the sixfold rotation $C_6$ about the dodecagon center maps $z\to x\to y\to z$, and the mirror $S:y\to-y$, which passes through the sites $\mu=1$ and $4$, exchanges $x$ and $y$. The operations that leave every color unchanged form the subgroup $p2$ generated by the translations and the twofold rotation $C_2=C_6^3$, and the quotient $p6mm/p2\cong S_3$ realizes each permutation of the colors exactly once. A lattice operation $g$ is therefore a symmetry of Eq.~\eqref{eq:HKH} only together with a spin rotation $R_g$ that maps each spin axis $\bm{e}_a$ onto $\pm\bm{e}_{\pi_g(a)}$, where $\pi_g$ is the color permutation of $g$. The resulting group is
\begin{equation}
G_{\rm KH}=D_2\rtimes p6mm,
\label{eq:GKH}
\end{equation}
where $D_2=\{\one,R_x,R_y,R_z\}$ consists of the global $\pi$ rotations of all spins about the three axes, and $p6mm$ is embedded into $G_{\rm KH}$ by assigning to its generators the spin rotations
\begin{equation}
R_{T_1}=R_{T_2}=\one,\qquad R_{C_6}=C_3^{[111]},\qquad R_S=C_2^{[1\bar10]}.
\label{eq:section}
\end{equation}
Here $C_3^{[111]}$ maps $\bm{e}_x\to\bm{e}_y\to\bm{e}_z\to\bm{e}_x$, and $C_2^{[1\bar10]}$ maps $\bm{e}_x\to-\bm{e}_y$, $\bm{e}_y\to-\bm{e}_x$, and $\bm{e}_z\to-\bm{e}_z$. These matrices satisfy all relations of $p6mm$, so Eq.~\eqref{eq:section} is a homomorphism onto the dihedral group $D_3$. Together with time reversal $\T$, $G_{\rm KH}\times\Z_2^\T$ has $96$ elements per unit cell. The stabilizer of a site, the subgroup of symmetries that leave it fixed, is generated by the mirror through it and by $D_2$; it is $D_4\times\Z_2^\T$, and its spin parts form the point group $422$ about the spin axis $\bm{e}_a$ of the color $a$ of the site's inter-triangle bond.

The symmetries of Eq.~\eqref{eq:HKH} on $3\times3$ and $4\times4$ tori, determined by an exhaustive search that assumes no lattice symmetry and allows site-dependent spin rotations, are for generic $K$ and $J$ exactly $48L^2$ unitary and $48L^2$ antiunitary operations, all of them elements of $G_{\rm KH}\times\Z_2^\T$. Completeness within the class of site permutations and on-site proper spin rotations also follows analytically for generic couplings $JK\neq0$ and $K^2\neq J^2$. Write the exchange matrix of an $a$ bond as $E_a=J\one+(K-J)\bm e_a\bm e_a^T$. If a symmetry maps it to a $b$ bond, its endpoint rotations obey $O_iE_aO_j^T=E_b$, and hence $O_iE_a^2O_i^T=E_b^2$. The nondegenerate color-axis eigenvalue forces $O_i\bm e_a=\pm\bm e_b$. Since every site has all three colors, $O_i$ is a signed permutation; $E_bO_i=O_iE_a$ and the invertibility of $E_a$ then give $O_j=O_i$. Connectivity makes the rotation global. Contracting the uniquely identifiable triangles reduces the graph automorphisms to those of the honeycomb lattice, $p6mm$; for each permitted color permutation the four proper spin rotations differ by $D_2$. This gives precisely $G_{\rm KH}$, with time reversal supplying the antiunitary coset. The coloring does not enlarge the unit cell. The symmetry is larger at special couplings. At $K=J$ the model is the Heisenberg model with $SU(2)\times p6mm\times\Z_2^\T$. At $K=-J$, that is $\phi=3\pi/4$ and $7\pi/4$, it has a hidden $SU(2)$ symmetry obtained from the Heisenberg one by the duality of Sec.~\ref{sec:duality}. At $J=0$ the Kitaev model has in addition the conserved physical flux operators. These are gauge-invariant integrals of motion, distinct from the $\Z_2$ gauge redundancy of the Majorana representation, and do not enter the generic-coupling symmetry classification.

The group $G_{\rm KH}\times\Z_2^\T$ is generated by $T_1$, $T_2$, $C_6$, $S$, $R_x$, $R_y$, and $\T$, where $C_6$ and $S$ now stand for the lattice operations combined with the spin rotations~\eqref{eq:section}. On the sites,
\begin{align}
T_1&:(x,y,\mu)\mapsto(x+1,y,\mu),\nonumber\\
T_2&:(x,y,\mu)\mapsto(x,y+1,\mu),\nonumber\\
C_6&:(x,y,\mu)\mapsto(-y,x+y,[\mu+1]),\nonumber\\
S&:(x,y,\mu)\mapsto(y,x,[2-\mu]),
\label{eq:action}
\end{align}
with $[k]=((k-1)\bmod6)+1$, while $R_x$, $R_y$, and $\T$ fix every site. The group is fully defined by its generators and $26$ relations (a complete presentation): the thirteen relations of $p6mm\times\Z_2^\T$,
\begin{align}
&T_2T_1T_2^{-1}T_1^{-1}=T_1C_6T_1^{-1}T_2C_6^{-1}=T_2C_6T_1^{-1}C_6^{-1}=e,\nonumber\\
&T_2ST_1^{-1}S^{-1}=T_1ST_2^{-1}S^{-1}=e,\nonumber\\
&C_6^6=S^2=(SC_6)^2=\T^2=e,\nonumber\\
&O\T O^{-1}\T^{-1}=e\quad(O=T_1,T_2,C_6,S),
\label{eq:rel-p6mm}
\end{align}
which are those of the SHD lattice~\cite{RajIqbalSHD}, and thirteen relations involving $D_2$,
\begin{align}
&R_x^2=R_y^2=(R_xR_y)^2=e,\nonumber\\
&T_aR_bT_a^{-1}R_b^{-1}=e\quad(a=1,2;\ b=x,y),\nonumber\\
&C_6R_xC_6^{-1}R_y^{-1}=C_6R_yC_6^{-1}(R_xR_y)^{-1}=e,\nonumber\\
&SR_xS^{-1}R_y^{-1}=SR_yS^{-1}R_x^{-1}=e,\nonumber\\
&\T R_x\T^{-1}R_x^{-1}=\T R_y\T^{-1}R_y^{-1}=e.
\label{eq:rel-D2}
\end{align}
Coset enumeration shows that this presentation is complete.
\section{Partons and Projective Symmetry with Spin--Orbit Locking}
\label{sec:parton}

\subsection{Abrikosov fermions and mean-field ans\"atze}

We write each spin in terms of Abrikosov fermions $f_{i\alpha}$, fermionic partons with one fermion per site~\cite{Abrikosov1965,BaskaranZouAnderson1987,AffleckZouHsuAnderson1988}, $\bm{S}_i=\frac12f^\dagger_{i\alpha}\bm{\sigma}_{\alpha\beta}f_{i\beta}$ with $f^\dagger_{i\alpha}f_{i\alpha}=1$, and collect them into
\begin{equation}
\psi_i=\begin{pmatrix} f_{i\uparrow} & f^\dagger_{i\downarrow}\\ f_{i\downarrow} & -f^\dagger_{i\uparrow}\end{pmatrix},
\label{eq:psi}
\end{equation}
on which spin rotations act from the left and the $SU(2)$ gauge transformations from the right, $\psi_i\to\psi_iW_i^\dagger$~\cite{AffleckZouHsuAnderson1988,DagottoFradkinMoreo1988,Wen2002}, in the convention of Ref.~\cite{LiuHalaszBalents2021}. Pauli matrices $\sigma^a$ act on spin and $\tau^a$ on the gauge index. The most general quadratic parton Hamiltonian is~\cite{LiuHalaszBalents2021}
\begin{equation}
H_{\rm MF}=\sum_{i,j}\sum_{\alpha=0,x,y,z}\Tr\big[\sigma^\alpha\psi_iu^{(\alpha)}_{ij}\psi_j^\dagger\big]+\sum_i\Tr\big[\psi_ia_i\psi_i^\dagger\big],
\label{eq:HMF}
\end{equation}
with the singlet and triplet link matrices, the $2\times2$ matrices of hopping and pairing amplitudes on the bond $ij$,
\begin{align}
u^{(0)}_{ij}&=\ii\chi^0_{ij}\tau^0+\chi^3_{ij}\tau^3+\Delta^1_{ij}\tau^1+\Delta^2_{ij}\tau^2,\nonumber\\
u^{(a)}_{ij}&=t^{a0}_{ij}\tau^0+\ii\big(t^{a1}_{ij}\tau^1+t^{a2}_{ij}\tau^2+t^{a3}_{ij}\tau^3\big),
\label{eq:links}
\end{align}
real coefficients, $u^{(\alpha)}_{ji}=u^{(\alpha)\dagger}_{ij}$, and Lagrange multipliers $a_i=a^k_i\tau^k$, which impose single occupancy on average; we call such a set of amplitudes and multipliers a mean-field ansatz. Singlet links describe spin-independent hopping and pairing; triplet links describe spin-dependent hopping and pairing and become allowed when the spin rotations are tied to the lattice operations, as in Eq.~\eqref{eq:section}~\cite{DoddsBhattacharjeeKim2013,LiuHalaszBalents2021,RajIqbalSHD}. Under a gauge transformation $u^{(\alpha)}_{ij}\to W_iu^{(\alpha)}_{ij}W_j^\dagger$ and $a_i\to W_ia_iW_i^\dagger$. The invariant gauge group (IGG) of an ansatz is the group of gauge transformations that leave it invariant: $\Z_2=\{\pm\tau^0\}$, $U(1)$, or $SU(2)$; it is the gauge group of the emergent gauge field of the spin liquid~\cite{Wen2002}.

\subsection{Projective symmetry}

An ansatz is symmetric under an element $g$ of $G_{\rm KH}\times\Z_2^\T$ with spin part $R_g\in SO(3)$ if there is a gauge transformation $G_g(i)$ with
\begin{equation}
u^{(b)}_{g(i)g(j)}=s_g\sum_a[R_g]_{ba}\,G_g(g(i))\,u^{(a)}_{ij}\,G_g^\dagger(g(j)),
\label{eq:psg}
\end{equation}
where $[R_g]_{00}=1$, the spin rotation acts on the triplet index only, and $s_g=-1$ for antiunitary $g$ and $+1$ otherwise~\cite{Wen2002,BieriLhuillierMessio2016,LiuHalaszBalents2021}; the multipliers transform in the same way. The gauge parts compose as $G_{g_1g_2}(i)=G_{g_1}(i)G_{g_2}(g_1^{-1}(i))$, without complex conjugation, and the time reversal of the Abrikosov fermions is $G_\T=\ii\tau^2$. The spin of a spinon transforms under the double cover of $R_g$; since the triplet amplitudes transform with $R_g$ itself and $-1$ in spin space equals $-\tau^0$ in gauge space, these signs are absorbed into the IGG, and the PSG, the group of combined symmetry and gauge operations that leave the ansatz invariant, is a projective extension of $G_{\rm KH}\times\Z_2^\T$ by the IGG. Every relation of Eqs.~\eqref{eq:rel-p6mm} and~\eqref{eq:rel-D2}, a product of generators equal to the identity, is then represented by an element of the IGG, which we call its relation value. Relation values in which every generator appears an even number of times are invariant under gauge transformations and under the sign change $G_g\to-G_g$ of a gauge part, which we call a re-lifting, and nine of them label a $\Z_2$ PSG:
\begin{align}
&n_{T_2}=[T_2T_1T_2^{-1}T_1^{-1}],\quad n_{\T S}=[S\T S^{-1}\T^{-1}],\nonumber\\
&n_{\T C_6}=[C_6\T C_6^{-1}\T^{-1}],\quad n_S=[S^2],\nonumber\\
&n_{SC_6}=[(SC_6)^2],\quad n_{C_6}=[C_6^6],\quad t=[\T^2],\nonumber\\
&n_R=[R_x^2],\quad n_{TR}=[T_1R_yT_1^{-1}R_y^{-1}].
\label{eq:labels}
\end{align}
We write a $\Z_2$ PSG as Z$n_{T_2}n_{\T S}n_{\T C_6}n_Sn_{SC_6}n_{C_6}t\,n_Rn_{TR}$, with the digit $1$ for a relation value $-1$. The first seven digits are the labels of the SHD lattice~\cite{RajIqbalSHD}. The relations of $G_{\rm KH}$ force $[R_x^2]=[R_y^2]=[(R_xR_y)^2]$ and $[T_1R_y T_1^{-1}R_y^{-1}]=[T_2R_xT_2^{-1}R_x^{-1}]$, while $[T_1R_xT_1^{-1}R_x^{-1}]=[T_2R_yT_2^{-1}R_y^{-1}]=+1$; the sixfold rotation alone gives only $[T_1R_xT_1^{-1}R_x^{-1}]=[T_2R_yT_2^{-1}R_y^{-1}]$ and $[T_1R_yT_1^{-1}R_y^{-1}]=[T_2R_xT_2^{-1}R_x^{-1}][T_2R_yT_2^{-1}R_y^{-1}]$, and the mirror $S$, which exchanges $T_1R_yT_1^{-1}R_y^{-1}$ and $T_2R_xT_2^{-1}R_x^{-1}$, forces the value $+1$.

\subsection{Spin-gauge locking}

Since $R_x$ and $R_y$ fix every site, $G_{R_x}(i)$ and $G_{R_y}(i)$ are on-site matrices. If $n_R=+1$, both square to $\tau^0$ and are therefore $\pm\tau^0$; if $n_R=-1$, they are traceless and anticommute. We call a PSG spin-gauge locked, or locked, if
\begin{equation}
G_{R_x}(i)G_{R_y}(i)=-G_{R_y}(i)G_{R_x}(i),
\label{eq:locked}
\end{equation}
which on the star lattice is equivalent to $n_R=-1$. In a locked class the gauge parts of $D_2$ form the quaternion group, the algebra of the matrices $\ii\sigma^a$ that represent the $\pi$ rotations on a spin $1/2$ (their spin-$1/2$ lifts). The full transformation of a spinon, the product of its spin lift and its gauge part, then represents $D_2$ linearly: the sign of $(\ii\sigma^x)(\ii\sigma^y)=-(\ii\sigma^y)(\ii\sigma^x)$ in spin space is compensated in gauge space. This is the structure of the exact Kitaev states~\cite{Kitaev2006}, in which each global $\pi$ spin rotation in $D_2$ can be accompanied by a compensating gauge transformation of the Majorana fermions~\cite{BurnellNayak2011,YouKimchiVishwanath2012}. The label $n_{TR}=-1$ means that $T_1$ anticommutes with $R_y$ and $T_2$ with $R_x$ in the gauge sector, as do the spin lifts of the Klein-duality translations $T_1R_x$ and $T_2R_y$ (Sec.~\ref{sec:duality}); the pairs $(T_1,R_x)$ and $(T_2,R_y)$ always commute.

\subsection{Chiral classes}
\label{sec:chiral}

A chiral spin liquid breaks $\T$, while every element of $G_{\rm KH}$ remains a symmetry either alone or combined with $\T$~\cite{WenWilczekZee1989,BieriLhuillierMessio2016}. The elements kept without $\T$ form the kernel of a homomorphism $\chi:G_{\rm KH}\to\Z_2$. The translations are commutators in $p6mm$, and $R_x$, $R_y$, $R_z$ are conjugate to each other with $R_z=R_xR_y$, so $\chi$ is trivial on them; a $\pi$ rotation is never combined with $\T$. Such a homomorphism, a character of $G_{\rm KH}$, is fixed by $\chi(C_6)$ and $\chi(S)$, which gives four chiral classes. The mirrors of $p6mm$ fall into two families: $\sigma_v$, conjugate to $S$, through the dodecagon centers, the triangle centers, and the sites, and $\sigma_d$, conjugate to $SC_6$, which bisect the $\Je$ bonds perpendicularly and contain no sites. In terms of these,
\begin{center}
\begin{tabular}{ccc}
class & $(\chi(C_6),\chi(S))$ & combined with $\T$\\\hline
A & $(+,+)$ & --\\
B & $(+,-)$ & $\sigma_v$, $\sigma_d$\\
C & $(-,+)$ & $C_6$, $\sigma_d$\\
D & $(-,-)$ & $C_6$, $\sigma_v$
\end{tabular}
\end{center}
Class B is the class of the Kalmeyer--Laughlin state, in which every mirror is a symmetry only together with $\T$. The symmetry groups of the chiral states, in which some elements are combined with $\T$, are isomorphic to $G_{\rm KH}$, so the algebraic PSGs of the four classes coincide; $\chi$ enters only through the sign $s_g$ of Eq.~\eqref{eq:psg}~\cite{BieriLhuillierMessio2016}. We label a chiral $\Z_2$ PSG by ZC$n_{T_2}n_Sn_{SC_6}n_{C_6}n_Rn_{TR}$ and add the class letter where needed, and the suffix -$a$ or -$b$ of Sec.~\ref{sec:counts}, which distinguishes two PSGs with equal relation values, for the $32$ pairs of non-locked chiral PSGs that share all relation values.

Only class B admits a nonzero Chern number of the spinon bands, by the argument of Ref.~\cite{BieriLhuillierMessio2016}. In class A the unitary mirrors reverse the orientation and force the Chern number to vanish. In classes C and D the twofold rotation $C_2=C_6^3$ is combined with $\T$; this antiunitary element maps $\bm{k}$ to itself and makes the Berry curvature odd at every momentum. In class B, $C_6$ and $D_2$ are unitary and preserve the orientation, and the antiunitary mirrors also preserve the sign of the Chern number. The same holds for the chiral $SU(2)$ and $U(1)$ states of the Heisenberg model in Sec.~\ref{sec:heisenberg}.
\section{Classification and the Locking Theorem}
\label{sec:classification}

\subsection{Algebraic PSGs}
\label{sec:counts}

We determined the $\Z_2$ PSGs of $G_{\rm KH}\times\Z_2^\T$ with three independent methods: by computing all consistent sets of relation values for sites without site symmetry (cohomology classes on a free $G$-set) and then imposing the on-site condition (at every site the stabilizer $D_4\times\Z_2^\T$ must be represented by gauge matrices whose products reproduce the relation values, and the inequivalent on-site representations of a class are counted modulo the characters of the group); by searching all explicit PSGs in a fixed standard gauge, the canonical gauge (Appendix~\ref{app:forms}), and identifying gauge-equivalent solutions; and by starting from the homomorphisms of the site stabilizer into $SO(3)$, inducing gauge parts from them, and solving the remaining sign conditions on finite tori. Each gives
\begin{equation}
352\ \text{algebraic}\ \Z_2\ \text{PSGs}.
\label{eq:count}
\end{equation}
These carry $192$ distinct sets of relation values. For $160$ of them two PSGs share all nine labels~\eqref{eq:labels} and differ in the ratio $\zeta=G_{R_x}(i_z)/G_{R_x}(i_x)=\pm1$ between the gauge parts of $R_x$ on a site $i_z$ whose inter-triangle bond is a $z$ bond and on a site $i_x$ of the same unit cell with an $x$ bond. In these PSGs $G_{R_x}(i)=\pm\tau^0$ is central, that is, it commutes with every gauge matrix, so $\zeta$ is unchanged by gauge transformations and by the re-lifting $G_{R_x}\to-G_{R_x}$, and it is the same for every cell and every such pair of sites. We write the suffix -$b$ for $\zeta=+1$ and -$a$ for $\zeta=-1$. The $32$ unpaired sets are exactly the locked classes, $n_R=-1$. Relation values alone therefore do not label the PSGs, as on lattices with nontrivial site symmetry generally~\cite{RajIqbalSHD}. On finite tori the canonical sign functions, which contain $x(x+1)/2$, have period four, so a torus with periodic boundary conditions for the spinons, $T_1^L=T_2^L=+1$, reproduces the classification only if its linear size is a multiple of four; a $6\times6$ torus gives $176$ classes. There are $64$ $SU(2)$ PSGs. For the chiral PSGs, the PSGs of $G_{\rm KH}$ alone, the same methods give $80$ algebraic $\Z_2$ PSGs, $16$ of them locked, and $32$ $SU(2)$ PSGs; as explained in Sec.~\ref{sec:chiral}, each of the four chiral classes A--D has these $80$ PSGs. Restricted to $p6mm\times\Z_2^\T$, the same procedure gives the $80$ $\Z_2$ PSGs of the Heisenberg model, the number obtained from the SHD lattice by restriction to the site stabilizer~\cite{RajIqbalSHD}. The $U(1)$ PSGs fall into eight sectors according to whether $C_6$, $S$, and $\T$ preserve or reverse the $U(1)$ axis. Exact integer elimination on the star sites gives $256$ discrete $U(1)$ PSGs and $16$ one-parameter families with a continuous flux, the families occurring only in the sector in which only $S$ reverses the axis; an explicit search in the canonical gauge gives the same number of discrete PSGs in each of the eight sectors (Supplemental Material~\cite{SM}). The $U(1)$ PSGs play no role below.

At every site the gauge parts of $(S,R_x,R_y,\T)$ take one of six forms in the canonical gauge, which we call on-site types I to VI (Table~\ref{tab:onsite} of Appendix~\ref{app:forms}). In five of them $G_{R_x}$ and $G_{R_y}$ are $\pm\tau^0$ on the site, so the $\pi$ rotations act on the spinons only through their spin-$1/2$ lifts; in the sixth, which occurs in the $32$ locked classes, $G_{R_x}$ and $G_{R_y}$ are $\ii(\tau^1\pm\tau^2)/\sqrt2$ and the mirror acts through the faithful two-dimensional representation of the double cover of $D_4$.

\subsection{Realized ans\"atze}

\begin{table}
\caption{Number of $\Z_2$ PSGs realized by generic ans\"atze with singlet and triplet amplitudes on all symmetry-allowed bonds up to a given distance shell (cumulative; lengths in units of the nearest-neighbor bond). For the $352$ time-reversal-symmetric PSGs the two columns count ans\"atze with bond amplitudes only and with the allowed on-site multipliers in addition; for the chiral classes the number in brackets is for bonds only. A PSG is realized if the generic ansatz is connected and has IGG $\Z_2$.}
\label{tab:realized}
\begin{ruledtabular}
\begin{tabular}{cccccc}
shell & length & bonds & $+$ on-site & A, C & B, D\\
\hline
1 & $1$ & $8$ & $24$ & $26\ [18]$ & $34\ [18]$\\
2 & $1.932$ & $184$ & $184$ & $80$ & $56$\\
3 & $2.732$ & $192$ & $192$ & $80$ & $64$\\
4 & $2.909$ & $200$ & $200$ & $80$ & $72$\\
5 & $3.346$ & $264$ & $264$ & $80$ & $80$\\
6 & $3.732$ & $264$ & $264$ & $80$ & $80$\\
7 & $3.864$ & $268$ & $268$ & $80$ & $80$
\end{tabular}
\end{ruledtabular}
\end{table}

For every PSG and bond orbit we computed the space of symmetry-allowed singlet and triplet amplitudes on the reference bond (Supplemental Material~\cite{SM}) and the IGG of generic ans\"atze, which we determine as the set of gauge matrices that commute with all bond amplitudes (their commutant) on a connected patch, with an absolute singular-value threshold. We call a PSG realized up to a given range if a generic ansatz with all allowed amplitudes up to that range connects all sites and has IGG $\Z_2$, so that it describes a $\Z_2$ spin liquid; Table~\ref{tab:realized} gives the numbers of realized PSGs. In $80$ of the $352$ time-reversal-symmetric PSGs no amplitude is allowed at any range. These are the PSGs of on-site types I, III, and VI with $n_{\T C_6}=+1$, in which $G_\T(i)$ is one and the same sign times $\tau^0$ on every site, so that $\T$ maps every bond amplitude and every on-site term to its negative. Four further PSGs, Z001011000-$a$/$b$ and Z101011001-$a$/$b$, have only ans\"atze with IGG $U(1)$ up to the seventh shell; they are realized as $\Z_2$ spin liquids from the ninth shell on (length $4.625$). Up to the fourth shell, $200$ PSGs are realized with singlet and triplet amplitudes, but only $46$ with singlet amplitudes alone.

With nearest-neighbor bonds only, eight time-reversal-symmetric PSGs are realized as connected $\Z_2$ spin liquids: Z100001100-$a$/$b$, Z100000101-$a$/$b$, Z000001101-$a$/$b$, and Z000000100-$a$/$b$. Two of them, Z100001100-$b$ and Z000000100-$b$, allow purely singlet amplitudes on both orbits and are the Heisenberg PSGs Z100001(1) and Z000000(1) of Sec.~\ref{sec:heisenberg}, embedded with trivial gauge parts of $D_2$. Their duals, Z000001101-$b$ and Z100000101-$b$, have purely triplet amplitudes on both orbits, as Eq.~\eqref{eq:dual-links} requires, and the four -$a$ PSGs combine a singlet amplitude on one orbit with a triplet amplitude on the other. The eight form four dual pairs (Sec.~\ref{sec:duality}). Sixteen further PSGs have nearest-neighbor bond amplitudes with IGG $U(1)$ that an allowed on-site multiplier would break to $\Z_2$; whether such a multiplier survives the constraint equations is a self-consistency question that we leave open.

\subsection{The locking theorem}
\label{sec:theorem}

The intra-triangle amplitude is allowed in $128$ of the $352$ PSGs, but in none of the $32$ locked ones, for a reason that does not depend on the lattice.

\begin{theorem}
\label{thm:locking}
Let the Hamiltonian contain only couplings that are diagonal in a common spin frame, $\sigma^a_iJ^a_{ij}\sigma^a_j$ of any range and bond dependence, so that its symmetry group $G$ contains $D_2\times\Z_2^\T$, and let $D_2$ be a normal subgroup of $G$. Consider a $\Z_2$ spin liquid that preserves $D_2$ and $\T$ and whose PSG is locked, that is, the gauge parts of $R_x$ and $R_y$ anticommute, Eq.~\eqref{eq:locked}. Then there are a unique $d\in D_2$, invariant under conjugation by $G$, and signs $\theta(i)=\pm1$ such that $\T d$ acts on the spinons as the gauge transformation $\theta(i)\tau^0$. Every amplitude of every ansatz in the class, of any range and in any channel, obeys
\begin{equation}
u_{ij}=-\theta(i)\theta(j)\,R_d\,u_{ij},
\label{eq:thm}
\end{equation}
where $R_d$ acts on the triplet index. In addition, (i) no on-site term is allowed, for any $d$; (ii) the singlet and the $d$-triplet amplitudes join only sites with $\theta(i)\theta(j)=-1$, the two other triplet amplitudes only sites with $\theta(i)\theta(j)=+1$; (iii) $\theta(g(i))=\chi(g)\theta(i)$ for a character $\chi$ of $G$, so $\theta(i)\theta(j)$ is constant on every bond orbit. If a symmetry permutes the three Kitaev bond types cyclically, then $d=1$, every amplitude joins sites of opposite sign, and the bonds with nonzero amplitude form a bipartite graph.
\end{theorem}

The existence of $d$, Eq.~\eqref{eq:thm}, and statements (i) and (ii) use only the anticommutation of the gauge parts of $R_x$ and $R_y$ on each site; the normality of $D_2$ enters only the $G$-invariance of $d$, statement (iii), and the conclusion $d=1$ for a cyclic permutation of the colors. Normality holds for generic Kitaev--Heisenberg couplings, $G=G_{\rm KH}\times\Z_2^\T$. It fails when the couplings have a continuous spin symmetry, as for XXZ or Heisenberg couplings; these three statements then apply with $G$ replaced by the normalizer of $D_2$ in the symmetry group.

The proof proceeds as follows. The relation value $[\T,b]$ of $\T b\T^{-1}b^{-1}$, $b\in D_2$, defines a homomorphism $D_2\to\Z_2$. In a locked PSG the commutator form $[a,b]$ on $D_2$ is nondegenerate, $[a,b]=-1$ for distinct nontrivial $a,b$, so this homomorphism equals $[d,\cdot]$ for exactly one $d$. Then $[\T d,b]=[\T,b][d,b]=+1$, so the gauge part of $\T d$ commutes with $G_{R_x}(i)$ and $G_{R_y}(i)$; these generate an irreducible representation of the quaternion group, and by Schur's lemma the gauge part of $\T d$ is a scalar, $\theta(i)\tau^0$. For the $G$-invariance of $d$, note that relation values are invariant under conjugation: for $g\in G$, $gbg^{-1}\in D_2$ because $D_2$ is normal, and since $\T$ commutes with $g$, $[\T,gbg^{-1}]=[\T,b]$ and $[gag^{-1},gbg^{-1}]=[a,b]$. Hence $[g^{-1}dg,b]=[d,gbg^{-1}]=[\T,gbg^{-1}]=[\T,b]=[d,b]$ for every $b\in D_2$, and nondegeneracy gives $g^{-1}dg=d$. Equation~\eqref{eq:thm} is the invariance condition~\eqref{eq:psg} for the antiunitary element $\T d$, and (ii) follows because $R_d$ is $+1$ on the singlet and the $d$-triplet and $-1$ on the two other triplets. For (i), the on-site singlet multiplier $a_i$ and the on-site triplet field $t^a_i\tau^0$, the only Hermitian on-site triplet term, must be invariant under $D_2$. Conjugation by $G_{R_x}(i)$ and $G_{R_y}(i)$ leaves $a_i$ invariant only if $a_i$ commutes with the quaternion group, so $a_i$ is a traceless scalar and vanishes, while $D_2$ acts on $t^a_i$ through the spin rotations alone, and no nonzero vector is invariant under all three $\pi$ rotations. For $d=1$ Eq.~\eqref{eq:thm} also gives this directly; for $d\neq1$ the two triplet components orthogonal to $d$ satisfy Eq.~\eqref{eq:thm} trivially and are removed by $D_2$ alone. Since $d$ is $G$-invariant, every $g$ commutes with $\T d$ in $G$, and (iii) follows by evaluating the relation $g(\T d)g^{-1}(\T d)^{-1}$ at the site $g(i)$; its relation value is $\chi(g)$. A cyclic permutation of the colors makes $[\T,\cdot]$ invariant under the threefold permutation of $R_x$, $R_y$, $R_z$, which forces it to be trivial and hence $d=1$; without such a symmetry $d=R_z$ is also possible (Appendix~\ref{app:general}).

For $d=1$ no odd loop carries amplitude on all of its bonds, in any channel, and a bond orbit that contains an odd cycle, such as the triangle edges of the star lattice, is forbidden entirely; other bonds of odd loops may still carry amplitude. For $d=R_z$ the bipartite structure holds only for the singlet and $z$-triplet channels, and odd loops can carry $x$- and $y$-triplet amplitude on all of their bonds, as on the pentaheptite lattice (Appendix~\ref{app:general}).

\begin{corollary}
\label{cor:chiral}
Let the Hamiltonian be as in Theorem~\ref{thm:locking}. A locked $\Z_2$ spin liquid that preserves $D_2$ and has a nonzero singlet amplitude on every bond of a graph that contains an odd cycle preserves none of $\T$, $\T R_x$, $\T R_y$, $\T R_z$. It can keep only products of $\T$ with lattice operations.
\end{corollary}

On the star lattice, $d=1$ because $C_6$ permutes the colors cyclically, and $\chi$ is fixed by $\chi(T_{1,2})=1$, $\chi(S)=n_{\T S}$, and $\chi(C_6)=n_{\T C_6}$. The mirror $S$ fixes the site $\mu=1$, so $n_{\T S}=+1$ in all locked classes. If $n_{\T C_6}=+1$, $\theta$ is uniform on the single site orbit and every amplitude vanishes; this is the case for $16$ of the $32$ locked PSGs (Appendix~\ref{app:locked}, Table~\ref{tab:locked}). If $n_{\T C_6}=-1$, $\theta(x,y,\mu)=(-1)^\mu$, the triangle edges and the $J_{3b}$ bonds join sites of the same sign and are forbidden, and the $\Je$, $J_2$, $J_{3a}$, and $J_4$ bonds are allowed by $\T$. Ten of these $16$ PSGs allow a $\Je$ amplitude. Odd loops therefore still carry amplitude on some of their bonds: a $J_2$ bond closes a triangle with one $\Jt$ and one $\Je$ bond, and that triangle keeps its $\Je$ and $J_2$ amplitudes. Because $d=1$ on the star lattice, the triangle edges carry no amplitude in any channel in a locked time-reversal-symmetric class. The Kitaev ansatz has a nonzero amplitude on every nearest-neighbor bond, including the triangle edges (Sec.~\ref{sec:kitaev}), so a Kitaev-type spin liquid on the star lattice must break time reversal as long as it is described by a locked $\Z_2$ PSG, in agreement with the exact solution~\cite{YaoKivelson2007}; what survives is a chiral PSG. This uses statement (ii) for $d=1$ rather than Corollary~\ref{cor:chiral}, whose assumption of a nonzero singlet amplitude fails for the Kitaev ansatz of Sec.~\ref{sec:kitaev-psg} at $\lambda_{ij}=K_{ij}$, where the singlet amplitude vanishes.

Related arguments are known for the exactly solvable points, where the flux operators of odd plaquettes have eigenvalues $\pm\ii$, which time reversal interchanges, so every flux eigenstate breaks time reversal~\cite{Kitaev2006,YaoKivelson2007,Peri2020,Cassella2023}; this uses the conserved fluxes of the Kitaev model. In the PSG of the honeycomb Kitaev spin liquid, time reversal acts with opposite gauge signs on the two sublattices, so that second-neighbor couplings are odd under $\T$~\cite{YouKimchiVishwanath2012}. In three-dimensional Kitaev models on bipartite lattices time reversal acts on the fermions as a sign that alternates between the two sublattices, and on the $(9,3)a$ lattice, whose elementary loops have odd length, $\T$ is broken spontaneously~\cite{OBrienHermannsTrebst2016}. Kimchi and Vishwanath noted that odd cycles force a Majorana mean-field state of the Kitaev type to break $\T$, possibly keeping its product with a lattice rotation~\cite{KimchiVishwanath2014}. Theorem~\ref{thm:locking} places these observations at the level of the PSG class: it applies to every ansatz of every locked $\Z_2$ PSG, with any range, channel, and Lagrange multipliers, and without exact solvability. It also gives the channel rule for $d\neq1$ and the orbit structure through $\chi$. That time reversal acts as a site sign is not special to locked classes: in the non-locked PSGs of on-site types I and III (Table~\ref{tab:onsite}), $G_\T(i)=\pm\tau^0$ on every site as well. The theorem shows that locking forces this structure, with $\T$ replaced by $\T d$, in every locked class; in gauge-invariant terms, for $d=1$ Eq.~\eqref{eq:thm} makes the Wilson loop of every odd loop, the product of the amplitudes around it, odd under $\T$, so that it vanishes. Appendix~\ref{app:general} applies the theorem to the honeycomb, square-octagon, and pentaheptite lattices.
\section{Kitaev Limit: One Chiral PSG for Abelian and Non-Abelian Liquids}
\label{sec:kitaev}

\subsection{Exact solution}

\begin{figure}
\centering
\includegraphics[width=\columnwidth]{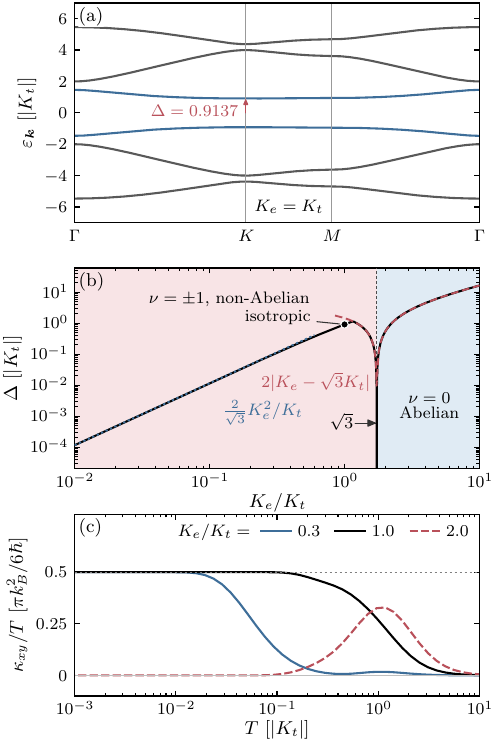}
\caption{Kitaev limit for $\Kt,\Ke>0$, in units of $|\Kt|$. (a) Majorana bands at $\Ke=\Kt$ along $\Gamma$--$K$--$M$--$\Gamma$ in the ground flux sector (triangle flux $+\pi/2$, dodecagon flux $\pi$), with the gap $0.9137$ at $K$; the labels of the momenta depend on the gauge. (b) Single-Majorana gap as a function of $\Ke/\Kt$, with the asymptote $(2/\sqrt3)\Ke^2/\Kt$ at small $\Ke$ and the gap at $\Gamma$, $2|\Ke-\sqrt3\Kt|$, which is the full gap from $\Ke/\Kt\approx1.155$ on. The gap closes at a single Majorana cone at $\Ke/\Kt=\sqrt3$, which separates the non-Abelian phase with Chern number $|\nu|=1$ from the Abelian phase with $\nu=0$; the dot marks the isotropic point of the KH model. All states in both phases, for either sign of $\Kt$ and $\Ke$, have the chiral PSG \YK\ of class B. (c) Free-Majorana thermal Hall conductivity $\kappa_{xy}/T$ in units of $\pi k_B^2/(6\hbar)$ (Kubo formula, Bloch phases with the physical site positions) in the $+\pi/2$ sector for $\Ke/\Kt=0.3$ and $1$ ($\nu=+1$) and $2$ ($\nu=0$); the dotted line is the plateau $+1/2$. Thermally excited fluxes, not included here, limit the plateau to temperatures well below the flux gaps~\cite{NasuMotome2015}.}
\label{fig:kitaev}
\end{figure}

At $J=0$ we write $H_K=-\sum_{\langle ij\rangle_a}K_{ij}\sigma^a_i\sigma^a_j$ with $K_{ij}=\Kt$ on the triangles and $\Ke$ between them. With $\sigma^a_j=\ii b^a_jc_j$ and the conserved bond variables $\mathsf{u}_{jk}=\ii b^a_jb^a_k=\pm1$, the Hamiltonian becomes $H_K=\frac{\ii}{4}\sum_{jk}A_{jk}c_jc_k$ with $A_{jk}=2K_{jk}\mathsf{u}_{jk}$~\cite{Kitaev2006}. We define the flux through a plaquette $p$ as the phase of the cyclic product of the matrix elements of $\ii A$ along its counterclockwise boundary; for a triangle with vertices $1,2,3$ in counterclockwise order, $\Phi_{\rm tri}=\arg[(\ii A)_{12}(\ii A)_{23}(\ii A)_{31}]$. Every triangle carries a flux $\pm\pi/2$ for every configuration of the bond variables (gauge sector), since it is a product of three imaginary numbers, and every dodecagon a flux $0$ or $\pi$. In terms of spin operators, $\Phi_{\rm tri}=\frac{\pi}{2}\,\mathrm{sgn}(\Kt)\,W_{\rm tri}$ with $W_{\rm tri}=\sigma^{a_1}_1\sigma^{a_2}_2\sigma^{a_3}_3$, where $a_k$ is the color of the inter-triangle bond at vertex $k$. Energies are given in units of $|K|$ at $\Ke=\Kt=K$ (as in Sec.~\ref{sec:pt}, with $|K|=1$) and of $|\Kt|$ when $\Ke/\Kt$ varies (Fig.~\ref{fig:kitaev}); Secs.~\ref{sec:meanfield} and~\ref{sec:vmc} use the normalization $|K|=2$ of Eq.~\eqref{eq:HKH} at the Kitaev points.

We enumerated all flux patterns of the infinite lattice that admit a periodic bond-variable gauge on supercells with up to four unit cells ($14{,}960$ distinct patterns, $15{,}460$ counted over the $15$ supercells), with energies integrated over the Brillouin zone of each supercell, at nine ratios $\Ke/\Kt$ between $0.1$ and $6$, and for $24$ sets of anisotropic triangle couplings. For all these couplings the lowest-energy pattern of this set has the same flux on all triangles, $+\pi/2$ or $-\pi/2$, and $\pi$ through every dodecagon, in agreement with Yao and Kivelson~\cite{YaoKivelson2007}. The margin to the next pattern is smallest at the ends of this range, $1.1\times10^{-5}|\Kt|$ per site at $\Ke/\Kt=0.1$ and $9\times10^{-8}|\Kt|$ per site at $\Ke/\Kt=6$. On finite tori the lowest flux sector, the set of gauge sectors with a given flux pattern, can differ; on the $2\times2$ torus of Sec.~\ref{sec:vmc} the triangle fluxes alternate. Lieb's bipartite-lattice theorem~\cite{Lieb1994} does not directly apply to the star lattice. For mirror-symmetric coupling magnitudes, its reflection argument has a single-plane Majorana counterpart~\cite{JaffePedrocchi2015}: the mirrors $\sigma_d$, which contain no sites, cut disjoint inter-triangle bonds, whose coefficients can be gauged to a common sign, as required by the reflection bound. Reflecting one half of the hopping problem can then produce a minimizing configuration symmetric under that reflection. A bisected dodecagon consists of two reflected five-edge paths and two crossing bonds, whose hopping-phase product is negative, so such a configuration has flux $\pi$ on these dodecagons. No site-free mirror crosses a triangle. This single-plane argument supports the computed dodecagon pattern; it does not by itself prove simultaneous minimization for all reflection planes or establish the ground flux sector for generic anisotropic couplings. The mirrors and $\T$ reverse the triangle flux, so each of the two ground states breaks $\T$ and all mirrors and keeps every product of a mirror with $\T$.

At the isotropic point $\Ke=\Kt$, the ground-state energy is $E_0=-0.7646302546|K|$ per site, consistent with the iDMRG energy of about $-0.77$ at $\phi=\pi/2$ in Fig.~7(a) of Ref.~\cite{dOrnellasKnolle2024}, whose energies are half of those in the units of Eq.~\eqref{eq:HKH} (it shows $-0.75$ at the FM Heisenberg point, where Eq.~\eqref{eq:HKH} gives $-1.5$), and the Majorana gap is $0.91370|K|$ [Fig.~\ref{fig:kitaev}(a)]. For nonzero couplings, as a function of $|\Ke|/|\Kt|$ the gap closes only at $|\Ke|/|\Kt|=\sqrt3$, at a single Majorana cone whose momentum depends on the gauge [Fig.~\ref{fig:kitaev}(b)]. For $0<|\Ke|/|\Kt|<\sqrt3$, a range that contains the isotropic point and the regime of dominant intra-triangle coupling, the Chern number of the Majorana bands is $\nu=\pm1$ and the phase is the non-Abelian Ising phase~\cite{Kitaev2006}; for $|\Ke|/|\Kt|>\sqrt3$, $\nu=0$ and the phase is Abelian with the anyons of the toric code~\cite{YaoKivelson2007,Kells2010}. On $L\times L$ tori with $L=6$ to $12$ and $\Kt\Ke>0$, and for even $L$ for either sign, of the Majorana vacua of the four holonomy sectors, which differ in the $\Z_2$ fluxes through the two noncontractible loops of the torus, three are physical (survive the projection onto physical states) in the non-Abelian phase, while the fourth requires a Majorana excitation; in the Abelian phase all four vacua are physical. Their energies approach threefold and fourfold degeneracy, respectively, in the thermodynamic limit for each chirality, with finite-size differences between holonomy sectors; for $\Kt\Ke<0$ and odd $L$ the odd number $3L^2$ of reversed inter-triangle bond variables changes the physical fermion parity, and in the non-Abelian phase the fourth sector becomes the unique lowest one. For $0<|\Ke|\ll|\Kt|$ the gap is $(2/\sqrt3)\Ke^2/|\Kt|$, the second-order mass of the triangle zero modes derived by Dusuel \emph{et al.}\ in the isolated-triangle limit~\cite{Dusuel2008}. The flux excitations are cheap in the same regime: at $\Ke/\Kt=0.1$ the lowest flux excitations, created by flipping one bond variable (a triangle and a dodecagon, or two dodecagons), cost $0.0093|\Kt|$ and $0.0096|\Kt|$ on a $20\times20$ torus ($0.0108|\Kt|$ and $0.0111|\Kt|$ on $12\times12$), comparable to the Majorana gap of $0.0115|\Kt|$. At the isotropic point a single vison (flipped flux) on a dodecagon or a triangle, defined as half the energy of a well-separated pair on a $12\times12$ torus, costs about $0.13|K|$ or $0.16|K|$.

For general couplings $J_x,J_y,J_z$ on the triangle edges and $Z$ on the inter-triangle bonds, we find that the exact phase diagram has, besides the critical sphere of Kells \emph{et al.}~\cite{Kells2010}, an Abelian region at strongly anisotropic triangles, $J_a^2>J_b^2+J_c^2+Z^2$, which Ref.~\cite{Kells2010} assigns to the non-Abelian phase and which for $Z\to0$ is the Abelian phase of the effective honeycomb model of the triangle zero modes, the anisotropic generalization of the model of Ref.~\cite{Dusuel2008} (Appendix~\ref{app:remarks}). It does not intersect the family of isotropic triangles studied here.

The signs of the couplings can be absorbed into the bond variables in the unprojected Majorana spectrum; the projection onto physical states on a finite torus imposes an additional fermion-parity condition. Reversing $\Kt$ is equivalent to $\mathsf{u}\to-\mathsf{u}$ on all triangle edges, which reverses the triangle fluxes and leaves the dodecagons unchanged, and reversing $\Ke$ changes no flux. On the $12$-site tori (the three index-two supercells, related by $C_6$) the complete spectra of all four sign combinations of $(\Kt,\Ke)$ coincide to $10^{-12}$, that is, $H_K$ and $-H_K$ are isospectral there; on tori with an odd number of inter-triangle bonds, such as the $6$-site torus, reversing $\Ke$, which flips an odd number of bond variables, changes the physical fermion parity and breaks the isospectrality, whereas reversing $\Kt$ does not. In the non-Abelian phase, the FM and AFM Kitaev liquids differ in the relation between the spin chirality and the Chern number, $\nu=\mathrm{sgn}(\Kt)\,W_{\rm tri}$; in the Abelian phase $\nu=0$ for either chirality. Here $\nu$ is the Chern number of the three negative-energy Majorana bands in the convention of Ref.~\cite{Kitaev2006}. Explicitly, with their projector $P(\bm k)$ and the positively oriented coordinates dual to $(\bm T_1,\bm T_2)$, $\nu=(2\pi\ii)^{-1}\int_{\rm BZ}\Tr\{P[\partial_{k_1}P,\partial_{k_2}P]\}\,\mathrm d^2k$; it is the negative of the Berry-curvature integral defined with connection $\mathcal A=\ii\langle u|\nabla_{\bm k}u\rangle$. In this convention $\kappa_{xy}/T=\frac{\nu}{2}\frac{\pi k_B^2}{6\hbar}$, and $\nu=+1$ in the non-Abelian $+\pi/2$ sector, where the edge mode circulates counterclockwise. The nearest-neighbor correlations at the isotropic point are $\langle\sigma^a_i\sigma^a_j\rangle=\mathrm{sgn}(K)\,0.4817$ on the triangles and $\mathrm{sgn}(K)\,0.5658$ between them, all other two-spin correlations vanishing as required by the conserved fluxes~\cite{BaskaranMandalShankar2007}. Each edge of a ribbon in the non-Abelian phase carries one net chiral Majorana mode, which circulates in the sense of the triangle flux, and the free-Majorana thermal Hall conductivity shows the plateau $\kappa_{xy}/T=\frac{\nu}{2}\frac{\pi k_B^2}{6\hbar}$ below about a tenth of the gap. Thermally excited fluxes limit this plateau: in quantum Monte Carlo simulations the quantized value appears only below the transition at which time reversal is broken~\cite{NasuMotome2015}; in the regime of dominant intra-triangle coupling the flux energies and the Majorana gap are comparable, so the quantized low-temperature limit requires temperatures below both scales as well as below the chirality-ordering transition. In the Abelian phase the in-gap edge states depend on the termination and have zero net chirality.

\subsection{Parton description and PSG}
\label{sec:kitaev-psg}

The Kitaev Majoranas are linear combinations of the Abrikosov fermions~\cite{BurnellNayak2011}. We use
\begin{align}
b^x_i&=f^\dagger_{i\uparrow}+f_{i\uparrow},& b^y_i&=-\ii(f^\dagger_{i\uparrow}-f_{i\uparrow}),\nonumber\\
b^z_i&=-(f^\dagger_{i\downarrow}+f_{i\downarrow}),& c_i&=\ii(f^\dagger_{i\downarrow}-f_{i\downarrow}),
\label{eq:majorana-map}
\end{align}
for which the projector onto single occupancy is $(1+b^x_ib^y_ib^z_ic_i)/2$ and $\ii b^a_ic_i$ equals $f^\dagger_i\sigma^af_i$ on the physical states. The identification given in Eqs.~(10) and~(A1) of Ref.~\cite{BurnellNayak2011} differs from Eq.~\eqref{eq:majorana-map} by an orthogonal transformation of determinant $-1$ of the four Majoranas; with it the projected states are Kitaev's eigenstates up to the global spin rotation $\exp(-\ii\frac{\pi}{2}\sum_jS^z_j)$, as used for the SHD lattice~\cite{RajIqbalSHD}, whereas with Eq.~\eqref{eq:majorana-map} no rotation is needed (Appendix~\ref{app:remarks}). Decoupling every bond term as $\sigma^a_i\sigma^a_j=-(\ii b^a_ib^a_j)(\ii c_ic_j)$ gives a quadratic Hamiltonian in which the $c$ Majoranas hop with amplitude $K_{ij}\mathsf{u}_{ij}$ and the $b^a$ Majoranas of each $a$ bond form a dimer with an arbitrary positive amplitude $\lambda_{ij}$. In the link matrices of Eq.~\eqref{eq:links}, the singlet part of an $a$ bond is pure hopping, $u^{(0)}_{ij}\propto(K_{ij}-\lambda_{ij})\ii\tau^0$, and each triplet channel $b$ is locked to the gauge direction $\tau^b$, with amplitude proportional to $K_{ij}-\lambda_{ij}$ for $b=a$ and to $K_{ij}+\lambda_{ij}$ for $b\neq a$. The Bogoliubov spectrum consists of the exact $c$-Majorana bands and flat $b$-dimer bands, and its Gutzwiller projection, when nonzero, is an exact Kitaev eigenstate in that flux sector.

Solving Eq.~\eqref{eq:psg} directly for this ansatz in the ground flux sector gives the same symmetry content for $\Ke/\Kt=0.3$, $1$, $\sqrt3$, and $2.5$, both chiralities, both signs of $K$, mixed signs, and two choices of $\lambda$. The translations, $C_6$, and $D_2$ are realized unitarily, every mirror only in combination with $\T$, and $\T$ itself is broken. The IGG is $\Z_2$. The state therefore belongs to chiral class B, and its PSG is
\begin{align}
\text{ZC011110}:\quad &n_{T_2}=n_{TR}=+1,\nonumber\\
&n_S=n_{SC_6}=n_{C_6}=n_R=-1.
\label{eq:yk-psg}
\end{align}
Among all chiral PSGs, the ones whose allowed amplitudes contain this ansatz carry the same label. The allowed space of \YK\ in class B contains, on each nearest-neighbor orbit, one singlet and two triplet directions and no on-site term; the singlet part alone has IGG $SU(2)$, and the color-locked triplets reduce it to $\Z_2$.

By Eq.~\eqref{eq:yk-psg}, the Abelian and the non-Abelian phases, the critical point between them, both signs of the Kitaev coupling, and both chiralities have one PSG. Their transition is a band inversion at a single Majorana cone within chiral class B, the only class in which a Chern number is allowed (Sec.~\ref{sec:chiral}). The single cone is a property of the band structure~\cite{YaoKivelson2007,Kells2010}; by the PSG, both sides of it carry the same projective action on the fermion. In the non-Abelian phase this action is not independent of the topological order, since for on-site symmetries the possible symmetry actions on the anyons (fractionalization classes) of Ising topological order are detected only by the non-Abelian anyon $\sigma$~\cite{Barkeshli2019}. On the honeycomb lattice the gapless phase of Kitaev's model is in class (I)(B) of Ref.~\cite{YouKimchiVishwanath2012}, whereas the gapped Abelian phases require anisotropic couplings that lower the lattice symmetry; on the star lattice the Abelian phase is reached at full symmetry, and the transition changes a Chern number within one chiral class. The two phases differ in the band topology of the spinon, and with it in the vison sector, threefold versus fourfold torus degeneracy. The PSG does not determine the topological order: within one PSG the Chern number $\nu$ of the Majorana bands can change at a gap closing, and the anyons follow from $\nu \bmod 16$ through Kitaev's sixteenfold way~\cite{Kitaev2006}, Ising anyons for $\nu=\pm1$ and toric-code anyons for $\nu=0$. The PSG allows $\nu\neq0$ only in class B, and it does not fix the symmetry action on the $\sigma$ anyon of the non-Abelian phase. The PSG does constrain how $\nu$ can change. Since $C_6^6=-1$ on the Majoranas in \YK, the $C_6$ eigenvalues are $e^{\ii\pi(2m+1)/6}$, and the $C_6$, $C_3$, and $C_2$ eigenvalues of the three occupied Majorana bands at $\Gamma$, $K$, and $M$ determine $\nu\bmod6$ through the rotation-eigenvalue formula for the Chern number of Ref.~\cite{FangGilbertBernevig2012}. For the eigenvalues $\xi_n$, $\zeta_n$, and $\theta_n$ of the active counterclockwise rotations $C_6$, $C_3$, and $C_2$, respectively, on the occupied Bloch states (sewing-matrix eigenvalues), our Chern convention gives $e^{-2\pi\ii\nu/6}=-\prod_{n=1}^3\xi_n(\Gamma)\zeta_n(K)\theta_n(M)$: the minus sign is $(-1)^{3F}$ with $F=1$, and the exponent accounts for the Berry-sign convention above. The computed eigenvalues satisfy this relation in both phases. In the translation-invariant gauge in which $C_6$ acts without a momentum shift, the single cone at $\Ke=\sqrt3\Kt$ lies at $\Gamma$ (in the other three translation-invariant gauges at one of the $M$ points) and exchanges a particle--hole pair of $C_6$ eigenvalues about the dodecagon center, $e^{\pm\ii\pi/6}$ for one sign of the lift of $C_6$ and $e^{\pm5\ii\pi/6}$ for the other, between occupied and empty bands, which changes $\nu$ by one. For $C_6^6=+1$ such an exchange of particle--hole partners at $\Gamma$ could change $\nu$ only by an even number, so the single-cone transition relies on the projective sign $n_{C_6}=-1$ of \YK.

In a translation-invariant gauge, every gauge part $G_g$ of \YK\ equals, up to a $\Z_2$ sign field, the $SU(2)$ lift of the spin rotation $R_g$ mapped into gauge space by a fixed isomorphism, for the threefold rotation $C_3^{[111]}$ and the mirror rotation $C_2^{[1\bar10]}$ as well as for $D_2$. On the SHD lattice the Kitaev coloring is invariant under all lattice operations and only $D_2$ is locked~\cite{RajIqbalSHD}; here the lattice operations permute the colors, and the locking extends to the whole point group. Combining spin lifts and gauge parts, every relation acts on the four Majoranas of a site as the identity except $C_6^6=-1$, for a suitable choice of the signs of the gauge parts. This sign is the $\pi$ flux through the dodecagon at the rotation center: the $c$ Majoranas carry half-odd-integer angular momentum about it, as on the SHD lattice. Unlike there, the mirrors square to $+1$ on the Majoranas and the combination of $C_2$ with $\T$ is broken, so there is no symmetry-enforced double degeneracy throughout the Brillouin zone; isolated band crossings can still occur.

The allowed space of \YK\ contains a pure-singlet nearest-neighbor hopping family proportional to $\ii\tau^0$, with flux $\pi/2$ through every triangle and $\pi$ through every dodecagon. This family is SL$[\frac\pi2,\frac\pi2,\pi]$ of Ref.~\cite{YangParamekantiKim2010}, with IGG $SU(2)$ and PSG SU0001 in class B (Sec.~\ref{sec:heisenberg}). Its branch with Chern number $C=\pm1$ per spin is the Kalmeyer--Laughlin candidate, with semion topological order expected if the projected state remains in the corresponding gapped phase. Adding the color-locked triplet amplitudes reduces the gauge group to $\Z_2$ and ties the gauge rotations to the spin rotations, giving the Yao--Kivelson family as a descendant in the space of ans\"atze. When the singlet hopping amplitudes are chosen proportional to the Kitaev matter hoppings, their $c$-Majorana band structures agree; the common flux pattern alone does not fix those amplitudes or their band topology. In particular, the pure-singlet family also has a $C=0$ branch, to which the semion identification does not apply. The parent family lies in the allowed space of \YK\ where the $c$ Majoranas, the $b^a$ Majoranas of the bond color $a$ (on-color), and the two other $b$ Majoranas (off-color) of each bond hop with equal amplitude. This embedding need not coincide with simply deleting the triplets of an arbitrary exact Kitaev representative, whose singlet amplitudes depend on $\lambda_{ij}$.

\subsection{Comparison with the honeycomb lattice}

On the honeycomb lattice the same construction gives $256$ $\Z_2$ PSGs of $G_{\rm KH}\times\Z_2^\T$, with $D_2$ included, and the exact Kitaev state lies in a locked class in which, by Theorem~\ref{thm:locking}, time reversal forbids second-neighbor amplitudes; the lattice is bipartite and the state keeps $\T$. You, Kimchi, and Vishwanath classified the PSGs of $p6mm\times\Z_2^\T$ with spin--orbit-coupled amplitudes on the honeycomb lattice and identified the Kitaev state in their class (I)(B)~\cite{YouKimchiVishwanath2012}. Our classification gives the same identification; remarks on the parametrization of their Appendix~A are collected in Appendix~\ref{app:remarks}.
\section{Klein Duality as a PSG Automorphism}
\label{sec:duality}

\subsection{The duality}

Let $P(x,y)=R_x^{\,x}R_y^{\,y}$ depend on the parity of the cell coordinates, and let $U=\prod_jV_j$, where $V_j$ is the spin-$1/2$ lift of $P$ on the cell of site $j$ [Fig.~\ref{fig:lattice}(b)]. Every nearest-neighbor bond of color $c$ joins cells with $P_iP_j=R_c$, so $U$ leaves $\sigma^c_i\sigma^c_j$ invariant and reverses the two off-color products, $UH(K,J)U^\dagger=H(K,-J)$~\cite{dOrnellasKnolle2024}. In the language of Kimchi and Vishwanath this is a Klein duality, which exists if and only if every cycle of the lattice contains equal numbers of $x$, $y$, and $z$ bonds modulo two~\cite{KimchiVishwanath2014}; on the honeycomb lattice it is the four-sublattice rotation of Chaloupka, Jackeli, and Khaliullin~\cite{ChaloupkaJackeliKhaliullin2010,ChaloupkaKhaliullin2015}, which Khaliullin had applied to the triangular lattice~\cite{Khaliullin2005}. On even tori the pattern is unique up to a global element of $D_2$, and the smallest compatible torus has $2\times2$ cells. In terms of $\phi$, as on the honeycomb lattice~\cite{ChaloupkaJackeliKhaliullin2013},
\begin{equation}
H(\tilde\phi)=\lambda\,UH(\phi)U^\dagger,\qquad \tan\tilde\phi=-\tan\phi-1,
\label{eq:dualphi}
\end{equation}
with $\cos\phi\cos\tilde\phi<0$ and $\lambda=|\cos\tilde\phi/\cos\phi|$. The map is an involution of the circle that reverses its orientation. Its only fixed points are the Kitaev points $\phi=\pi/2$ and $3\pi/2$: as Kimchi and Vishwanath showed for the KH family on general lattices, the two pure Kitaev Hamiltonians are the only self-dual ones, and at these points the duality is an additional Klein $\Z_2$ symmetry~\cite{KimchiVishwanath2014}. A point with $K=0$ is mapped to $\phi+\pi$, for instance $0.85242\pi\leftrightarrow1.85242\pi$. The map takes the FM Heisenberg point $\phi=0$ onto the exactly solvable stripy point $3\pi/4$, and the AFM Heisenberg point $\phi=\pi$ onto $7\pi/4$, both with $\lambda=1/\sqrt2$. At these two points the model has the hidden $SU(2)$ symmetry generated by $\tilde{\bm{S}}=\frac12\sum_jP_j\bm{\sigma}_j$.

In ED on a $12$-site open cluster that contains all four cell parities, all $4096$ levels at $\phi$ and $\tilde\phi$ agree after rescaling to $4\times10^{-13}$, and $[H,\tilde{\bm{S}}]$ vanishes to $10^{-14}$ at $3\pi/4$ and $7\pi/4$. On the $24$-site $2\times2$ torus, $E_0(\pi)=-44.833293453722$ and $E_0(7\pi/4)=E_0(\pi)/\sqrt2$ to $3\times10^{-13}$; the ground state at $7\pi/4$ is a singlet of $\tilde{\bm{S}}$.

\subsection{Action on the PSGs}

These statements about the coupling circle follow from Ref.~\cite{KimchiVishwanath2014}, with the angle form of Ref.~\cite{ChaloupkaJackeliKhaliullin2013}. Below we work out the action of the duality on the PSGs: the automorphism~\eqref{eq:automorphism}, the label rule~\eqref{eq:labelrule}, the identification of the self-dual classes with the locked ones, Theorem~\ref{thm:duality}, and the duals of specific PSGs. Parton mean-field theories of the honeycomb KH model in the rotated basis, with triplet decoupling channels, were constructed by Schaffer, Bhattacharjee, and Kim, who noted that a PSG classification with triplet channels was still open~\cite{SchafferBhattacharjeeKim2012}; here the rotation is applied to the classification itself.

Since $U$ acts on the spin index only, it maps an ansatz $u$ to an ansatz $Uu$ with the same IGG, the same spinon spectrum, and the same Chern numbers, but with different symmetry labels. On a bond whose cells have the same parity, the singlet stays a singlet and the triplets change sign according to $P$. On a bond with $P_iP_j=R_c$, which includes all nearest-neighbor bonds and the $J_2$ and $J_4$ bonds, the singlet and the $c$ triplet are exchanged,
\begin{equation}
u'^{(c)}_{ij}=\epsilon_{ij}\,\ii u^{(0)}_{ij},\qquad u'^{(0)}_{ij}=\epsilon'_{ij}\,\ii u^{(c)}_{ij},
\label{eq:dual-links}
\end{equation}
with signs $\epsilon_{ij},\epsilon'_{ij}=\pm1$ fixed by the spin lifts $V_i$ and $V_j$ of the two cells: for $V_i=\one$ and $V_j=\ii\sigma^c$ both are $-1$, whereas for $V_i=\ii\sigma^a$ and $V_j=\ii\sigma^b$ with $R_aR_b=R_c$ they are opposite, a difference that no choice of lifts removes. The two off-color triplets are rotated into each other. A Heisenberg ansatz with singlet amplitudes only is thus mapped to an ansatz of color-locked triplet amplitudes on the nearest-neighbor bonds.

If $u$ is invariant under $g$ with gauge part $G_g$, then $Uu$ is invariant under $\varphi(g)=UgU^{-1}$ with the same gauge part up to a sign field, the sign of the spin double cover. The induced map on the symmetry group is
\begin{equation}
\varphi:\ T_1\mapsto T_1R_x,\quad T_2\mapsto T_2R_y,\quad C_6,S,R_x,R_y,\T\ \text{fixed},
\label{eq:automorphism}
\end{equation}
an involutive automorphism of $G_{\rm KH}\times\Z_2^\T$ that maps all $26$ relations to identities. It is outer, since no conjugation adds a spin rotation to a translation. Expressing the relation values of the dual PSG through those of the original PSG gives the label rule
\begin{equation}
n_{T_2}'=-n_Rn_{T_2},\qquad n_{TR}'=-n_Rn_{TR},
\label{eq:labelrule}
\end{equation}
with the other seven labels and the suffix unchanged. To derive it, note that conjugation by $U$ maps the operator that implements $\varphi(g)$ on $u$, the product of its spin lift and its gauge part, onto the operator that implements $g$ on $Uu$. A relation $r$ of the group, that is, a product of generators equal to the identity $e$ in $G_{\rm KH}\times\Z_2^\T$, therefore has in the dual PSG the full value, spin lifts included, of the product $\varphi(r)$ in the original PSG, obtained by replacing each generator in $r$ by its image under $\varphi$. The labels~\eqref{eq:labels} are gauge values, the full value divided by the value of the spin lifts. The words of the seven labels without a translation are fixed by $\varphi$. For $n_{T_2}$ and $n_{TR}$ the spin lifts of $r$ give $+1$, since the translations carry no spin rotation, so the dual label is the full value of $\varphi(r)$. Writing $[a,b]$ for the full value of $aba^{-1}b^{-1}$, which is central, so that $[ab,c]=[a,c][b,c]$,
\begin{align}
n_{TR}'&=[T_1R_x,R_y]=[T_1,R_y]\,[R_x,R_y]=n_{TR}\cdot n_R\cdot(-1),\nonumber\\
n_{T_2}'&=[T_2R_y,T_1R_x]=[T_2,T_1]\,[T_2,R_x]\,[R_y,T_1]\,[R_y,R_x]\nonumber\\
&=n_{T_2}\cdot n_{TR}^2\cdot n_R\cdot(-1).
\label{eq:labelrule-derivation}
\end{align}
Here $[T_2,R_x]=[T_1,R_y]=n_{TR}$, as forced by the relations of $G_{\rm KH}$ (Sec.~\ref{sec:parton}), and $[R_y,T_1]=[T_1,R_y]^{-1}=n_{TR}$; the gauge parts of $R_x$ and $R_y$ commute for $n_R=+1$ and anticommute for $n_R=-1$, so their commutator is $n_R$; and the spin lifts give $(\ii\sigma^x)(\ii\sigma^y)(\ii\sigma^x)^{-1}(\ii\sigma^y)^{-1}=-1$. The factor $-1$ in Eq.~\eqref{eq:labelrule} is thus the commutator of the spin lifts: the dual translations $T_1R_x$ and $T_2R_y$ anticommute on a spin-$1/2$ spinon unless the gauge parts of $D_2$ anticommute as well, $n_R=-1$, and cancel the sign. For the ans\"atze of the $200$ time-reversal-symmetric $\Z_2$ PSGs realized up to the fourth shell and for $120$ chiral ans\"atze, the PSG of the image under $U$, mapped bond by bond, agrees with Eq.~\eqref{eq:labelrule}.

The duality therefore permutes the PSGs. Of the $352$ time-reversal-symmetric $\Z_2$ PSGs, exactly the $32$ locked ones are self-dual, and the other $320$ form $160$ dual pairs. Of the $80$ chiral PSGs, the $16$ locked ones are self-dual and the others form $32$ pairs; the chiral class is preserved, because $\varphi$ fixes $C_6$ and $S$. No $SU(2)$ PSG is self-dual, since its gauge parts of $D_2$ are central; the same argument applies to $U(1)$. More generally, self-duality requires locking under the following condition.

\begin{theorem}
\label{thm:duality}
Let a Klein duality $U$ normalize the symmetry group $G$, and let $\varphi(g)=gd_g$ with $d_g\in D_2$ be the induced automorphism. If some $t\in G$ with $d_t\neq1$ and some $b\in D_2\setminus\{1,d_t\}$ satisfy $tbt^{-1}b^{-1}=e$ in $G$, then every self-dual $\Z_2$ PSG is locked.
\end{theorem}

The proof applies the argument of Eq.~\eqref{eq:labelrule-derivation} to the relation $tbt^{-1}b^{-1}$; since $U$ is a product of on-site elements of the abelian group $D_2$, $\varphi(b)=b$, and the gauge value of the relation changes by $[d_t,b]$ times the commutator $-1$ of the spin lifts of two distinct nontrivial elements of $D_2$; self-duality requires $[d_t,b]=-1$ for a nontrivial $b\neq d_t$, so the commutator form on $D_2$ is nonzero and hence nondegenerate. On the star lattice $t=T_1$, $d_t=R_x$, and $b=R_y$. The converse is not general: on the square-octagon lattice only $512$ of the $2048$ locked time-reversal-symmetric classes are self-dual, and on the honeycomb lattice restricted to the group $p6$ with $D_2$ only $16$ of $32$ (Appendix~\ref{app:general}). For the full groups of the star and honeycomb lattices all $32$ locked classes are self-dual.

Self-duality is a property of the class. A generic ansatz in a locked class is mapped to an ansatz in the same class that is in general not gauge equivalent to it; random ans\"atze of all realized locked classes, time-reversal symmetric or chiral, including random nearest-neighbor ans\"atze of \YK, are all mapped to gauge-inequivalent ans\"atze. Only where $U$ is a symmetry of the Hamiltonian, at the Kitaev points, is the state itself constrained. There the ground state is degenerate, between the two chiralities and, on a torus, between holonomy sectors, but $U$ leaves every bond operator $\sigma^a_i\sigma^a_j$ of the Kitaev model invariant and hence commutes with every plaquette and loop operator; it acts within each flux and holonomy sector and maps the ground state of a sector onto itself up to a phase when that state is unique. The Yao--Kivelson ansatz is invariant: $Uu$ is gauge equivalent to $u$, and the triangle fluxes are invariant. It follows that if a $\Z_2$ spin liquid at a Kitaev point is symmetric and mapped by $U$ to a state with the same PSG, its PSG is locked: spin-gauge locking of Kitaev spin liquids follows from the Klein duality alone, without the exact solution. Moreover, because Eq.~\eqref{eq:dualphi} reverses the orientation of the circle and has only the Kitaev points as fixed points, any interval of $\phi$ that is mapped onto itself contains $\pi/2$ or $3\pi/2$; conversely, Kimchi and Vishwanath observed that the self-dual Kitaev Hamiltonians lie in the interior of a phase rather than on a phase boundary, rigorously on the honeycomb lattice and within the Luttinger--Tisza approximation for classical ground states on the other lattices they studied, so that such a phase is mapped onto itself~\cite{KimchiVishwanath2014}. Together with the PSG result, a connected phase interval mapped onto itself must contain a Kitaev point and, if it is a $\Z_2$ spin liquid, be described by a locked PSG. The converse does not hold: a phase whose PSG is locked need not contain a Kitaev point, since its dual may be a phase of the same class on a disjoint interval. A phase described by a non-locked PSG cannot be self-dual, and its dual partner is a different spin liquid at the dual coupling.

\subsection{Dual spin liquids}

Kimchi and Vishwanath pointed out that a spin liquid at a Heisenberg point has a Klein dual spin liquid at the dual coupling, which is likely to respond differently to external fields~\cite{KimchiVishwanath2014}. The label rule identifies the PSGs of these dual liquids. The Heisenberg PSGs of $p6mm\times\Z_2^\T$ embed into the KH classification with trivial gauge parts of $D_2$, as the $80$ classes Z$\cdots$00-$b$. Their duals Z$\cdots$01-$b$, with $n_{T_2}$ reversed, are the PSGs of the conjugated symmetry group $U[SU(2)\times p6mm\times\Z_2^\T]U^{-1}$ at the hidden-$SU(2)$ points. At $7\pi/4$, inside the triplet-VBS phase of Ref.~\cite{dOrnellasKnolle2024}, they are the exact duals of the Heisenberg spin liquids at $\phi=\pi$; at $3\pi/4$ the dual of the FM product state is the exact stripy state. In particular, the fermionic partners of the two bosonic spin liquids of Choy and Kim (Sec.~\ref{sec:heisenberg}) map as
\begin{align}
\text{Z100001100-}b\ &\to\ \text{Z000001101-}b,\nonumber\\
\text{Z000000100-}b\ &\to\ \text{Z100000101-}b,
\label{eq:ck-duals}
\end{align}
and likewise for the suffix $a$; the eight time-reversal-symmetric $\Z_2$ PSGs realized with nearest-neighbor bonds form exactly these four dual pairs. On the dual side the singlet nearest-neighbor amplitudes become color-locked triplet hopping, as Eq.~\eqref{eq:dual-links} requires. The chiral Kalmeyer--Laughlin candidate SL$[\frac\pi2,\frac\pi2,\pi]$ is mapped to a chiral state of the same class B with $n_{T_2}$ and $n_{TR}$ reversed, the same spinon bands, and the same Chern number, $C=+1$ per spin for triangle flux $+\pi/2$ ($C=2\Phi_{\rm tri}/\pi$); both have IGG $SU(2)$. It is not self-dual, whereas its locked $\Z_2$ descendant, the Yao--Kivelson state, is.

The phase boundaries of Ref.~\cite{dOrnellasKnolle2024}, extracted from its figures with an uncertainty of $\pm0.005\pi$, are mutually dual within this uncertainty for the Kitaev spin-liquid windows of iDMRG, within $0.008\pi$ for the other iDMRG boundaries, and within the $0.02\pi$ angular resolution of the Majorana mean field (Appendix~\ref{app:remarks}).
\section{Self-Consistent Mean Field Across the Phase Circle}
\label{sec:meanfield}

\subsection{Abrikosov and Kitaev forms of the spin operator}

We consider Gaussian states, which are fixed by their two-point correlations and obey Wick's theorem, of the four Majoranas $(b^x_i,b^y_i,b^z_i,c_i)$ of every site, equivalently Bogoliubov states of the Abrikosov fermions, with all on-site and bond bilinears: the three spin densities, the three constraint generators, and the sixteen bond channels of Eq.~\eqref{eq:links}. The energy is the expectation value of Eq.~\eqref{eq:HKH} evaluated with Wick's theorem, including the magnetic Hartree terms, and the single-occupancy constraint is imposed on every site for all three gauge generators. The remaining freedom is the form of the spin operator, which on the physical subspace can be written in two equivalent ways,
\begin{align}
\text{Abrikosov:}\quad &\sigma^a_i=f^\dagger_i\sigma^af_i=\tfrac12\big(\ii b^a_ic_i-\tfrac12\epsilon^{ade}\,\ii b^d_ib^e_i\big),\nonumber\\
\text{Kitaev:}\quad &\sigma^a_i=\ii b^a_ic_i,
\label{eq:reps}
\end{align}
which differ by constraint generators. The Kitaev form is the one used in the Majorana mean-field theory of Ref.~\cite{dOrnellasKnolle2024}. The energies that the two forms give with Wick's theorem (their Wick functionals) differ for states outside the physical subspace, and so do their unprojected mean-field results.

At the Kitaev points the Kitaev form reproduces the exact solution: the Yao--Kivelson ansatz is a self-consistent solution with $E=-0.7646302546|K|$ per site and gap $0.91370|K|$, and its Chern number is $\nu=+1$ at $\phi=\pi/2$ and $-1$ at $3\pi/2$ for the same $\Z_2$ gauge field (for $W_{\rm tri}=+1$). In the Abrikosov form the same state is also self-consistent, but its energy is exactly a quarter of the exact one, $-0.3823151273$ at $|K|=2$, because in the Wick decoupling only the term $\frac14(\ii b^a_ic_i)(\ii b^a_jc_j)$ of the Abrikosov form of $\sigma^a_i\sigma^a_j$ survives; the exact energy is recovered only by Gutzwiller projection. At the Heisenberg points the roles reverse: the Abrikosov form reduces to the $SU(2)$ singlet mean-field theory of Sec.~\ref{sec:heisenberg}, whereas the Kitaev form is not invariant under $SU(2)$ gauge transformations and does not. No unprojected functional is exact at both points, and we report both.

\subsection{Kitaev representation}

\begin{figure*}
\centering
\includegraphics[width=\textwidth]{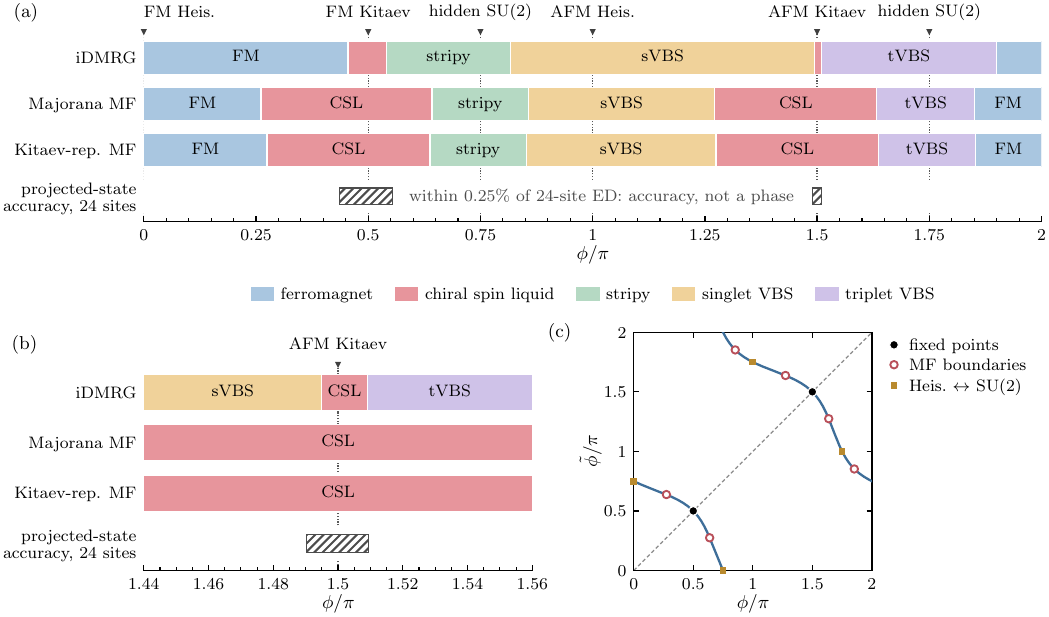}
\caption{(a) Phases of the KH model along the circle of Eq.~\eqref{eq:phi}: iDMRG and Majorana mean field of Ref.~\cite{dOrnellasKnolle2024} (boundaries from Figs.~7 and~8 there, with an uncertainty of $\pm0.005\pi$), the self-consistent mean field (MF) in the Kitaev representation of this work (Kitaev-rep.\ MF), and, hatched in the bottom row, the intervals in which the Gutzwiller-projected nearest-neighbor state with alternating triangle fluxes (class-D PSG ZC110110, the lowest flux sector of the Kitaev model on the $24$-site torus; Sec.~\ref{sec:vmc}) reproduces the $24$-site ED energy to $0.25\%$, $|J/K|\le0.093$ (FM) and $|J/K|\le0.0152$ (AFM). The hatched intervals measure the accuracy of a variational state on $24$ sites at an arbitrary threshold; they are not phases or phase boundaries. CSL, chiral spin liquid; sVBS and tVBS, singlet and triplet valence-bond solids. (b) Magnification around the AFM Kitaev point. (c) The duality map $\phi\to\tilde\phi$ of Eq.~\eqref{eq:dualphi}, with its fixed points at the Kitaev points, the boundaries of the Kitaev-representation mean field (open circles), which map onto one another, and the Heisenberg and hidden-$SU(2)$ pairs (squares). Special points: Heisenberg ($0$, $\pi$), Kitaev ($\pi/2$, $3\pi/2$), and hidden $SU(2)$ ($3\pi/4$, $7\pi/4$).}
\label{fig:phase}
\end{figure*}

The class-B state of the PSG \YK\ is continued from both Kitaev points in steps of $0.01\pi$ on $24\times24$ and $48\times48$ momentum meshes, and the energies of its converged stationary points are compared with those of the product states and dimer coverings below and of further trial states obtained by randomly perturbed relaxations and on $2\times2$ supercells. States that violate the on-site constraints are excluded; a trial state that satisfies them provides a valid energy comparison even if the self-consistency iteration has not converged. These energy comparisons establish neither stationarity nor local stability. The competitors, product states with FM, stripy, or other classical order and coverings of singlet and triplet dimers, are represented exactly in this functional. Within this set of competitors, the resulting phase diagram [Fig.~\ref{fig:phase}(a)] is
\begin{center}
\begin{tabular}{cc}
phase & $\phi/\pi$\\\hline
FM & $[-0.14758,\,0.27462]$\\
FM chiral spin liquid & $[0.27462,\,0.63757]$\\
stripy & $[0.63757,\,0.85242]$\\
singlet VBS & $[0.85242,\,1.27462]$\\
AFM chiral spin liquid & $[1.27462,\,1.63757]$\\
triplet VBS & $[1.63757,\,1.85242]$
\end{tabular}
\end{center}
with the boundaries that do not involve the spin liquid at the exact crossings $K=0$; since the dual of a point with $K=0$ is $\phi+\pi$, the crossings at $0.85242\pi$ and $1.85242\pi\equiv-0.14758\pi$ form a dual pair. The unrounded boundaries are mutually dual under Eq.~\eqref{eq:dualphi} to $10^{-5}\pi$, and the spin-liquid energies at dual angles obey $E(\tilde\phi)=\lambda E(\phi)$ to $10^{-15}$. Throughout both windows the $c$ Majoranas see flux $\pm\pi/2$ through the triangles and $\pi$ through the dodecagons, the Chern number is $+1$ on the FM side and $-1$ on the AFM side (for $W_{\rm tri}=+1$), and the single-Majorana gap stays open: from $1.8274$ at the Kitaev points ($|K|=2$) it first rises to $1.894$ near $0.476\pi$ (and $1.476\pi$) and then falls to $1.054$ at the lower edge, while it decreases monotonically to $0.679$ at the upper edge of each window. The Heisenberg coupling enters the class-B ans\"atze through off-color hopping of the $b$ Majoranas. This phase diagram agrees with the Majorana mean field of Ref.~\cite{dOrnellasKnolle2024} within the $0.02\pi$ angular resolution of that calculation, the largest difference, $0.0136\pi$, being at the lower edge of the FM window. Both spin-liquid windows are much wider than the iDMRG windows, and they have equal widths.

The widths are equal because both the spin liquid and its lowest competitor have the same energy at $\phi$ and $\phi+\pi$. For any Gaussian state without mixed $b$--$c$ correlations, which includes every nearest-neighbor state of class B and the dimer coverings, changing the sign of all $c$--$c$ correlations, $\Gamma_c\to-\Gamma_c$, leaves the constraints unchanged and maps the Kitaev-form energy at $(K,J)$ to that at $(-K,-J)$. Since $\phi\to\phi+\pi$ reverses both couplings,
\begin{equation}
E_{\rm kit}(\phi+\pi;\Gamma_b,-\Gamma_c)=E_{\rm kit}(\phi;\Gamma_b,\Gamma_c)
\label{eq:pi-symmetry}
\end{equation}
for every such state, so that the lowest energies of these states at $\phi$ and $\phi+\pi$ coincide. The competitors also come in pairs with equal energies at $\phi$ and $\phi+\pi$. Magnetically ordered product states carry the on-site correlation $\langle\ii b^a_ic_i\rangle=\langle\sigma^a_i\rangle\neq0$ and are not covered by Eq.~\eqref{eq:pi-symmetry}. Their energies per site, which are exact in this functional, nevertheless coincide with those of the dimer coverings of the inter-triangle bonds at $\phi+\pi$,
\begin{align}
E_{\rm FM}(K,J)&=-\tfrac12(K+2J)=E_{\rm sVBS}(-K,-J),\nonumber\\
E_{\rm stripy}(K,J)&=-\tfrac12(K-2J)=E_{\rm tVBS}(-K,-J),
\label{eq:competitor-pairs}
\end{align}
where $E_{\rm sVBS}=\frac12(K+2J)$ and $E_{\rm tVBS}=\frac12(K-2J)$ for singlet dimers and for triplet dimers with $\langle\sigma^a_i\sigma^a_j\rangle=-1$ on a bond of color $a$. These coincidences are exact properties of the two product states and the two dimer coverings, not consequences of Eq.~\eqref{eq:pi-symmetry}. No product state can lie below the lower envelope of these competitor energies: for normalized Bloch vectors, $-c m_i^a m_j^a\ge-|c|[(m_i^a)^2+(m_j^a)^2]/2$ gives $E_{\rm prod}/N\ge-(|K|+2|J|)/2$. For $K\ge0$ the FM or stripy state saturates this bound, while for $K<0$ the lower dimer-covering energy equals it. In numerical searches on supercells of up to $4\times4$ cells, the lowest product state is the FM or the stripy state on the FM side, and on the AFM side, over $[1.20,1.70]\pi$, it lies at least $0.21$ per site above the lower of the dimer coverings and the spin liquid, for instance at $-1.41$ against $-1.735$ for the singlet covering at $1.2746\pi$, and at $-1.00$ against $-1.529$ for the spin liquid at $3\pi/2$. The lowest competitor therefore has the same energy at $\phi$ and $\phi+\pi$, and with Eq.~\eqref{eq:pi-symmetry} the two windows have equal widths. Equation~\eqref{eq:pi-symmetry} is not a symmetry of the spin model. The operation acts on the $c$ sector alone, it maps physical singlet dimers onto states outside the physical subspace, and it constrains neither projected states nor the exact spectra, which differ at $\phi$ and $\phi+\pi$. It shows that an unprojected mean field in the Kitaev representation yields FM and AFM spin-liquid windows of equal width whenever its competitors come in such pairs, Eq.~\eqref{eq:competitor-pairs}, as they do here and in Ref.~\cite{dOrnellasKnolle2024}; the asymmetry of the iDMRG windows lies outside this class of theories. Fig.~8 of Ref.~\cite{dOrnellasKnolle2024} is $\pi$-periodic to $0.011\pi$, consistent with this statement.

\subsection{Abrikosov representation}

In the Abrikosov form, product states are exact, with energies close to $-1$ near $\phi=\pi/2$, whereas the class-B branch lies between $-0.27$ and $-0.43$ inside the FM and between $-0.38$ and $-0.59$ inside the AFM window of the Kitaev representation ($-0.382$ at both Kitaev points). Among all stationary states and competitors found, a product state is lowest at every sampled $\phi$, and no spin-liquid window is found in this unprojected functional. The class-B branch itself remains informative about the response to the Heisenberg coupling. As $J\to0$ we measure the growth of the off-color amplitude $u_\perp$ relative to the on-color amplitude $u_a$ by the slope $d(u_\perp/u_a)/d(J/K)$. In the Kitaev form this slope is the same on both sides. In the Abrikosov form its magnitude is $123$ times larger on the AFM side than on the FM side for triangle bonds and $194$ times larger for inter-triangle bonds; the triangle slopes have opposite signs on the two sides; the self-consistency map is already unstable at $J=0$. Independently, positive and negative eigenvalues of the constrained Wick-energy Hessian establish that the class-B branch is a saddle and not a local minimum of the functional at both Kitaev points: small perturbations of the off-color correlations relax toward lower, dimer-like Gaussian states, and on the AFM side the singlet-dimer covering, at $-0.5$ for $|K|=2$, lies below the branch energy $-0.382$ already at $J=0$. Following the branch by Newton's method for $J>0$, the converged states at $1.524\pi$ and $1.528\pi$ have Majorana-band Chern numbers $-1$ and $-4$, respectively, for $W_{\rm tri}=+1$, consistent with a gap closing between them. Near $1.560\pi$ the amplitudes approach a form related to the equal-amplitude Kalmeyer--Laughlin ansatz by sublattice-dependent spin rotations. Their bond-dependent relative signs in the original spin frame do not restore global spin-$SU(2)$ symmetry. For $J<0$, the converged states at $1.470\pi$ and $1.472\pi$ have Chern numbers $-4$ and $-1$, respectively, consistent with another gap closing between them; the smallest sampled gap occurs near $1.4715\pi$. These features are specific to the unprojected functional, and we do not relate them to the iDMRG boundaries.

The two unprojected functionals thus give contradictory answers. The Kitaev form describes the Kitaev points exactly and produces symmetric windows that are about four (FM) and twenty to forty (AFM) times wider than those of iDMRG, and the Abrikosov form describes product states exactly and yields no spin-liquid window among the states found. The question of the asymmetry has to be addressed with projected wave functions.
\section{Projected Wave Functions near the Kitaev Points}
\label{sec:vmc}

\subsection{Method}

The Gutzwiller projection of a Bogoliubov state of the Abrikosov fermions onto singly occupied sites has amplitudes given by Pfaffians of the pairing function restricted to the occupied site and spin labels~\cite{BeccaSorella2017}. We evaluate them by Metropolis sampling with single-spin and pair flips, since the Kitaev-type states do not conserve $S^z$, and compute the local energy of the full Hamiltonian~\eqref{eq:HKH}~\cite{Gros1989,BeccaSorella2017}. The variational states are the nearest-neighbor ans\"atze of the Kitaev form in the Majorana frame of Eq.~\eqref{eq:majorana-map}, in the uniform-flux sector, the PSG \YK, and in the sector with alternating triangle fluxes, the class-D PSG ZC110110 (listed in Table~S3 of the Supplemental Material~\cite{SM}), each with six real parameters (one singlet and two triplet amplitudes on each of the two orbits) and no on-site term, which locking forbids in both; the exact Kitaev states are special points of this manifold. Projected states with spin-dependent amplitudes constrained by the PSG have been used in this way for the honeycomb $K$-$J$-$\Gamma$ model~\cite{WangNormandLiu2019}. Parameters are optimized by stochastic reconfiguration~\cite{BeccaSorella2017}, a natural-gradient method, with a trust region, separately for each of the four boundary conditions of the partons, and the energies quoted are averages over independent Monte Carlo simulations. The $96$-site energies at fixed parameters quoted below are sampled with the inverse pairing matrix recomputed after every accepted move, and their errors are obtained by binning a single Markov chain for each boundary condition. On $24$ sites we enumerate all $2^{24}$ amplitudes exactly. For the curves in Fig.~\ref{fig:vmc} this exact energy is optimized further with the deterministic L-BFGS-B and Nelder--Mead methods, with five independent amplitude ratios after fixing the overall scale. At fixed parameters the exact projected energies agree with the sampled ones within the statistical errors, and at the Kitaev points the projected energies equal the exact Majorana energies with vanishing variance on $24$, $54$, and $96$ sites. The reference energies on the $24$-site $2\times2$ torus are obtained by Lanczos diagonalization in the full Hilbert space.

On the $2\times2$ torus the lowest Kitaev flux sector has triangle fluxes of alternating sign, a finite-size effect; at the Kitaev points the projected state in this sector is the exact ground state. On $96$ sites the uniform sector of the infinite lattice is lower by $0.0148$ per site. We therefore compare with ED in the alternating sector on $24$ sites and use the uniform sector, the PSG \YK\ itself, on $96$ sites.

\subsection{Results}

\begin{figure}
\centering
\includegraphics[width=\columnwidth]{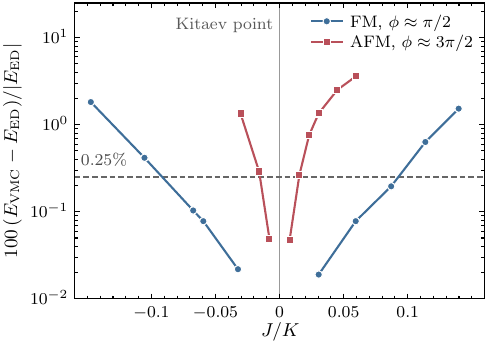}
\caption{Relative deviation of the energy of the Gutzwiller-projected nearest-neighbor state with alternating triangle fluxes (class-D PSG ZC110110, the lowest flux sector of the Kitaev model on this cluster) from the exact ground-state energy on the $24$-site $2\times2$ torus, as a function of $J/K$ near the FM ($\phi\approx\pi/2$) and AFM ($\phi\approx3\pi/2$) Kitaev points; at $J=0$ the projected state is exact. The deviation is positive (variational) and even in $J/K$, since $\pm J/K$ are dual angles. The dashed line marks the $0.25\%$ threshold of the accuracy intervals hatched in Fig.~\ref{fig:phase}.}
\label{fig:vmc}
\end{figure}

Figure~\ref{fig:vmc} shows the relative deviation of the optimized projected energy from the exact one on $24$ sites. The projected state in the alternating sector reproduces the exact ground-state energy to better than $0.25\%$ for $|J/K|\lesssim0.09$ on the FM side but only for $|J/K|\lesssim0.015$ on the AFM side. The deviations at dual angles agree, as the exact self-duality of the class and of the spectrum requires, for instance $0.078\%$ at $\phi=0.46\pi$ and its dual $0.5356\pi$; the deviation is therefore an even function of $J/K$, and interpolating both sides together gives $|J/K|\le0.093$ and $|J/K|\le0.0152$, that is, $[0.4356,0.5536]\pi$ and $[1.4902,1.5095]\pi$, each mapped onto itself by Eq.~\eqref{eq:dualphi} [hatched in Fig.~\ref{fig:phase}(a) and (b)]. These intervals measure the accuracy of one variational state on one cluster at an arbitrary threshold; they are not phase boundaries. The projected state's own response to $J$ is asymmetric as well: finite-difference estimates of the coefficient $c$ in $E_2=cNJ^2$ at fixed $|K|=2$ are approximately $c_{\rm FM}=-0.26$ and $c_{\rm AFM}=-0.86$ per site, suggesting a response about three times stronger on the AFM side. On $96$ sites, with the parameters optimized on $24$ sites, the energies at $\phi=0.46\pi$ and $1.49\pi$ are consistent with this trend in both flux sectors; size-converged $J\to0$ curvatures require optimization on the larger clusters. The magnitude of its off-color correlation on the inter-triangle bonds grows about $12$ to $15$ times faster with $|J/K|$ on the AFM side. The unprojected Kitaev-representation mean field of Sec.~\ref{sec:meanfield}, by contrast, is exactly symmetric. On $96$ sites, with the parameters optimized on $24$ sites, the projected \YK\ state lies below the projected state with alternating triangle fluxes (class-D PSG ZC110110), both at $\phi=1.49\pi$, $-1.55339(11)$ versus $-1.53842(14)$ per site, and at $0.46\pi$, $-1.61719(15)$ versus $-1.60156(18)$, and the product states lie far above both.

The ratio of the $24$-site window widths depends on the measure: energy thresholds between $0.05\%$ and $1\%$ give ratios of the FM to the AFM width between $6.2$ and $4.7$. On this cluster the AFM edge of the window also coincides with a level crossing in the exact spectrum, and the exact response itself differs between $24$-site clusters: it is quadratic in $J$ on the AFM side of the $2\times2$ torus but linear in $|J|$ on a $4\times1$ torus. The $4\times1$ torus does not admit the Klein duality, so its spectrum need not be even in $J$; an explanation of its nonanalytic response requires an analysis of its different low-energy degeneracies and matrix elements, which we do not attempt here. By contrast, the $2\times2$ torus is the smallest cluster on which the duality holds. The exact curvatures on the same cluster are $c_{\rm FM}=-0.340$ and $c_{\rm AFM}=-3.52$ per site, a ratio of $10.4$, compared with approximately $3.3$ for the projected-state response estimates, corresponding to about $77\%$ of the exact curvature on the FM side and $24\%$ on the AFM side. Most of the asymmetry of the $24$-site windows therefore comes from the exact spectrum, and the projected state accounts for a factor of about three in its own response. Measured against the thermodynamic-limit values of Sec.~\ref{sec:pt}, $-0.4286$ and $-4.6998$, the projected response estimates are about $61\%$ and $18\%$. The finite-size data therefore support a projected-state response to $J$ that is about three times stronger on the AFM side, which an unprojected Kitaev-representation mean field whose competitors come in pairs of equal energy (Sec.~\ref{sec:meanfield}) cannot produce; on $24$ sites the projected state is in addition accurate over a much wider range of $J$ on the FM side, mostly because of the exact spectrum. A quantitative comparison with the iDMRG window widths requires size-converged benchmarks and the perturbative analysis of Sec.~\ref{sec:pt}.
\section{Perturbation Theory about the Kitaev Points}
\label{sec:pt}

The projected states of Sec.~\ref{sec:vmc} show an asymmetric response to $J$ but are limited to nearest-neighbor amplitudes and small clusters. Near the Kitaev points the response can be computed exactly. We expand Eq.~\eqref{eq:HKH} about $J=0$, with $H_K$ the Kitaev part and $V=-J\sum_{\langle ij\rangle_a}\sum_{b\neq a}\sigma^b_i\sigma^b_j$ the off-color part, and write $V=J\mathcal W$. We evaluate in the thermodynamic limit both the second-order energy and the fidelity susceptibility with respect to $J$~\cite{ZanardiPaunkovic2006,YouLiGu2007}, which measures how strongly $V$ mixes excited states into the ground state. Throughout this section $|K|=1$, except for the coefficients $c$ of Eq.~\eqref{eq:cpt}.

\subsection{Vison creation and interference blocks}

Each off-color term $\sigma^b_i\sigma^b_j$ on an $a$ bond anticommutes with the flux operators of the four plaquettes that touch the sites $i$ and $j$, as on the honeycomb lattice (Supplemental Material of Ref.~\cite{SongYouBalents2016}), so it creates four visons. Hence $\langle0|V|0\rangle=0$, and the leading corrections to the energy and to the ground state are
\begin{align}
E_2&=-J^2\sum_P\int_0^\infty d\tau\,G_P(\tau),\nonumber\\
\chi_P&=\int_0^\infty d\tau\,\tau\,G_P(\tau)=\sum_n\frac{|\langle n|\mathcal W_P|0\rangle|^2}{(E_n-E_0)^2},\nonumber\\
G_P(\tau)&=\langle0|\mathcal W_P^\dagger e^{-(H_K-E_0)\tau}\mathcal W_P|0\rangle,
\label{eq:E2}
\end{align}
where $V_P=J\mathcal W_P$ collects all terms that create the same flux pattern $P$, which we call an interference block, and only terms within one $V_P$ interfere. The fidelity susceptibility $\chi_P$, in the spectral and imaginary-time forms of Ref.~\cite{YouLiGu2007}, has units of inverse energy squared; $J^2\chi_P$ is the weight of the excited states admixed by $V_P$ to this order. The imaginary-time correlator $G_P$ is a matrix element between Majorana vacua of two gauge sectors with the flipped bond variables; we evaluate it exactly by writing the vacuum of the flipped sector as the exponential of a fermion-pair operator acting on the original vacuum (Thouless representation), including the projection onto the physical fermion parity, on tori up to $24\times24$ unit cells, where $E_2$ and $\chi_P$ are converged to all digits shown. On the star lattice the interference blocks are the following. The two off-color terms of an inter-triangle bond flip the same four plaquettes, its two triangles and its two dodecagons, because their product is $-\sigma^a_i\sigma^a_j$, which commutes with every flux; and all six off-color terms on the edges of a triangle flip that triangle and its three dodecagons. No other terms interfere. For an inter-triangle bond the lowest state reached is a four-vison state at $\Delta_e=0.131|K|$, about a seventh of the Majorana gap. The lowest triangle-flipped state, at $0.123|K|$, receives exactly zero weight (Sec.~\ref{sec:pt-results}), so the lowest states reached by a triangle block are two-Majorana states above $1.394|K|$. On a $24$-site torus the same calculation reproduces the exact curvature of the ED ground-state energy to $10^{-6}$ on both sides.

\subsection{Results}
\label{sec:pt-results}

\begin{figure}
\centering
\includegraphics[width=0.94\columnwidth]{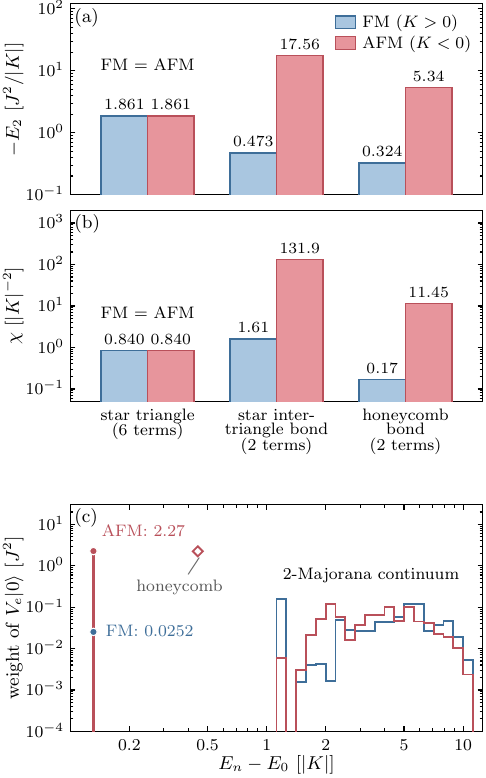}
\caption{Exact second-order perturbation theory in $J$ about the Kitaev points, $|K|=1$. (a) Second-order energy $-E_2$ and (b) fidelity susceptibility $\chi$ with respect to $J$, in units of $|K|^{-2}$, per interference block at the FM ($K>0$) and AFM ($K<0$) Kitaev points: the six off-color terms of a triangle and the two terms of an inter-triangle bond of the star lattice, and the two terms of a bond of the honeycomb lattice (extrapolated from odd $L\times L$ tori). The triangle values are identical for both signs of $K$. (c) Spectral weight of $V_e|0\rangle$ for one inter-triangle bond (units of $J^2$) versus excitation energy, on an $8\times8$ torus. Sticks: lowest four-vison state at $0.131|K|$, with weight $0.0252$ (FM) and $2.27$ (AFM). Histograms: two-Majorana continuum above $1.19|K|$ in logarithmic bins of $0.05$ decades, with total weight $0.812$ (FM) and $0.830$ (AFM); states with four or more quasiparticles carry a further $0.03$. Open diamond: lowest four-flux state of the honeycomb model at $0.446|K|$ (AFM weight $2.238$; the FM weight vanishes).}
\label{fig:pt}
\end{figure}

With $|K|=1$, a triangle block contributes $E_2=-1.86124J^2/|K|$ at both Kitaev points, whereas an inter-triangle bond contributes $-0.47349J^2/|K|$ at the FM point and $-17.5585J^2/|K|$ at the AFM point, a factor of $37$ [Fig.~\ref{fig:pt}(a)]. In the convention of Sec.~\ref{sec:vmc}, $E_2=cNJ^2$ at fixed $|K|=2$, so that $c$ is half the curvature $\partial^2E/\partial J^2$ per site, the second-order coefficients per site are
\begin{equation}
c_{\rm FM}=-0.4286,\qquad c_{\rm AFM}=-4.6998,
\label{eq:cpt}
\end{equation}
a ratio of $11.0$. The triangles contribute $-0.3102$ on both sides, and the inter-triangle bonds $-0.1184$ (FM) and $-4.3896$ (AFM): the entire asymmetry sits on the inter-triangle bonds. The fidelity susceptibilities show the same selectivity more sharply [Fig.~\ref{fig:pt}(b)]: $\chi=0.840/K^2$ for a triangle on both sides, and $1.61/K^2$ (FM) versus $131.9/K^2$ (AFM) for an inter-triangle bond, a factor of $82$. Per site the values are $1.09$ and $66.2$ (in units of $|K|^{-2}$), a factor of $61$. At the AFM point $99.9\%$ of the $\chi$ of an inter-triangle bond comes from the lowest four-vison state, at the FM point $91\%$.

The spectra do not distinguish the two sides: reversing $K$ while relabeling $\mathsf{u}\to-\mathsf{u}$ leaves the Majorana hopping matrix unchanged, so the corresponding flux sectors have the same Majorana spectrum. This relabeling reverses the triangle-flux labels and is not a gauge transformation within one flux sector. The correlator of each single off-color term is identical for FM and AFM coupling. Only the interference within a block changes. On an inter-triangle bond of color $a$, let $X_1=\sigma^b_i\sigma^b_j$ and $X_2=\sigma^c_i\sigma^c_j$ with $(a,b,c)$ a cyclic permutation of $(x,y,z)$. Since $X_2=-X_1\sigma^a_i\sigma^a_j$, the two terms reach the lowest four-vison state with amplitudes that differ by a phase,
\begin{align}
\langle{\rm low}|V_e|0\rangle&=-J\langle{\rm low}|X_1|0\rangle(1-t),\nonumber\\
t&=\mathrm{sgn}(K)\,0.97801-0.20857\,\ii,
\label{eq:phase}
\end{align}
in the ground state with $W_{\rm tri}=+1$; for $W_{\rm tri}=-1$, $t\to t^*$. The modulus $|t|=1$ is exact: the product of $\T$ with the mirror that contains the inter-triangle bond leaves each chiral ground state and the four-vison pattern invariant and exchanges $X_1$ and $X_2$. Only the relative phase depends on $\mathrm{sgn}\,K$. The interference is nearly destructive at the FM point and nearly constructive at the AFM point, and the weight of the lowest state is $2\times0.5737\times(1\mp0.978)=0.0252$ (FM) and $2.270$ (AFM) in units of $J^2$, a ratio of $90$ [Fig.~\ref{fig:pt}(c)]. This single state supplies $98.5\%$ of the energy of an inter-triangle bond at the AFM point and $41\%$ at the FM point, where the rest comes from two-Majorana states above $1.19|K|$ (on the $8\times8$ torus of Fig.~\ref{fig:pt}(c)). The total weights obey the exact sum rule $(\sigma^b_i\sigma^b_j+\sigma^c_i\sigma^c_j)^2=2(1-\sigma^a_i\sigma^a_j)$, which gives $\langle V_e^2\rangle=2J^2(1\mp0.5658)$ with the nearest-neighbor correlation of Sec.~\ref{sec:kitaev}; this ratio of $3.6$ is much smaller than that of the energies because the weight shifts toward the soft four-vison state.

The triangles are exactly blind to the sign of $K$. Label the vertices $1,2,3$ of a triangle clockwise, so that the edge $(12)$ has the color $a_3$, let $X=\sigma^{a_1}_1\sigma^{a_1}_2$ be one of its two off-color terms, $W=W_{\rm tri}$ the triangle flux operator of Sec.~\ref{sec:kitaev}, and $S_{12}=\sigma^{a_3}_1\sigma^{a_3}_2$, $S_{23}=\sigma^{a_1}_2\sigma^{a_1}_3$ the Kitaev bond operators of two edges. The six terms $Y_n$ of the triangle satisfy the operator identity
\begin{equation}
\sum_{n=1}^{6}Y_n=X(1-\ii W)(1+\ii Ws),\quad s=S_{23}-S_{12},
\label{eq:triangle-identity}
\end{equation}
where $s^2=2$; for counterclockwise labels the identity holds with $W$ replaced by $-W$. The ground states at $K$ and $-K$ are related by $\mathsf{u}\to-\mathsf{u}$ at fixed $c$-Majorana state, under which the eigenvalue of $W$ and the operator $s$ change sign, so $1+\ii Ws$ is invariant and only the prefactor $1\mp\ii W$ changes, whose modulus is $\sqrt2$ in both cases. The entire spectral function of a triangle block, not only its contribution to $E_2$ and $\chi$, is therefore the same at both Kitaev points, and $\langle V_t^2\rangle=6J^2$. The lowest triangle-flipped state receives no weight at all: the six single-term amplitudes onto it have equal moduli and, within each orbit of the threefold rotation about the triangle combined with the cyclic permutation of the spin axes, phases $1$, $\omega$, $\omega^2$ with $\omega=e^{\pm2\pi\ii/3}$, so they cancel. The triangle energy comes almost entirely from two-Majorana states above $1.394|K|$.

\subsection{Relation to earlier work and to the phase diagram}

The interference of the two off-color terms of a bond, with a relative sign set by the sign of $K$, was identified by Zhang, Hal\'asz, Zhu, and Batista as the reason for the asymmetric stability of the honeycomb Kitaev spin liquids: in their variational treatment of flux-pair excitations, the two Heisenberg contributions to the hopping of a flux pair interfere constructively for one sign of $K$ and destructively for the other, making the AFM Kitaev spin liquid more fragile against $J$~\cite{ZhangHalaszZhuBatista2021}. The relation $X_2=-X_1\sigma^a_i\sigma^a_j$ behind Eq.~\eqref{eq:phase} is the algebra of their Eq.~(22). They also noted that the matrix elements that create two flux pairs from the flux-free sector [their Fig.~2(a)] renormalize the ground-state energy and lower the flux-pair energy, but neglected them in their variational treatment~\cite{ZhangHalaszZhuBatista2021}. An analogous interference governs the flux-pair hopping in a $[111]$ magnetic field~\cite{ZhangHalaszBatista2022}. Joy and Rosch argued that the hopping of single visons at order $J^2$ cancels for both signs of $K$ and related the sensitivity of the AFM liquid to $J$ to the tunneling of vison pairs~\cite{JoyRosch2022}. Here we compute the exact second-order ground-state energy and fidelity susceptibility through the process of Eq.~\eqref{eq:E2}, in which $V$ creates four visons from the ground state, where the same interference appears with exactly equal moduli, and find that on the decorated star lattice it depends on the bond type. On the honeycomb lattice we evaluate the same expressions on odd $L\times L$ tori, $L=17$ to $43$ with $L$ not divisible by $3$, in the lowest physical holonomy sector of the flux-free sector; even $L$ oscillate more strongly with $L$ and are not used. Fits in $1/L^2$ and $1/L^3$ to the two subsequences $L\equiv1$ and $5\pmod 6$ give, per bond, $E_2=-0.324(2)J^2/|K|$ (FM) and $-5.338(2)J^2/|K|$ (AFM), a ratio of $16.5(1)$, and $\chi=0.170(3)/K^2$ and $11.45(1)/K^2$, a ratio of $67$; the errors cover the spread of the fits and the largest tori. Here $t=\pm1$ exactly, because $\T$ is unbroken and the mirror through the bond is itself a symmetry, and the lowest four-flux state, at $0.446|K|$, receives exactly zero weight at the FM point and supplies $94.0(2)\%$ of $E_2$ at the AFM point. The interference thus occurs on both lattices studied here, in line with the asymmetric honeycomb phase diagrams~\cite{ChaloupkaJackeliKhaliullin2013,Gohlke2017} and with exact diagonalization of amorphous Kitaev--Heisenberg clusters, where the reported thresholds are $|J_H/J_K|\approx0.12$ (FM) and $\approx0.01$ (AFM) in that paper's convention, $H=\sum[J_K\sigma_i^a\sigma_j^a+J_H\bm\sigma_i\cdot\bm\sigma_j]$; the identification of the AFM phase there is limited by finite-size effects~\cite{GrushinRepellin2023}. On the star lattice the interference in addition depends on the bond type: the triangles are exactly sign blind, and the inter-triangle bonds carry the whole asymmetry through a four-vison state that is softer by a factor of three than on the honeycomb lattice.

For the projected nearest-neighbor states of Sec.~\ref{sec:vmc}, finite-difference response estimates give coefficients of approximately $-0.26$ (FM) and $-0.86$ (AFM) per site. On $96$ sites, with the parameters optimized on $24$ sites, the energies support the same trend but do not determine size-converged $J\to0$ curvatures. To compare with Eq.~\eqref{eq:cpt} at $|K|=2$, we use $c_2(\phi)=|K(\phi)|[E(\phi)/N-|K(\phi)|e_K/2]/[2J(\phi)^2]$, where $e_K$ is the Kitaev energy per site at $|K|=2$. This both subtracts the running Kitaev baseline and rescales the energy and perturbation to fixed $|K|$. Fits of the exact $24$-site projected energies near $J/K=0$ give approximately $-0.2614$ (FM) and $-0.8622$ (AFM); the $96$-site data probe only the response at finite angles. They correspond to about $61\%$ of the thermodynamic-limit coefficient on the FM side and $18\%$ on the AFM side, where the exact response is dominated by the soft four-vison state of the inter-triangle bonds. The selectivity may also be related to the phases that border the Kitaev spin liquids. On the AFM side the neighboring phases in Ref.~\cite{dOrnellasKnolle2024} are the singlet VBS ($J<0$) and the triplet VBS ($J>0$), both dimerized on the inter-triangle bonds, and both the second-order energy and the fidelity susceptibility of the AFM Kitaev spin liquid are concentrated on exactly these bonds; the FM side, whose triangles are sign blind and whose inter-triangle bonds are nearly decoupled from the lowest flux state, borders ferromagnetic and stripy order. This connection is suggestive; perturbation theory about the spin liquid does not select either VBS.

A larger $|E_2|$ is an energy gain of the spin liquid and does not by itself make it fragile; in an energy-crossing picture it even works against the observed asymmetry, since at fixed competitor energies a more negative $E_2$ moves the first-order boundary of the AFM spin liquid outward. Whether the soft four-vison state itself approaches zero energy would require its own second-order shift, which we have not computed. Fragility is the rate at which the ground state changes, measured by $\chi$, and here it is the admixture of the soft four-vison state. A heuristic stability scale per block follows from setting the admixed weight $J^2\chi$ to one: $|J/K|\approx(K^2\chi)^{-1/2}=0.087$ for the inter-triangle bonds at the AFM point, $0.79$ for them at the FM point, and $1.09$ for the triangles on both sides. Criteria based on the lowest state actually reached give the same ordering: equating the second-order shift or the matrix element with the energy of that state gives $0.086$--$0.087$ (AFM inter-triangle bonds), $0.53$--$0.83$ (FM inter-triangle bonds), and $0.87$ for the triangles, whose lowest reached states lie at $1.394|K|$. All of these scales are heuristic. The iDMRG boundaries, $|J/K|\approx0.008$--$0.014$ (AFM) and $0.066$--$0.069$ (FM), lie six to eleven times (AFM) and eleven to twelve times (FM) inside the scales $(K^2\chi)^{-1/2}$, consistent with first-order transitions into competing phases that a perturbative treatment of the spin liquid does not describe. Neither the ratios $11.0$ and $16.5$ of the second-order coefficients nor the ratio $9$ of the heuristic scales is a ratio of window widths.
\section{The Heisenberg Model: Fermionic Partners of Bosonic Spin Liquids and Valence-Bond Crystals}
\label{sec:heisenberg}

\subsection{Fermionic PSGs of the \texorpdfstring{$\Jt$--$\Je$}{Jt-Je} model}

For the Heisenberg model~\eqref{eq:HH} the symmetry group is $SU(2)\times p6mm\times\Z_2^\T$ and only singlet amplitudes are allowed. Its $\Z_2$ PSGs are labeled by the first seven digits of Eq.~\eqref{eq:labels}, written Z$n_{T_2}n_{\T S}n_{\T C_6}n_Sn_{SC_6}n_{C_6}(t)$, and they follow from those of the SHD lattice by restriction to the site stabilizer~\cite{RajIqbalSHD}. There are $80$ algebraic $\Z_2$ PSGs, of which $64$ are realizable, that is, allow a nonzero amplitude: every realizable PSG has an allowed amplitude within the second shell, and a generic ansatz is a connected $\Z_2$ spin liquid at the latest from the ninth shell (length $4.625$, the PSG Z001011(0)). The $16$ others are among the $32$ PSGs in which $\T$ acts on every site by a gauge matrix $\pm\tau^0$, that is, only by a sign, and they allow no amplitude on any bond. In the $48$ PSGs with $t=-1$ time reversal squares to $-1$ on the fermionic spinon, which is therefore a Kramers doublet. Restricted to the space group alone there are $16$ $\Z_2$ PSGs, as in the bosonic classification of Choy and Kim~\cite{ChoyKim2009}. There are $16$ $SU(2)$ PSGs, of which $8$ are realizable, none of them with an amplitude on the triangle edges: an $SU(2)$ amplitude on an odd loop breaks $\T$. For the IGG $U(1)$, the on-site condition reduces $208$ discrete classes and $16$ families of the cohomology to $64$ PSGs and $4$ one-parameter families, of which $48$ PSGs and $2$ families are realizable (Table~S4 of the Supplemental Material~\cite{SM}).

Yang, Paramekanti, and Kim studied six nearest-neighbor flux states of this model, two time-reversal-symmetric and four chiral ones, in mean field and with Gutzwiller projection~\cite{YangParamekantiKim2010}. We label flux states SL$[\Phi_\triangle,\Phi_\triangledown,\Phi_{12}]$ by the fluxes through the two triangles and the dodecagon, in their notation. With nearest-neighbor amplitudes on both orbits, four time-reversal-symmetric $U(1)$ flux states exist: SL$[0,0,0]$ and SL$[0,0,\pi]$ of Ref.~\cite{YangParamekantiKim2010}, and SL$[0,\pi,\pi]$ and SL$[\pi,0,0]$, in which the two triangles of a unit cell carry different fluxes and $C_6$ acts with a particle-hole gauge element; SL$[0,0,\pi]$ and SL$[\pi,0,0]$ need a $12$-site cell. At equal amplitudes on the two orbits, SL$[0,0,0]$ has a quadratic band touching at the Fermi level together with a flat band there, SL$[0,0,\pi]$ has a single-spinon gap (half the particle-hole gap) $0.197$, and SL$[0,\pi,\pi]$ and SL$[\pi,0,0]$ have gaps $0.382$, in units of the hopping amplitude. The four chiral states have triangle fluxes $\pm\pi/2$. Reference~\cite{YangParamekantiKim2010} describes them as $U(1)$ states, with a Chern--Simons term for SL$[\frac\pi2,\frac\pi2,\pi]$, but their IGG is $SU(2)$: a hopping loop of length $n$ and flux $\Phi$ is proportional to $\tau^0$ when $\Phi-n\pi/2$ is a multiple of $\pi$, which holds for every triangle and dodecagon of these states. The Kalmeyer--Laughlin candidate SL$[\frac\pi2,\frac\pi2,\pi]$ has the PSG SU0001 in class B, a single-spinon gap of $0.457$, and Chern number $C=\pm1$ per spin; SL$[\frac\pi2,\frac\pi2,0]$ has the class-B PSG SU1000, and the two states with staggered chirality, SL$[-\frac\pi2,\frac\pi2,0]$ and SL$[-\frac\pi2,\frac\pi2,\pi]$, are class-D states with a spinon Fermi surface. Here SU$n_{T_2}n_Sn_{SC_6}n_{C_6}$ labels the $SU(2)$ PSGs of $p6mm$. As Sec.~\ref{sec:kitaev} showed, the Kalmeyer--Laughlin candidate is the $SU(2)$ parent of the Yao--Kivelson liquid.

\subsection{Correspondence with Schwinger-boson PSGs}

Choy and Kim found $16$ algebraic PSGs of the space group for Schwinger-boson mean-field (SBMF) states on the star lattice, labeled by $(p_1,p_2,p_3,p_4)$, and showed that only $(p_1,1,0,1)$ admits nonzero amplitudes on all nearest-neighbor bonds: the zero-flux state ($p_1=1$) and the $\pi$-flux state ($p_1=0$), named after the flux of the bosonic ansatz through the dodecagon~\cite{ChoyKim2009}. A $\Z_2$ spin liquid can be described by bosonic or fermionic spinons, which differ by a vison, the $\Z_2$ flux excitation. Lu, Cho, and Vishwanath showed that the symmetry fractionalization classes of the two spinons, the sets of projective signs with which the symmetry relations act on them, are related by
\begin{equation}
\omega_f=\omega_b\,\omega_v\,\Omega,
\label{eq:lcv}
\end{equation}
where $\omega_v$ is the fractionalization class of the vison and $\Omega$ a set of twist factors, equal to $-1$ on the relations $C_6^6=e$, $m^2=e$, and $m\T m^{-1}\T^{-1}=e$ for a mirror $m$ and to $+1$ on the others~\cite{LuChoVishwanath2017}. For the kagome lattice they showed that the vison of every SBMF state transforms trivially, except that its translations anticommute because the unit cell contains an odd number of spins. For gapped, uncondensed quadratic Schwinger-boson ans\"atze, these arguments carry over to the star lattice: the unit cell contains six spins, so the vison translations commute; the absence of protected edge states in SBMF states makes the vison transform trivially under both mirror families and time reversal; and the $C_6$ axis passes through a dodecagon, so the disclination argument applies to $C_6$. These arguments, which take the limit in which the on-site boson chemical potential dominates, do not involve the gauge part of $\T$ on the bosons and therefore apply also to the staggered time reversal introduced below. Hence $\omega_v$ is trivial in all seven labels, and
\begin{equation}
\omega_f=\omega_b\times(+,-,+,-,-,-,+)
\label{eq:twist}
\end{equation}
in the order of the labels Z$n_{T_2}n_{\T S}n_{\T C_6}n_Sn_{SC_6}n_{C_6}(t)$. We computed $\omega_b$ from the explicit PSG of Ref.~\cite{ChoyKim2009}, after mapping their generators onto ours. Choy and Kim let time reversal act on the Schwinger bosons as complex conjugation with trivial gauge part, so that the amplitudes are real; this is consistent with their PSG because the gauge factors of their Eq.~(13) on nearest-neighbor bonds are $\pm1$. It is not the only choice: time reversal need only map the ansatz onto a gauge-equivalent one~\cite{WangVishwanath2006,MessioLhuillierMisguich2013}. In the gauge of their Eq.~(13), twice the gauge phase of every generator is constant, so the relations $X\T X^{-1}\T^{-1}=\pm1$ require the gauge phase $\varphi_\T$ of time reversal to change by $0$ or $\pi$ under each generator; it is periodic, because $C_6$ forces $T_i\T T_i^{-1}\T^{-1}=+1$. A uniform $\varphi_\T$ is removed by a global gauge transformation. The only other solution changes sign under $C_6$ and $\sigma_d$, which exchange the two triangles of the unit cell: $G_\T=+\ii$ on one triangle and $-\ii$ on the other. This staggered time reversal reverses $C_6\T C_6^{-1}\T^{-1}$ and $\sigma_d\T\sigma_d^{-1}\T^{-1}$ and no other relation, so it gives $16$ further bosonic $\Z_2$ PSGs, gauge inequivalent to those of Choy and Kim, each realized by a symmetric $\Z_2$ ansatz. In the gauge with $G_\T=\pm\ii$ their triangle amplitudes are imaginary; the gauge transformation with phases $\pi/4$ and $3\pi/4$ on the two triangles makes them real, and they are then the states of Choy and Kim with the sign of the triangle amplitudes reversed on one of the two triangles. In $(p_1,1,0,1)$ they are nonzero on all nearest-neighbor bonds; compared with the states of Choy and Kim, the bosonic flux through the dodecagon is shifted by $\pi$, and so is the flux through the eight-bond loop around two adjacent triangles, so these are two further nearest-neighbor $\Z_2$ states. Choy and Kim describe $R$ as a rotation about a lattice site; the star lattice has no site-centered sixfold axis, and their $R$ ($R^6=e$, sublattices $a\to d\to b\to f\to c\to e$) is the $60^\circ$ rotation about a dodecagon center. Their mirror $\sigma$ belongs to the family $\sigma_d$. The same mapping reproduces the correspondence of Ref.~\cite{LuChoVishwanath2017} for the eight SBMF states of the kagome lattice.

\begin{table}
\caption{Schwinger-boson PSGs $(p_1p_2p_3p_4)$ of Ref.~\cite{ChoyKim2009} and their fermionic partners from Eq.~\eqref{eq:twist}, for the time reversal of Ref.~\cite{ChoyKim2009} and for the staggered time reversal with $G_\T=\pm\ii$ on the two triangles (stag.\ $\T$), whose partner differs in $n_{\T C_6}$; the remaining columns refer to the first partner. $(\Jt,\Je)$: real dimensions of the allowed singlet amplitudes on the triangle edges and the inter-triangle bonds of the fermionic PSG. IGG of a generic nearest-neighbor ansatz with bond amplitudes only, and with the allowed on-site multipliers $a_i$ in addition; whether these multipliers survive the constraint equations is a self-consistency question (Sec.~\ref{sec:classification}). Rows $0101$ and $1101$ are the $\pi$-flux and zero-flux states. Asterisk: the nearest-neighbor bonds form disconnected trimers, and the IGG is that of one trimer.}
\label{tab:ck}
\begin{ruledtabular}
\begin{tabular}{cccccc}
 & & & & \multicolumn{2}{c}{IGG}\\
$p_1p_2p_3p_4$ & partner & stag.\ $\T$ & $(\Jt,\Je)$ & bonds & $+a_i$\\
\hline
$0000$ & Z010111(1) & Z011111(1) & $(1,0)^*$ & $U(1)$ & $U(1)$\\
$0001$ & Z010110(1) & Z011110(1) & $(1,1)$ & $U(1)$ & $U(1)$\\
$0010$ & Z001010(1) & Z000010(1) & $(2,0)^*$ & $U(1)$ & $\Z_2$\\
$0011$ & Z001011(1) & Z000011(1) & $(2,1)$ & $U(1)$ & $\Z_2$\\
$0100$ & Z000001(1) & Z001001(1) & $(2,0)^*$ & $U(1)$ & $\Z_2$\\
$0101$ & Z000000(1) & Z001000(1) & $(2,2)$ & $\Z_2$ & $\Z_2$\\
$0110$ & Z011100(1) & Z010100(1) & $(1,1)$ & $U(1)$ & $U(1)$\\
$0111$ & Z011101(1) & Z010101(1) & $(1,1)$ & $U(1)$ & $U(1)$\\
$1000$ & Z110110(1) & Z111110(1) & $(1,0)^*$ & $U(1)$ & $U(1)$\\
$1001$ & Z110111(1) & Z111111(1) & $(1,1)$ & $U(1)$ & $U(1)$\\
$1010$ & Z101011(1) & Z100011(1) & $(2,0)^*$ & $U(1)$ & $\Z_2$\\
$1011$ & Z101010(1) & Z100010(1) & $(2,1)$ & $U(1)$ & $\Z_2$\\
$1100$ & Z100000(1) & Z101000(1) & $(2,0)^*$ & $U(1)$ & $\Z_2$\\
$1101$ & Z100001(1) & Z101001(1) & $(2,2)$ & $\Z_2$ & $\Z_2$\\
$1110$ & Z111101(1) & Z110101(1) & $(1,1)$ & $U(1)$ & $U(1)$\\
$1111$ & Z111100(1) & Z110100(1) & $(1,1)$ & $U(1)$ & $U(1)$
\end{tabular}
\end{ruledtabular}
\end{table}

Table~\ref{tab:ck} gives the result. Each of the $32$ bosonic classes has a distinct fermionic partner among the $64$ realizable PSGs, all with Kramers spinons. The $\pi$-flux state corresponds to Z000000(1), the PSG in which no relation of the space group carries a projective sign, and the zero-flux state to Z100001(1), in which the spinon translations anticommute and $C_6^6=-1$. The two nearest-neighbor states with staggered time reversal correspond to Z001000(1) ($p_1=0$) and Z101001(1) ($p_1=1$). In these fermionic PSGs the triangle edges allow two real singlet amplitudes and the inter-triangle bonds one, and a generic ansatz with nearest-neighbor bond amplitudes alone has the IGG $U(1)$: it is SL$[0,\pi,\pi]$ in Z001000(1) and SL$[\pi,0,0]$ in Z101001(1). The allowed on-site multipliers or second-neighbor amplitudes reduce the IGG to $\Z_2$. On the kagome lattice the fermionic PSG without projective signs pairs instead with the bosonic class $(p_1,p_2,p_3)=(1,1,0)$ of Ref.~\cite{LuChoVishwanath2017}, whose spinon translations anticommute ($p_1=1$). The difference comes from the vison, whose translations anticommute on the kagome lattice and commute here, so that on the star lattice the fermionic and bosonic translation signs agree. The partners Z000000(1) and Z100001(1) of the two states of Choy and Kim are realized as $\Z_2$ spin liquids with nearest-neighbor bond amplitudes alone and are gapped at generic amplitudes. Each of them contains two $U(1)$ states. Amplitudes collinear on the two orbits give SL$[0,0,0]$ in Z000000(1) and SL$[0,0,\pi]$ in Z100001(1); amplitudes rotated by $\pi/2$ against each other, hopping on one orbit and pairing on the other, give SL$[0,\pi,\pi]$ and SL$[\pi,0,0]$, respectively. SL$[0,0,\pi]$ is the lowest projected spin liquid of Ref.~\cite{YangParamekantiKim2010} (Sec.~\ref{sec:heis-energy}).

The remaining $32$ realizable fermionic $\Z_2$ PSGs have no SBMF counterpart. Sixteen of them have non-Kramers spinons ($t=+1$), whereas the bosonic spinons of SBMF states are Kramers doublets. Among the $48$ Kramers PSGs the partners are fixed by a simple rule. A mirror $\sigma_v$ maps sites on its line onto themselves, and at such a site the gauge phases give $\sigma_v^2$ and $\sigma_v\T\sigma_v^{-1}\T^{-1}$ the same value; Eq.~\eqref{eq:twist} flips both, so the partners are exactly the $32$ Kramers PSGs in which $\sigma_v^2$ and $\sigma_v\T\sigma_v^{-1}\T^{-1}$ take equal values. The mirrors $\sigma_d$ contain no sites, and the two time reversals realize all four combinations of their values. The other $16$ Kramers PSGs have $\sigma_v^2=-1$ and $\sigma_v\T\sigma_v^{-1}\T^{-1}=+1$, the criterion of Ref.~\cite{LuChoVishwanath2017} for mirror-protected edge modes; $4$ of them satisfy it also for $\sigma_d$. Eight further PSGs satisfy the criterion for $\sigma_d$ alone and are partners of SBMF states with staggered time reversal. Since SBMF states carry no protected edge modes, an ansatz with $\sigma_d$-protected edge modes in these classes would have a nontrivial vison PSG and no SBMF counterpart. Whether a given ansatz in these $24$ classes that satisfy an edge-mode criterion actually carries protected edge modes depends on its band topology, which we have not computed. The chiral states, among them the Kalmeyer--Laughlin candidate, break $\T$ and have no counterpart in the time-reversal-symmetric bosonic classification.

\subsection{Energetics before and after projection}
\label{sec:heis-energy}

\begin{table}
\caption{Unprojected self-consistent energies per site of the Heisenberg model~\eqref{eq:HH} with $\Je=1$, in the Wick decoupling, in which a singlet dimer has energy $-3J/8$ instead of the exact $-3J/4$. Upper part: symmetric nearest-neighbor spin liquids at self-consistent stationary points in the interior of their manifolds (dashes: no such solution), which need not be the global minima of the manifolds; for SL$[0,0,\pi]$ at $\Jt=1.5$, for instance, the global minimum is the dimer limit; SL$[0,\pi,\pi]$ and SL$[\pi,0,0]$ require Lagrange multipliers of opposite sign on the up and down triangles to keep $\langle n_i\rangle=1$. Lower part: products of $\Je$ dimers ($-3/16$) and of bowtie units of two $\Jt$ dimers and one $\Je$ dimer [$-(2\Jt+\Je)/16$].}
\label{tab:energies}
\begin{ruledtabular}
\begin{tabular}{lcccc}
state & $\Jt=1$ & $1.25$ & $1.5$ & $2$\\
\hline
SL$[\frac\pi2,\frac\pi2,\pi]$ & -- & $-0.16986$ & $-0.19834$ & $-0.25878$\\
SL$[\frac\pi2,\frac\pi2,0]$ & $-0.13861$ & $-0.16687$ & $-0.19668$ & $-0.25779$\\
SL$[0,\pi,\pi]$ & -- & -- & $-0.18524$ & $-0.22985$\\
SL$[\pi,0,0]$ & -- & -- & $-0.18422$ & $-0.22915$\\
SL$[0,0,\pi]$ & -- & -- & $-0.17687$ & $-0.21738$\\
SL$[0,0,0]$ & -- & -- & $-0.17662$ & $-0.21737$\\
\hline
$\Je$ dimers & $-0.1875$ & $-0.1875$ & $-0.1875$ & $-0.1875$\\
bowties & $-0.1875$ & $-0.21875$ & $-0.25$ & $-0.3125$
\end{tabular}
\end{ruledtabular}
\end{table}

Before projection, mean-field energies do not decide between spin liquids and valence-bond states, but they show which states are self-consistent. We located the self-consistent stationary points of the Wick energy inside the two-parameter manifolds of the symmetric nearest-neighbor states with exact constraints and compared with unconstrained searches on cells of $6$, $18$, and $24$ sites (Table~\ref{tab:energies}). The chiral class-B states are the lowest symmetric spin liquids wherever they exist, as in the mean-field theory of Ref.~\cite{YangParamekantiKim2010}, where the $\pi/2$-flux states lie below the zero-flux ones. The Kalmeyer--Laughlin candidate SL$[\frac\pi2,\frac\pi2,\pi]$ is self-consistent from $\Jt/\Je=1.25$ on, but not at $1.1$, and lies slightly below SL$[\frac\pi2,\frac\pi2,0]$, which exists already at $\Jt=\Je$. The single-spinon gap of the Kalmeyer--Laughlin candidate decreases from $0.0555$ at $\Jt=1.25$ to $0.0135$ at $\Jt=2$, in the self-consistent units, toward the flat bands of decoupled chiral triangles. Among the sampled unprojected mean-field states, a product of decoupled clusters lies lower at every coupling considered: $\Je$ dimers for $\Jt<\Je$ and coverings of the triangle network by bowtie units for $\Jt>\Je$. This crossing is a property of the decoupling and does not locate a phase transition. Products of decoupled triangles are not included: no Gaussian singlet state of an isolated triangle satisfies all three on-site constraints. The symmetric spin liquids are saddle points of the full problem; at $\Jt=2$, for instance, the energy of the Kalmeyer--Laughlin candidate decreases along a degenerate pair of directions out of its manifold. Unprojected mean-field theory therefore cannot rank the spin liquids against the valence-bond phases, as on the square-kagome lattice~\cite{RajIqbalSQK}. In the large-$N$ theory of Choy and Kim, the limit $\kappa\to0$ of small boson density has a phase dimerized on the $\Je$ bonds for $\Jt<\Je$, the zero-flux spin liquid exactly at $\Jt=\Je$, and for $\Jt>\Je$ a phase with spin correlations on the triangles only; its translation-invariant ansatz, with one amplitude on each bond type, does not contain bowtie coverings~\cite{ChoyKim2009}. At finite $\kappa$ this spin liquid occupies a window around $\Jt=\Je$ whose boundaries are the same for all four nearest-neighbor bosonic states, but its interior is not occupied by the zero-flux state: in the same large-$N$ theory, with nearest-neighbor pairing, the staggered state with partner Z001000(1) lies below both states of Choy and Kim and below the other staggered state at all sampled points inside the window for $0.1\le\kappa\le0.5$, and the $\pi$-flux state lies below the zero-flux state for $\kappa\gtrsim0.15$ at $\Jt=\Je$. The differences are small, at most about $1.3\times10^{-4}(\Jt+\Je)$ per site and flavor for $\kappa\le0.5$, and follow the bosonic fluxes, as expected from flux expulsion, which favors zero flux through the shortest loops at small $\kappa$~\cite{TchernyshyovMoessnerSondhi2006}: the staggered state has zero flux both through the dodecagon and through nine of the fifteen fourteen-bond loops around it, whereas the zero-flux state has flux $\pi$ through all fourteen-bond loops.

Projection changes the ordering. Yang, Paramekanti, and Kim found by variational Monte Carlo on $216$ spins that SL$[0,0,\pi]$ is the lowest projected spin liquid for $0\le\Jt/\Je\le3$, with $-0.462$ per site at $\Jt=\Je$. At $\Jt=2\Je$ its energy is $-0.654\Je$ per spin, below the product of bowtie singlets, $-\Jt/4-\Je/8=-0.625\Je$, whereas in their unprojected mean field, above the dimer product, the same state lies at $-0.498\Je$~\cite{YangParamekantiKim2010}. With triplet fluctuations included in a bond-operator treatment, their $18$-site valence-bond crystal crosses SL$[0,0,\pi]$ only near $\Jt/\Je\approx2.4$~\cite{YangParamekantiKim2010}. For the two Choy--Kim partners we computed projected energies at the isotropic point on the $24$-site torus, optimizing the four nearest-neighbor amplitudes and the boundary conditions of the partons and evaluating the best states by exact enumeration. Z000000(1) reaches $-0.46200$ per site and Z100001(1) $-0.46208$, $1.07\%$ and $1.06\%$ above the exact ground-state energy of the same cluster, $-0.46701$, and degenerate with each other to $8\times10^{-5}$; on this cluster the optimum of Z100001(1) is its $U(1)$ limit SL$[0,0,\pi]$. In Ref.~\cite{YangParamekantiKim2010} the $U(1)$ limit SL$[0,0,0]$ of Z000000(1) lies about $0.011$ above SL$[0,0,\pi]$ at $\Jt=\Je$ on $216$ spins; whether the $\Z_2$ amplitudes remove this difference or the $24$-site torus does not resolve it requires larger clusters. This comparison refers to one finite cluster, whose ground-state energy lies about $0.6\%$ below the infinite-lattice estimates quoted below.

\subsection{Symmetry of the valence-bond phases and their PSGs}

At the isotropic point, iPEPS~\cite{JahromiOrus2018,Ghosh2025} and ED~\cite{Richter2004} place the ground state in the phase of singlets on the $\Je$ bonds, with energies per site of $-0.464114$~\cite{JahromiOrus2018} and $-0.464299$~\cite{Ghosh2025}; ED gives $-0.4649$ on $42$ sites and $-0.4637$ after extrapolation~\cite{Richter2004}. The location and the nature of the transition out of this phase are disputed. Jahromi and Or\'us found a continuous transition at $\Jt/\Je\approx1.1$~\cite{JahromiOrus2018}; exact diagonalization in the nearest-neighbor singlet subspace suggests a transition between $1.1$ and $1.3$~\cite{MisguichSindzingre2007}; Yang, Paramekanti, and Kim estimated $2$ to $2.5$~\cite{YangParamekantiKim2010}; and Ghosh \emph{et al.}\ found a first-order transition at $\Je/\Jt=0.185(5)$, that is, $\Jt/\Je\approx5.4$~\cite{Ghosh2025}. We do not choose among these estimates. By that of Ghosh \emph{et al.}, the whole range of Table~\ref{tab:energies} lies inside the $\Je$-dimer phase. For the phase at large $\Jt$, Jahromi and Or\'us proposed a resonating VBS of triangle dimers that keeps the six-site unit cell, with three different bond energies on each triangle~\cite{JahromiOrus2018}. Ghosh \emph{et al.}\ found a $\sqrt3\times\sqrt3$ VBC of bowtie units, the $18$-site crystal proposed in Refs.~\cite{MisguichSindzingre2007,YangParamekantiKim2010}, nearly degenerate with a columnar bowtie crystal, and noted that the product of bowtie singlets already lies below the large-$\Jt$ energy of Jahromi and Or\'us~\cite{Ghosh2025}.

The symmetry of these states fixes which PSGs can describe their fractionalized neighbors. Each of the three iPEPS patterns in Fig.~4 of Ref.~\cite{JahromiOrus2018} keeps only the mirror perpendicular to the $\Je$ bond of the drawn cell, so its wallpaper group is $cm$, with a single mirror. The three patterns share this mirror and are not related by any element of $p6mm$, so their threefold degeneracy is not enforced by symmetry. This concerns the individual bond patterns. Their equal-weight average, associated with the resonating state discussed in Ref.~\cite{JahromiOrus2018}, has the same energy on every triangle edge. Uniform averaged bond energies alone do not establish symmetry restoration of a coherent superposition: its symmetry also depends on the component wave functions and their relative phases, which have not been reconstructed here. Ghosh \emph{et al.}\ suggested that the symmetry breaking in the earlier iPEPS may come from the square-lattice contraction scheme~\cite{Ghosh2025}; the blocks of the setup used there at large $\Jt$, two triangle edges and one $\Je$ bond, are not $C_3$ symmetric either. The columnar bowtie crystal of Ref.~\cite{Ghosh2025} has the group $cmm$. Their $\sqrt3\times\sqrt3$ crystal is the Kekul\'e covering of the honeycomb network of triangles by bowtie units, and its $18$-site cell has the full point group $6mm$ (it breaks the threefold axes through the triangles of the star lattice, as noted in Ref.~\cite{Ghosh2025}): the sixfold axis passes through the dodecagons that carry no singlet on their $\Je$ bonds, threefold axes pass through the other dodecagons, which carry three $\Je$ singlets each, and twofold axes through the bowtie centers. Its space group is $p6mm$ on the $\sqrt3\times\sqrt3$ superlattice, a subgroup of index $3$, and both crystals are threefold degenerate. The $18$ sites of the cell form two orbits, the $6$ sites of the $\Je$ singlets, which lie on mirrors, and $12$ sites in general position. For these groups, with time reversal, we find $80$ algebraic $\Z_2$ PSGs for $cm$, $448$ for $cmm$, and $112$ for the $\sqrt3\times\sqrt3$ group; respectively $76$, $416$, and $96$ admit nonzero bond amplitudes, the last count being reproduced by an independent method. The $\sqrt3\times\sqrt3$ group has $32$ algebraic $SU(2)$ PSGs, of which $24$ admit nonzero bond amplitudes. These counts refer to the existence of allowed amplitudes, not to realization by a connected ansatz with exactly the named IGG. The $80$ PSGs of the symmetric phase restrict to $80$ distinct $\sqrt3\times\sqrt3$ PSGs, and $32$ of the $96$ classes admitting nonzero amplitudes have no symmetric parent. Unprojected mean-field searches from a range of initial states within the $cm$ and $cmm$ nearest-neighbor manifolds at $\Jt/\Je=1$, $1.5$, and $2$ find dimer coverings, including the columnar bowtie covering, and decoupled chains. Connected stationary ans\"atze also occur: at $\Jt=\Je=1$, a $cmm$ solution has energy approximately $-0.1520$ per site, above the $\Je$-dimer energy $-3/16=-0.1875$. Its on-site multipliers vanish, and its loop holonomies have an $SU(2)$ commutant, so its gauge group is enhanced from the nominal $\Z_2$ class. The solution remains connected upon refining the momentum mesh and lowering the auxiliary temperature, and its energy Hessian within the restricted manifold is nonnegative within numerical precision. This establishes a higher-energy connected stationary ansatz, not a competing spin-liquid phase or a minimum against unrestricted perturbations. Remarks on the energies printed in Ref.~\cite{JahromiOrus2018} are collected in Appendix~\ref{app:remarks}.
\section{Conclusions and Outlook}
\label{sec:conclusion}

We have classified the fermionic parton spin liquids of the Kitaev--Heisenberg model on the star lattice by their PSGs, including the projective action of the $\pi$ rotations $D_2$ and spin--orbit-coupled triplet amplitudes, and used the classification to address the questions raised by the phases found in Refs.~\cite{dOrnellasKnolle2024,JahromiOrus2018,ChoyKim2009}.

The chirality of the Kitaev spin liquid is a property of its PSG class. Whenever the gauge parts of the $\pi$ rotations anticommute and time reversal is kept, time reversal combined with a fixed element $d$ of $D_2$ acts on the spinons only as a sign on each site (a site sign), and every ansatz in the class, of any range, obeys the channel rule of Theorem~\ref{thm:locking}: singlet amplitudes and triplet amplitudes along $d$ join sites of opposite sign, and the other two triplet channels join sites of the same sign. When a symmetry permutes the bond colors cyclically, $d=1$ and every amplitude joins sites of opposite sign. On the star lattice this sign alternates between the two triangles of the unit cell or is uniform, so a time-reversal-symmetric locked spin liquid either has no amplitude at all or none on the triangles, and a Kitaev-type spin liquid must break time reversal, away from the solvable point as well as at it. This places the odd-loop arguments of Refs.~\cite{Kitaev2006,YaoKivelson2007,YouKimchiVishwanath2012,KimchiVishwanath2014,OBrienHermannsTrebst2016,Peri2020,Cassella2023} on the level of the PSG. Appendix~\ref{app:general} shows how the channel rule works on lattices without a cyclic permutation of the bond colors, where $d\neq1$ is possible and odd loops can carry amplitude in the two triplet channels that join sites of the same sign.

The Abelian and non-Abelian Yao--Kivelson liquids, for both signs of the Kitaev coupling and both chiralities, have the single chiral PSG \YK\ of class B, the only chiral class that admits a Chern number. Their transition, a change of the Majorana Chern number at a single Majorana cone~\cite{YaoKivelson2007,Kells2010}, is therefore a band inversion within one PSG at full lattice symmetry, permitted by its projective sign $n_{C_6}=-1$; the topological order changes with the Chern number, as dictated by the sixteenfold way~\cite{Kitaev2006}. For the nearest-neighbor honeycomb Kitaev model at zero field, the gapless phase is in class (I)(B)~\cite{YouKimchiVishwanath2012}, whereas the gapped phases exist only for anisotropic couplings, which lower the symmetry. The PSG is the locked $\Z_2$ descendant of the Kalmeyer--Laughlin candidate of the same lattice, with the gauge parts of all point-group operations tied to their spin rotations.

The Klein duality acts on PSGs through an outer automorphism of the symmetry group, which is not a conjugation by a group element. Self-dual PSGs are locked whenever some symmetry element $t$ that the duality maps onto $td_t$ with $d_t\neq1$ commutes with a $\pi$ rotation $b\neq d_t$, $tbt^{-1}b^{-1}=e$ (Theorem~\ref{thm:duality}), as $T_1$ and $R_y$ do on the star lattice; on the star lattice the $32$ locked classes are exactly the self-dual ones, while all other classes, among them the fermionic partners of the Choy--Kim states and the Kalmeyer--Laughlin candidate, come in dual pairs in which singlet and color-locked triplet amplitudes are exchanged. Since the duality reverses the orientation of the circle of couplings $\phi$ and, as noted by Kimchi and Vishwanath~\cite{KimchiVishwanath2014}, its only self-dual points are the Kitaev points, a connected phase interval mapped onto itself must contain a Kitaev point and, if it is a $\Z_2$ spin liquid, be locked. Self-duality is a property of the class. At the Kitaev points the duality is a symmetry of the Hamiltonian; the Yao--Kivelson ansatz itself is invariant up to a gauge transformation.

The FM/AFM asymmetry of the windows lies beyond unprojected mean fields in the Kitaev representation whose competitors come in pairs of equal energy as in Eq.~\eqref{eq:competitor-pairs}. It maps every state without mixed Majorana correlations at $\phi$ onto a state with exactly the same energy at $\phi+\pi$, and the ordered product states, which carry such correlations, have exactly the energies of the dimer coverings at $\phi+\pi$; the gauge-invariant Abrikosov representation favors product states among all solutions found. Projection lifts this artificial symmetry, but the projected nearest-neighbor state of the chiral PSG, and its alternating-flux counterpart on $24$ sites, reproduce only part of the exact asymmetric response: the ratio of their AFM to FM finite-difference energy-response estimates is about $3.3$, with $96$-site energies evaluated at the parameters optimized on $24$ sites, against $10.4$ for the exact $24$-site spectrum and $11.0$ in the thermodynamic limit. On a $24$-site torus the alternating-flux state matches exact energies to $0.25\%$ for $|J/K|\lesssim0.09$ on the FM side but only for $|J/K|\lesssim0.015$ on the AFM side; most of this asymmetry of the accuracy ranges comes from the exact spectrum of the cluster, which on the AFM side shows a level crossing, and the ratio of the window widths depends on the measure. The exact second-order energy about the Kitaev points locates the asymmetric response of the spin liquid, although it does not determine the window widths. The two off-color terms of a bond create the same four visons with a relative phase set by the sign of $K$, the interference found by Zhang \emph{et al.}\ for flux-pair hopping~\cite{ZhangHalaszZhuBatista2021}; including the creation of four visons from the vacuum, it gives second-order coefficients of $-0.43$ and $-4.70$ per site at the FM and AFM points ($|K|=2$). On the star lattice the effect depends on the bond type. The triangles are exactly sign blind by an operator identity and a threefold selection rule, whereas the inter-triangle bonds carry the whole asymmetry through a low-energy four-vison state at $0.131|K|$, which they admix with a fidelity susceptibility $82$ times larger at the AFM point. The same calculation on the honeycomb lattice gives a ratio of $16.5$ per bond, so the interference itself is not specific to the star lattice, whereas its concentration on one bond type is. Table~\ref{tab:ratios} lists what each of these ratios compares. Only the last row concerns window widths, which are set by first-order transitions into the competing phases; the explanation of Ref.~\cite{dOrnellasKnolle2024}, that the VBS phases lie lower than their FM counterparts, addresses that side of the competition.

\begin{table}
\caption{Ratios of AFM to FM quantities quoted in this paper; the last row is the ratio of the FM to the AFM window width. $E_2$: coefficient of $J^2$ in the energy about the Kitaev points; $\chi_F$: fidelity susceptibility; TL: thermodynamic limit, exact second order; the weight is that of the lowest four-vison state in $E_2$ of an inter-triangle bond; the correlation is the off-color correlation of an inter-triangle bond, whose growth with $J/K$ is compared. Only the last row compares phase boundaries, for self-dual windows consistent with the iDMRG edges of Ref.~\cite{dOrnellasKnolle2024} (about $4$ to $12$ if the uncertainty of $\pm0.005\pi$ of these edges is included); the other rows compare responses at or near the Kitaev points; the projected-state row is a finite-difference estimate, not a size-converged curvature.}
\label{tab:ratios}
\begin{ruledtabular}
\begin{tabular}{lll}
ratio & quantity & Sec./App.\\
\hline
$1$ & $E_2$ of a triangle, TL & \ref{sec:pt}\\
$37$ & $E_2$ of an inter-triangle bond, TL & \ref{sec:pt}\\
$90$ & weight of the lowest four-vison state & \ref{sec:pt}\\
$82$ & $\chi_F$ of an inter-triangle bond, TL & \ref{sec:pt}\\
$11.0$ & $E_2$ per site, TL & \ref{sec:pt}\\
$61$ & $\chi_F$ per site, TL & \ref{sec:pt}\\
$16.5$ & $E_2$ per bond, honeycomb lattice, TL & \ref{sec:pt}\\
$10.4$ & $E_2$ per site, ED, $24$ sites & \ref{sec:vmc}\\
$\approx3.3$ & projected response estimate & \ref{sec:vmc}\\
$12$--$15$ & off-color correlation, projected state & \ref{sec:vmc}\\
$4.6$--$8.9$ & iDMRG window width (FM/AFM) & \ref{app:remarks}\\
\end{tabular}
\end{ruledtabular}
\end{table}

For the Heisenberg model, the two nearest-neighbor Schwinger-boson spin liquids of Choy and Kim correspond to the fermionic PSGs Z000000(1) ($\pi$ flux) and Z100001(1) (zero flux), with a trivial vison PSG as on the kagome lattice, except that the six spins per unit cell make the vison translations commute, and time reversal can also act on the Schwinger bosons with phases $\pm\ii$ on the two triangles, which doubles the bosonic $\Z_2$ PSGs to $32$ and adds two nearest-neighbor bosonic states, whose fermionic partners Z001000(1) and Z101001(1) require on-site multipliers or second-neighbor amplitudes to be $\Z_2$ spin liquids; in the large-$N$ limit the first of these bosonic states lies below both states of Choy and Kim at the sampled points inside their spin-liquid window for $0.1\le\kappa\le0.5$. Of the $64$ realizable fermionic $\Z_2$ PSGs, $32$ have no SBMF counterpart: the $16$ with non-Kramers spinons and the $16$ that satisfy the edge-mode criterion of Lu, Cho, and Vishwanath for $\sigma_v$. Unprojected mean-field states, the Kalmeyer--Laughlin candidate among them, lie above decoupled cluster states at the sampled couplings, but projection overturns this ordering: the projected SL$[0,0,\pi]$ state of Ref.~\cite{YangParamekantiKim2010}, a $U(1)$ limit of the zero-flux partner Z100001(1), lies below the product of valence bonds at $\Jt=2\Je$, and our projected states of both partners lie about $1\%$ above the $24$-site ED energy at the isotropic point. Each iPEPS pattern of the large-$\Jt$ state of Ref.~\cite{JahromiOrus2018} keeps a single mirror, and we gave the PSGs compatible with it and with the columnar and $\sqrt3\times\sqrt3$ bowtie crystals of Ref.~\cite{Ghosh2025}. In DiMACuS (Appendix~\ref{app:dimacus}), the large-$\Jt$ physics reduces at first order in $\Je$ to an unfrustrated model of pseudospins, the lowest Kramers doublets of the triangles, on the honeycomb lattice of triangles; the illustrative exchange scales from the susceptibility and specific-heat fits suggest pseudospin order near $0.07$ to $0.12$~K if the classical-to-quantum transition-temperature estimate applies. This is not a confidence interval; stochastic and model uncertainties remain, and the same model does not reproduce the magnetization curve.

Without lattice symmetry the Majorana bands of a gapped chiral $\Z_2$ liquid belong to the Altland--Zirnbauer symmetry class D (unrelated to the chiral class D of Sec.~\ref{sec:chiral}), whose strong invariant is the integer $\nu$~\cite{SchnyderRyuFurusakiLudwig2008,Kitaev2009}; coupling the fermions to the $\Z_2$ gauge field makes the anyons depend only on $\nu\bmod16$, while $\nu$ itself still sets the edge thermal transport~\cite{Kitaev2006}. With lattice symmetry the bands carry the projective representation fixed by the PSG, and their stable free-fermion band classification for that specified symmetry action is described by twisted equivariant $K$-theory~\cite{FreedMoore2013,ShiozakiSatoGomi2016}. This is a classification after stabilization by trivial bands, not a classification of all interacting Gutzwiller-projected states. For \YK\ the projective relation $C_6^6=-1$ restricts the $C_6$ eigenvalues to $e^{\ii\pi(2m+1)/6}$; the rotation eigenvalues of the occupied bands then determine $\nu\bmod6$, and the relation permits the single-cone change of $\nu$ by one at $\Gamma$ (Sec.~\ref{sec:kitaev-psg})~\cite{FangGilbertBernevig2012}. These eigenvalues fix the parity of $\nu$, and with it whether the anyons are Abelian, but not $\nu\bmod16$; since the PSG encodes only the projective action on the fermion, the symmetry action on the visons and on the $\sigma$ anyon is additional data~\cite{EssinHermele2013,Barkeshli2019}.

The classification constrains mean-field amplitudes but does not select a ground state. The projected states contain nearest-neighbor amplitudes only and no Jastrow factors, and their comparison with exact results is limited to $24$ sites; their finite-difference energy-response estimates are about $61\%$ (FM) and $18\%$ (AFM) of the thermodynamic-limit perturbative coefficients, with $96$-site energies evaluated at parameters optimized on $24$ sites rather than obtained by an independent size-converged optimization. Perturbation theory identifies which bonds make the AFM Kitaev spin liquid sensitive to the Heisenberg coupling, but not the positions of the transitions. The projected energies of the Choy--Kim partners are limited to $24$ sites and nearest-neighbor amplitudes. The $U(1)$ classification was used only qualitatively, and the protected edge modes of the $24$ $\Z_2$ classes that satisfy the edge-mode criterion of Lu, Cho, and Vishwanath were not computed. The DiMACuS analysis rests on a first-order pseudospin model and on the sign of the Dzyaloshinskii--Moriya (DM) vectors obtained from density-functional theory. These limitations call for size-converged benchmarks of projected chiral states with longer-range amplitudes, a direct comparison of the AFM Kitaev spin liquid with the singlet and triplet VBSs that border it, projected energies of the Choy--Kim partners on larger clusters, and low-temperature measurements on DiMACuS.

\begin{acknowledgments}
Y.I. thanks Chunxiao Liu, Julien Vidal and Atanu Maity for helpful discussions.
H.R. acknowledges the Council of Scientific and Industrial Research (CSIR), Government of India, for financial support through a Senior Research Fellowship (File No.~09/0084(16360)/2023-EMR-I).
The work of Y.I. was performed, in part, at the Aspen Center for Physics, which is supported by a grant from the Simons Foundation (1161654, Troyer). This research was also supported in part by grant NSF PHY-2309135 to the Kavli Institute for Theoretical Physics (KITP). Y.I. acknowledges support from the Abdus Salam International Centre for Theoretical Physics (ICTP) through the Associates Programme, from the Simons Foundation through Grant No.~284558FY19, from IIT Madras through the Institute of Eminence (IoE) program for establishing QuCenDiEM (Project No.~SP22231244CPETWOQCDHOC), and thanks the International Centre for Theoretical Sciences (ICTS), Bengaluru, for participation in the Discussion Meeting ``Fractionalized Quantum Matter'' (code: ICTS/DMFQM2025/07).  We acknowledge the use of the AI tools Claude Opus 5.5 and Claude Fable 5.1 (Anthropic) for assistance with numerical calculations, verifying some of the PSG algebra, and preparing parts of the manuscript. The authors reviewed all results and changes and take full responsibility for the content.
\end{acknowledgments}

\section*{Data availability}
The data that support the findings of this article are available from the authors upon reasonable request.

\makeatletter\let\SM@hangfrom@section\@hangfrom@section\let\SM@sectioncntformat\@sectioncntformat\makeatother
\appendix
\section{Canonical Forms of the \texorpdfstring{$\Z_2$}{Z2} PSGs}
\label{app:forms}

\subsection{Canonical gauge}

We fix the gauge in the standard way, translations first~\cite{Wen2002,RajIqbalSHD}. Evaluating $T_2T_1T_2^{-1}T_1^{-1}$ gives
\begin{equation}
G_{T_1}(x,y,\mu)=\tau^0,\qquad G_{T_2}(x,y,\mu)=n_{T_2}^{\,x}\,\tau^0 .
\label{eq:canon-T}
\end{equation}
For every other generator $O\in\{C_6,S,R_x,R_y,\T\}$ the relations $OT_aO^{-1}W_a^{-1}=e$, where $W_a$ is the translation equal to $OT_aO^{-1}$, become recursions in $x$ and $y$ whose general solution is
\begin{equation}
G_O(x,y,\mu)=(-1)^{\sigma_O(x,y,\mu)}\,g_O(\mu),
\label{eq:canon-O}
\end{equation}
with $\sigma_O$ an affine function of the bits $n_{T_2}$ and of the relation values of $OT_aO^{-1}W_a^{-1}$, obtained by propagating these relations from $\sigma_O(0,0,\mu)=0$; for $C_6$ it contains the term $x(x+1)/2$, as on the SHD lattice~\cite{RajIqbalSHD}, so its period is four. The sublattice gauge fixes $g_{C_6}(\mu)=\tau^0$ for $\mu=2,\dots,6$, and the remaining matrices $g_O(\mu)$ can be conjugated into a finite subgroup of $SU(2)$, for the star lattice the dicyclic group of order $48$. All $26$ relations are then imposed on every site of a patch of cells, and every solution also satisfies them on a larger patch. Two solutions are equivalent if they are related by a gauge transformation or a re-lifting; each class is identified by the minimum, over the $2^7$ re-liftings, of the relation values together with the on-site character at the reference site. This search finds exactly the $352$ classes of Eq.~\eqref{eq:count}.

\subsection{On-site types}

\begin{table}
\caption{On-site gauge parts of $(S,R_x,R_y,\T)$ at the site $(0,0,1)$, whose stabilizer contains $S$, in the canonical gauge, and the number of time-reversal-symmetric $\Z_2$ PSGs of each type. Type VI contains exactly the locked PSGs. For the chiral PSGs the types I, III, and VI occur, without the entry for $\T$, with $32$, $32$, and $16$ PSGs.}
\label{tab:onsite}
\begin{ruledtabular}
\begin{tabular}{cccccc}
type & $G_S$ & $G_{R_x}$ & $G_{R_y}$ & $G_\T$ & PSGs\\
\hline
I & $\tau^0$ & $\tau^0$ & $\tau^0$ & $\tau^0$ & $64$\\
II & $\tau^0$ & $\tau^0$ & $\tau^0$ & $\ii\tau^1$ & $64$\\
III & $\ii\tau^1$ & $\tau^0$ & $\tau^0$ & $\tau^0$ & $64$\\
IV & $\ii\tau^1$ & $\tau^0$ & $\tau^0$ & $\ii\tau^1$ & $64$\\
V & $\ii\tau^1$ & $\tau^0$ & $\tau^0$ & $\ii\tau^2$ & $64$\\
VI & $\ii\tau^1$ & $\frac{\ii(\tau^1+\tau^2)}{\sqrt2}$ & $\frac{\ii(\tau^1-\tau^2)}{\sqrt2}$ & $\tau^0$ & $32$
\end{tabular}
\end{ruledtabular}
\end{table}

At a site the gauge parts of its stabilizer form a representation of the double cover of $D_4\times\Z_2^\T$ with the prescribed relation values. Table~\ref{tab:onsite} lists the six types that occur. In types I to V the $\pi$ rotations are represented by central matrices, and $G_{R_x}$ and $G_{R_y}$ are $\pm\tau^0$ on every site; the ratio $\zeta=G_{R_x}(i_z)/G_{R_x}(i_x)=\pm1$ of these central matrices on two sites of one cell with $z$ and $x$ inter-triangle bonds is the invariant that separates the PSGs -$b$ ($\zeta=+1$) and -$a$ ($\zeta=-1$). Gauge transformations leave central matrices unchanged, a re-lifting of $R_x$ flips both factors, and the relations $T_aR_xT_a^{-1}R_x^{-1}$ multiply $G_{R_x}$ by the same sign on all sites of a cell, so $\zeta$ does not depend on the cell. The Heisenberg PSGs embed as -$b$ classes, with trivial gauge parts of $D_2$. In type VI the $\pi$ rotations act through the quaternion group, and the stabilizer is represented through the faithful two-dimensional representation of the double cover of $D_4$; since the two faithful two-dimensional representations of this group are exchanged by the re-lifting $G_S\to-G_S$, the locked PSGs are not doubled.

\subsection{The Yao--Kivelson PSG}

For the Yao--Kivelson ansatz of Sec.~\ref{sec:kitaev-psg} the PSG solved from the ansatz has, in a translation-invariant gauge, $G_{T_1}=G_{T_2}=\tau^0$ and
\begin{align}
G_{C_6}(\mu)&=\pm\tfrac12\big(\tau^0+\ii\tau^1-\ii\tau^2+\ii\tau^3\big),\nonumber\\
G_{S\T}(\mu)&=\pm\tfrac{\ii}{\sqrt2}\big(\tau^1+\tau^2\big),\nonumber\\
G_{R_x}&=\ii\tau^1,\quad G_{R_y}=\ii\tau^2,\quad G_{R_z}=-\ii\tau^3,
\label{eq:yk-gauge}
\end{align}
with sublattice-dependent signs. Each gauge part is the $SU(2)$ lift of the corresponding spin rotation, $C_3^{[111]}$, $C_2^{[1\bar10]}$, or a $\pi$ rotation, mapped into gauge space by the isomorphism $(\sigma^x,\sigma^y,\sigma^z)\to(\tau^1,-\tau^2,\tau^3)$. The relation values are those of Eq.~\eqref{eq:yk-psg}; in addition $SR_zS^{-1}R_z^{-1}=-1$ in the gauge sector, which is the commutator of the spin lifts of $C_2^{[1\bar10]}$ and $R_z$.
\section{The Locked Classes}
\label{app:locked}

\begin{table}
\caption{The $32$ spin-gauge-locked time-reversal-symmetric $\Z_2$ PSGs. Column $n_{\T C_6}$ gives the character $\chi(C_6)$ of Theorem~\ref{thm:locking}; the next columns give the real dimensions singlet/triplet of the allowed amplitudes on the reference bonds of Table~\ref{tab:bonds} ($0$: forbidden). The $\Jt$ and $J_{3b}$ amplitudes and all on-site terms vanish in every locked class and are omitted. Last column: shortest range $R_1$ ($\Jt,\Je$), $R_2$ ($+J_2$), $R_3$ ($+J_{3a},J_{3b}$), or $R_4$ ($+J_4$) at which a generic ansatz with singlet and triplet amplitudes is connected with IGG $\Z_2$.}
\label{tab:locked}
\begin{ruledtabular}
\begin{tabular}{rlccccc c}
No. & PSG & $n_{\T C_6}$ & $\Je$ & $J_2$ & $J_{3a}$ & $J_4$ & range\\
\hline
1 & Z000100010 & $+1$ & 0 & 0 & 0 & 0 & --\\
2 & Z000100011 & $+1$ & 0 & 0 & 0 & 0 & --\\
3 & Z000101010 & $+1$ & 0 & 0 & 0 & 0 & --\\
4 & Z000101011 & $+1$ & 0 & 0 & 0 & 0 & --\\
5 & Z000110010 & $+1$ & 0 & 0 & 0 & 0 & --\\
6 & Z000110011 & $+1$ & 0 & 0 & 0 & 0 & --\\
7 & Z000111010 & $+1$ & 0 & 0 & 0 & 0 & --\\
8 & Z000111011 & $+1$ & 0 & 0 & 0 & 0 & --\\
9 & Z001100010 & $-1$ & 0 & 1/3 & 1/2 & 0 & $R_2$\\
10 & Z001100011 & $-1$ & 1/1 & 1/3 & 1/2 & 1/1 & $R_2$\\
11 & Z001101010 & $-1$ & 1/2 & 1/3 & 1/2 & 1/3 & $R_2$\\
12 & Z001101011 & $-1$ & 0/1 & 1/3 & 1/2 & 0/2 & $R_2$\\
13 & Z001110010 & $-1$ & 0 & 1/3 & 0/1 & 0 & $R_2$\\
14 & Z001110011 & $-1$ & 0 & 1/3 & 0/1 & 1/1 & $R_2$\\
15 & Z001111010 & $-1$ & 0/1 & 1/3 & 0/1 & 1/3 & $R_2$\\
16 & Z001111011 & $-1$ & 0/1 & 1/3 & 0/1 & 0/2 & $R_2$\\
17 & Z100100010 & $+1$ & 0 & 0 & 0 & 0 & --\\
18 & Z100100011 & $+1$ & 0 & 0 & 0 & 0 & --\\
19 & Z100101010 & $+1$ & 0 & 0 & 0 & 0 & --\\
20 & Z100101011 & $+1$ & 0 & 0 & 0 & 0 & --\\
21 & Z100110010 & $+1$ & 0 & 0 & 0 & 0 & --\\
22 & Z100110011 & $+1$ & 0 & 0 & 0 & 0 & --\\
23 & Z100111010 & $+1$ & 0 & 0 & 0 & 0 & --\\
24 & Z100111011 & $+1$ & 0 & 0 & 0 & 0 & --\\
25 & Z101100010 & $-1$ & 1/2 & 1/3 & 1/2 & 1/3 & $R_2$\\
26 & Z101100011 & $-1$ & 0/1 & 1/3 & 1/2 & 0/2 & $R_2$\\
27 & Z101101010 & $-1$ & 0 & 1/3 & 1/2 & 0 & $R_2$\\
28 & Z101101011 & $-1$ & 1/1 & 1/3 & 1/2 & 1/1 & $R_2$\\
29 & Z101110010 & $-1$ & 0/1 & 1/3 & 0/1 & 1/3 & $R_2$\\
30 & Z101110011 & $-1$ & 0/1 & 1/3 & 0/1 & 0/2 & $R_2$\\
31 & Z101111010 & $-1$ & 0 & 1/3 & 0/1 & 0 & $R_2$\\
32 & Z101111011 & $-1$ & 0 & 1/3 & 0/1 & 1/1 & $R_2$\\
\end{tabular}
\end{ruledtabular}
\end{table}
\begin{table}
\caption{The $16$ locked chiral $\Z_2$ PSGs in class B, with the real dimensions singlet/triplet of the allowed amplitudes on the reference bonds and the shortest range at which a generic ansatz is connected with IGG $\Z_2$ (notation of Table~\ref{tab:locked}). The Yao--Kivelson PSG is \YK. On-site terms are forbidden in all of them. The other chiral classes are listed in the Supplemental Material~\cite{SM}.}
\label{tab:locked-chiral}
\begin{ruledtabular}
\begin{tabular}{rlccccccc}
No. & PSG & $\Jt$ & $\Je$ & $J_2$ & $J_{3a}$ & $J_{3b}$ & $J_4$ & range\\
\hline
1 & ZC010010 & 1/2 & 0 & 1/3 & 0/1 & 1/2 & 0 & $R_2$\\
2 & ZC010011 & 1/2 & 0 & 1/3 & 0/1 & 1/2 & 1/1 & $R_2$\\
3 & ZC010110 & 1/2 & 0/1 & 1/3 & 0/1 & 1/2 & 1/3 & $R_1$\\
4 & ZC010111 & 1/2 & 0/1 & 1/3 & 0/1 & 1/2 & 0/2 & $R_1$\\
5 & ZC011010 & 1/2 & 0 & 1/3 & 1/2 & 1/2 & 0 & $R_2$\\
6 & ZC011011 & 1/2 & 1/1 & 1/3 & 1/2 & 1/2 & 1/1 & $R_1$\\
7 & ZC011110 & 1/2 & 1/2 & 1/3 & 1/2 & 1/2 & 1/3 & $R_1$\\
8 & ZC011111 & 1/2 & 0/1 & 1/3 & 1/2 & 1/2 & 0/2 & $R_1$\\
9 & ZC110010 & 1/2 & 0/1 & 1/3 & 0/1 & 1/2 & 1/3 & $R_1$\\
10 & ZC110011 & 1/2 & 0/1 & 1/3 & 0/1 & 1/2 & 0/2 & $R_1$\\
11 & ZC110110 & 1/2 & 0 & 1/3 & 0/1 & 1/2 & 0 & $R_2$\\
12 & ZC110111 & 1/2 & 0 & 1/3 & 0/1 & 1/2 & 1/1 & $R_2$\\
13 & ZC111010 & 1/2 & 1/2 & 1/3 & 1/2 & 1/2 & 1/3 & $R_1$\\
14 & ZC111011 & 1/2 & 0/1 & 1/3 & 1/2 & 1/2 & 0/2 & $R_1$\\
15 & ZC111110 & 1/2 & 0 & 1/3 & 1/2 & 1/2 & 0 & $R_2$\\
16 & ZC111111 & 1/2 & 1/1 & 1/3 & 1/2 & 1/2 & 1/1 & $R_1$\\
\end{tabular}
\end{ruledtabular}
\end{table}

Table~\ref{tab:locked} lists the $32$ locked time-reversal-symmetric $\Z_2$ PSGs with their allowed amplitudes. All of them have $n_{\T S}=+1$, as required by Theorem~\ref{thm:locking} for the mirror that fixes a site. The $16$ classes with $n_{\T C_6}=+1$ have a uniform sign $\theta$ and allow no amplitude at all. In the $16$ classes with $n_{\T C_6}=-1$, $\theta(x,y,\mu)=(-1)^\mu$; the triangle edges, the $J_{3b}$ bonds, and the on-site terms are forbidden, the $J_2$ bond is allowed in all of them, and a generic ansatz is a connected $\Z_2$ spin liquid from the second shell on. With singlet amplitudes alone these ans\"atze have IGG $SU(2)$; the triplet amplitudes reduce it to $\Z_2$. All $32$ locked classes are the self-dual time-reversal-symmetric classes of Sec.~\ref{sec:duality}, and none of them contains a state with Kitaev-like hopping on all nearest-neighbor bonds.

Table~\ref{tab:locked-chiral} lists the $16$ locked chiral PSGs in class B. Here no time-reversal constraint acts, the triangle edges carry amplitudes in every class, and ten of them are realized as $\Z_2$ spin liquids with nearest-neighbor amplitudes alone, among them the Yao--Kivelson PSG \YK. All $16$ are self-dual.
\section{Locking and Klein Duality on Other Kitaev Lattices}
\label{app:general}

\begin{table*}
\caption{Theorems~\ref{thm:locking} and~\ref{thm:duality} on four Kitaev-colored lattices with the full symmetry group $G_{\rm KH}\times\Z_2^\T$ of a nearest-neighbor KH model. Color image: the permutations of the bond colors induced by the space group ($S_3$ contains a cyclic permutation, $\Z_2$ exchanges two colors). Locked: number of locked time-reversal-symmetric $\Z_2$ PSGs, split by the element $d$ of Theorem~\ref{thm:locking}. $G_{\T d}\propto\tau^0$: locked classes in which the gauge part of $\T d$ is $\pm\tau^0$ on every site of a $7\times7$-cell patch for exactly one $d\in D_2$, the same on all sites; Eq.~\eqref{eq:thm} then follows from the invariance condition~\eqref{eq:psg}. All NN: locked classes in which every nearest-neighbor orbit carries a nonzero symmetry-allowed amplitude ($d=1$ / $d=R_z$). Self-dual: locked classes fixed by the Klein duality.}
\label{tab:general}
\begin{ruledtabular}
\begin{tabular}{lccccccc}
lattice & group & color image & Klein pattern & locked ($d=1$/$d=R_z$) & $G_{\T d}\propto\tau^0$ & all NN & self-dual\\
\hline
star & $p6mm$ & $S_3$ & $2\times2$ cells & $32$ ($32$/$0$) & $32$ & $0$/$0$ & $32$\\
honeycomb & $p6mm$ & $S_3$ & $2\times2$ cells & $32$ ($32$/$0$) & $32$ & $10$/$0$ & $32$\\
square-octagon & $p4mm$ & $\Z_2$ & within the cell & $2048$ ($1024$/$1024$) & $2048$ & $140$/$288$ & $512$\\
pentaheptite & $cmm$ & $\Z_2$ & none & $1024$ ($512$/$512$) & $1024$ & $0$/$96$ & --
\end{tabular}
\end{ruledtabular}
\end{table*}

The proofs of Theorems~\ref{thm:locking} and~\ref{thm:duality} use only the anticommutation of the gauge parts of $R_x$ and $R_y$ on each site, the normality of $D_2$, and the action of the duality on the symmetry group. Beyond the star lattice, we classified the locked $\Z_2$ PSGs of the nearest-neighbor KH model on the honeycomb lattice, on the square-octagon lattice $4.8.8$ with the coloring of Ref.~\cite{YangZhouSun2007}, and on the pentaheptite lattice of Ref.~\cite{Peri2020}, a honeycomb lattice with one Stone--Wales defect per $2\times2$ cell and faces with five and seven edges, using the coloring that is invariant under its full space group. For each lattice the symmetry group was built from the geometry and the coloring and agrees with the graph automorphisms of finite periodic clusters that preserve the coloring up to a global color permutation. Table~\ref{tab:general} summarizes the results.

In every locked class on the four lattices the gauge part of $\T d$ is a scalar on every site for exactly one $d$, as Theorem~\ref{thm:locking} states; Eq.~\eqref{eq:thm} is then part of the invariance conditions, and the allowed amplitudes on all bond orbits up to graph distance three obey it. No locked class allows an on-site term, also for $d=R_z$, where $D_2$ rather than Eq.~\eqref{eq:thm} removes the on-site triplet field, and the singlet amplitude vanishes on every bond with $\theta(i)\theta(j)=+1$. On the star and honeycomb lattices the cyclic color permutation forces $d=1$. On the square-octagon and pentaheptite lattices the color image is $\Z_2$, and half of the locked classes have $d=R_z$. In these classes the odd loops of the pentaheptite lattice can carry amplitude on all of their bonds. In $96$ of the $512$ classes with $d=R_z$ every nearest-neighbor orbit is allowed, and the singlet amplitude vanishes on the bonds with $\theta(i)\theta(j)=+1$, which carry only $x$- and $y$-triplet amplitude. Every odd loop contains an odd number of such bonds, because the product of $\theta(i)\theta(j)$ around a loop is $+1$; the other bonds of pentagons and heptagons can carry singlet amplitude, and in $72$ of the $96$ classes some pentagon edges do. No locked time-reversal-symmetric class on the star or pentaheptite lattice allows singlet amplitudes on all nearest-neighbor bonds, whereas on the bipartite honeycomb and square-octagon lattices such classes exist, as the time-reversal-symmetric Kitaev solutions require.

For Theorem~\ref{thm:duality}, Klein patterns exist on the star and honeycomb lattices with $2\times2$ cells and on the square-octagon lattice within a single cell, where the induced automorphism fixes the translations and maps $C_4\to C_4R_x$ (for a suitable global frame of the pattern); it is outer nonetheless. On the pentaheptite lattice no Klein pattern exists for any coloring and period, because consecutive edges of a cycle have different colors, so a pentagon contains two colors twice and the third once, which violates the parity criterion of Ref.~\cite{KimchiVishwanath2014}. The direction ``self-dual $\Rightarrow$ locked'' holds on all three lattices with a Klein pattern, as it must. The converse holds for the full groups of the star and honeycomb lattices but fails in general. On the square-octagon lattice $512$ of the $2048$ locked classes are self-dual: in all $1024$ classes with $d=R_z$ the relation value of $C_4\T C_4^{-1}\T^{-1}$ changes, in $512$ of them together with that of $(SC_4)^2$, and in $512$ of the classes with $d=1$ only that of $(SC_4)^2$ changes. On the honeycomb lattice with the group $p6$ instead of $p6mm$, only $16$ of the $32$ locked classes are self-dual, the others changing the relation value of $T_2T_1T_2^{-1}T_1^{-1}$, and with $p3$ only $8$ of $16$; the mirrors are needed for the equivalence. The counts on finite tori agree with these infinite-lattice counts only if periodic boundary conditions for the spinons, $T_1^L=T_2^L=+1$, are imposed.
\section{The Spin-\texorpdfstring{$1/2$}{1/2} Star Magnet DiMACuS}
\label{app:dimacus}\label{sec:dimacus}

The only spin-$1/2$ star-lattice magnet known to us is the copper sulfate $[(\mathrm{CH_3})_2\mathrm{NH_2}]_3$\allowbreak$[\mathrm{Cu_3}(\mu_3\text{-}\mathrm{OH})$\allowbreak$(\mu_3\text{-}\mathrm{SO_4})$\allowbreak$(\mu_3\text{-}\mathrm{SO_4})_3]$\allowbreak$\cdot0.24\mathrm{H_2O}$ (DiMACuS)~\cite{Sorolla2020,Ishikawa2024}; the other star-lattice magnets known to us contain high-spin Fe(III)~\cite{Zheng2007,Ji2022}. The Cu triangles of DiMACuS are equilateral and rotated, and the Cu atoms of a layer lie at two heights: each $\Je$ bond joins triangles $1.22$~\AA\ apart along $c$ and has an inversion center at its midpoint~\cite{Sorolla2020,Ishikawa2024}. The layer group is therefore $p\bar3$; its projection onto the plane has the wallpaper group $p6$, in which the apparent twofold rotation is the inversion. Inversion leaves spins invariant, whereas a twofold rotation about $c$ rotates them, so a description with spin-orbit coupling must use $p\bar3\times\Z_2^\T$ with this spin action. For spin-rotation-invariant models only the action on the sites matters, and $p6\times\Z_2^\T$ has $16$ $\Z_2$ PSGs, $12$ of them realizable. Density-functional theory (DFT; full-potential local-orbital code FPLO, $U_d=9.5$~eV) gives $\Jt=81.5$~K, $\Je=5.4$~K (called $J_D$ in Ref.~\cite{Ishikawa2024}), and $J_H=0.3$~K between the triangles; relativistic calculations give a Dzyaloshinskii--Moriya (DM) interaction on the triangle bonds, with $D_\parallel/\Jt=0.493$ along $c$ and $D_\perp/\Jt=0.274$ in the plane [Eq.~(S15) of the Supplemental Material of Ref.~\cite{Ishikawa2024}], whereas inversion forbids it on the $\Je$ bonds; the DM interaction on the $J_H$ bonds was not computed and is neglected here. These ratios refer to the relativistic $\Jt=66.9$~K; we apply them to $\Jt=81.5$~K. We call these values, with the DM vectors of Eqs.~(S12)--(S14) of the Supplemental Material of Ref.~\cite{Ishikawa2024}, the DFT set; the fit set is $\Jt=57$~K, $|D|/\Jt=0.42$ from the susceptibility, which it describes down to about $10$--$15$~K~\cite{Ishikawa2024}; for it we keep the DFT ratio $D_\perp/D_\parallel=0.556$ and the DFT directions (the ratio $1/\sqrt3$ used in the fit of Ref.~\cite{Ishikawa2024} gives $36.3$~K, $11.3^\circ$, and $2.9\%$ instead of the values quoted below). No order is found down to $0.1$~K, and a $1/3$ magnetization plateau appears for fields along $c$ only~\cite{Ishikawa2024}.

\emph{Pseudospin model.} In this large-$\Jt$ regime ($\Je/\Jt\approx0.07$) the DM interaction isolates the lowest Kramers doublet of each triangle (gap $70.3$~K for the DFT set, $36.6$~K for the fit set), which defines a pseudospin $\bm s=\bm\sigma/2$. Projecting $\Je$ and $J_H$ onto the product space of these doublets gives, to first order in $\Je$ and $J_H$, on the honeycomb lattice of triangles,
\begin{equation}
H_{\rm eff}=\sum_{\langle ab\rangle}\big[J_\perp(s_a^xs_b^x+s_a^ys_b^y)+J_\parallel s_a^zs_b^z+\Gamma(s_a^ys_b^z+s_a^zs_b^y)\big],
\label{eq:dimacus}
\end{equation}
with the same coupling on all three bond directions when the doublet of a $B$ triangle is taken as the inversion image of the $A$ doublet. Here $s^z$ is the component that couples to the $c$ axis: with $P$ the projector onto the doublet, $PS^z_{\rm tot}P=1.022\,s^z$ and $P\bm S^\perp_{\rm tot}P=0$. For the DFT set $J_\perp=2.347$~K, $J_\parallel=0.705$~K, and $|\Gamma|=0.429$~K; the principal values are $0.600$, $2.347$, and $2.453$~K, i.e., $\Je/9$ and $(4\Je/9)(1\mp0.022)$. Including $J_H$ gives $J_\perp=2.217$~K, $J_\parallel=0.770$~K, and $|\Gamma|=0.405$~K (principal values $0.665$, $2.217$, $2.323$~K). In the principal-axis XXZ approximation to Eq.~\eqref{eq:dimacus}, used below, $J_z$ denotes the hard-axis principal value and $J_{xy}$ the mean of the two in-plane ones, $J_{xy}=4\Je/9=2.40$~K for the DFT set; the first-order principal values give $J_z/J_{xy}=0.250$, and the exact two-triangle projection gives $0.247$, which we use below. For $J_H=0$, exact projections on clusters of two to four triangles change the nearest-neighbor principal values by less than $1.9\%$ for the DFT set and $2.4\%$ for the fit set and generate second-neighbor couplings below $0.5\%$ of $J_{xy}$; three-pseudospin terms vanish by time reversal. The identical, symmetric, positive-definite bond matrices here can be diagonalized by a common pseudospin rotation; the resulting bipartite nearest-neighbor model is unfrustrated. For the opposite DM sign and $J_H=0$, the projected model is XXZ, with $J_\perp=4\Je/9$ and $J_\parallel=\Je/9$. With both $J_H$ and transverse DM retained, small direction-dependent corrections remain; for the sign-reversed DFT set with $J_H=0.3$~K, a projected bond has principal values $0.66480$, $2.26368$, and $2.27003$~K. We use the vectors of Eqs.~(S12)--(S14) of Ref.~\cite{Ishikawa2024} with $H_{\rm DM}=\bm D_{ij}\cdot(\bm S_i\times\bm S_j)$ and $A^{\alpha\beta}=\epsilon^{\alpha\beta\gamma}D^\gamma$, the convention used throughout that Supplemental Material (Appendix~\ref{app:remarks}); reversing all DM vectors gives the XXZ case when $J_H=0$. Ishikawa \textit{et al.} estimated the inter-triangle scale as $J_D/9$ and suggested that this reduction may explain the absence of order~\cite{Ishikawa2024}. That value is the hard-axis coupling; the in-plane coupling $4\Je/9$, which sets the ordering temperature, is four times larger. For the Heisenberg model without DM interaction, Reingruber \emph{et al.}\ derived first-order effective models of the trimers on the honeycomb lattice, which keep the chirality of each triangle, and argued that the absence of order in DiMACuS is expected~\cite{Reingruber2024}; the DM interaction lifts the chirality degeneracy and leaves the unfrustrated pseudospin model of Eq.~\eqref{eq:dimacus}.

\emph{Ordered state.} The easy axis of Eq.~\eqref{eq:dimacus} is tilted $13.8^\circ$ out of the plane ($14.6^\circ$ with $J_H$; $11.0^\circ$ for the fit set), and the two in-plane principal values differ by $4.4\%$. A twofold anisotropy is relevant along the whole line of Berezinskii--Kosterlitz--Thouless (BKT) fixed points~\cite{KosterlitzThouless1973,JoseKadanoffKirkpatrickNelson1977}, so the transition is expected to be Ising-like, with true long-range order below $T_c$. Classical Monte Carlo simulations of the principal-axis model ($J_z/J_{xy}=0.247$, $4.4\%$ splitting, up to $2\times48^2$ sites) give $T_c/T_{\rm BKT}\approx1.2$, where $T_{\rm BKT}$ refers to the model without splitting. This factor assumes the $4.4\%$ splitting of the DFT set; for the fit set the splitting is $2.8\%$, and the factor is then an upper estimate. For the quantum XXZ model, stochastic series expansion simulations~\cite{Sandvik1999,SyljuasenSandvik2002} on five system sizes at $J_z/J_{xy}=0.253$ give the estimate $T_{\rm BKT}\approx0.21J_{xy}$; fits with different logarithmic finite-size corrections give approximately $0.205$--$0.213$ in units of $J_{xy}$, with additional sampling uncertainty. Simulations at $J_z/J_{xy}=0.247$ on $L=6,12,24$ tori agree within that uncertainty. We therefore use $T_c\approx0.25J_{xy}$ if the classical ratio carries over. In the ordered state the pseudospins form a N\'eel state along the tilted easy axis. The in-plane components of the Cu spins form a $120^\circ$-type pattern with zero net in-plane moment per triangle. The tilt gives each triangle a moment along $c$ of $1.022\sin13.8^\circ=0.24$ of the pseudospin order parameter, opposite on the $A$ and $B$ triangles. For the opposite DM sign, the $J_H=0$ projected XXZ model has an easy plane and a BKT transition. For the DFT set the model orders at $T_c\approx0.6$~K ($T_{\rm BKT}=0.50$~K), well above the $0.1$~K bound.

\begin{figure}
\centering
\includegraphics[width=\columnwidth]{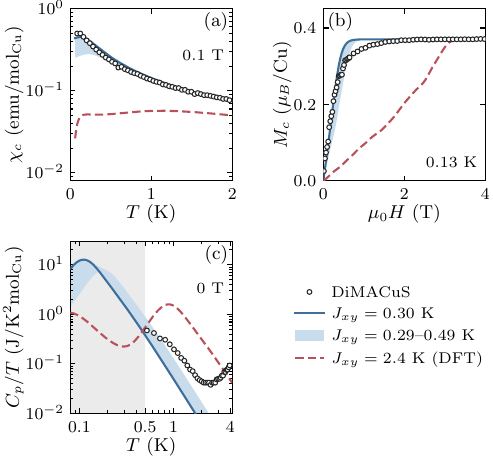}
\caption{DiMACuS: (a) susceptibility along $c$ at $0.1$~T, (b) magnetization along $c$ at $0.13$~K, and (c) specific heat divided by temperature in zero field. Circles: data of Ref.~\cite{Ishikawa2024}. Lines: finite-temperature Lanczos results on $24$ triangles, using $64$ random trace vectors per non-exact magnetization sector and $220$ Lanczos steps (small sectors treated exactly), for the XXZ model in principal axes ($J_z/J_{xy}=0.247$; the $4.4\%$ in-plane splitting and the tilt of the hard axis from $c$ are neglected), with pseudospin $g$ factor $2.22$ along $c$, fixed by the measured plateau magnetization of $0.37\,\mu_B$/Cu~\cite{Ishikawa2024}, for the DFT coupling $J_{xy}=4\Je/9=2.40$~K (dashed; the steps in (b) are finite-size effects at $T/J_{xy}=0.05$) and for $J_{xy}=0.30$~K (solid); the illustrative band covers $J_{xy}=0.29$ to $0.49$~K and is not a confidence interval. No data exist below $0.5$~K in (c) (shaded), and the model contains the magnetic contribution only.}
\label{fig:dimacus}
\end{figure}

\emph{Comparison with experiment.} Figure~\ref{fig:dimacus} compares finite-temperature Lanczos results~\cite{JaklicPrelovsek1994} of the principal-axis XXZ approximation to Eq.~\eqref{eq:dimacus} with the data of Ref.~\cite{Ishikawa2024}. With the DFT coupling the finite-cluster calculation underestimates $\chi_c$ at $0.1$~K (by a factor of about $11$ in the calculation shown in Fig.~\ref{fig:dimacus}, a factor sensitive to cluster size and stochastic trace sampling because $0.1$~K lies below the finite-size gap) and $M_c$ at $0.5$~T by a factor of $6$ to $7$, and it overestimates $C_p/T$ at $1$~K by a factor of $7$. Fitting $J_{xy}$ instead gives approximately $0.30$~K from $\chi_c$ and $0.49$~K from $C_p/T$ between $0.5$ and $1$~K in the $24$-triangle calculation shown. The best susceptibility fit, determined to $0.005$~K, is $0.295$~K; a $400$-resample stratified bootstrap of the trace vectors gives $2.5$th and $97.5$th percentiles of $0.290$ and $0.305$~K. These percentiles describe random-vector sampling only, conditional on this cluster and on the experimental data as extracted from Ref.~\cite{Ishikawa2024}; they exclude finite-size uncertainty, the uncertainty of the data extraction, and model uncertainty. Calculations with lower stochastic sampling shift the susceptibility-fit scale by several hundredths of a kelvin. Taking into account the uncertainty in extracting the $C_p$ data between $0.5$ and $1$~K from Ref.~\cite{Ishikawa2024}, the specific-heat fit gives $0.395$ to $0.640$~K, at the same $0.005$~K resolution. This range is separate from the stochastic sampling uncertainty. The $M_c(H)$ curve is fitted best by $0.34$~K, but its shape is not reproduced: the model reaches $95\%$ of the plateau at $0.52$~T for $J_{xy}=0.30$~K, the data at $1.1$~T, which would require $J_{xy}\approx0.8$--$0.9$~K. For illustration we use $J_{xy}=0.29$--$0.49$~K, i.e., an effective $\Je=0.65$--$1.1$~K, five to eight times below the DFT value; this interval is not a confidence interval for the exchange. With $T_c\approx0.25J_{xy}$, these values suggest order at $0.07$--$0.12$~K (up to $0.16$~K at the upper edge of the $C_p$ band), a range that overlaps the $0.1$~K bound. Within this model, the couplings fitted to $\chi_c$ and $C_p/T$ place the ordering temperature near the lowest measured temperature, whereas the field scale of $M_c(H)$ would place it near $0.2$~K. Since the model also misses $C_p/T$ above $0.6$~K and the in-plane magnetization, the present data do not discriminate between an ordered ground state below $0.1$~K and a spin liquid; Ishikawa \textit{et al.}\ noted that the upturn of $C_p/T$ may indicate long-range order below $0.1$~K~\cite{Ishikawa2024}. Weak unfrustrated interlayer coupling compatible with the layer order would tend to stabilize it; the sign and geometry of such couplings have not been determined here (the interlayer Cu--Cu separation is $8.76$~\AA, with dimethylammonium cations and water molecules between the layers~\cite{Ishikawa2024}). The origin of the discrepancy between the DFT exchange $\Je$, mediated by two sulfate bridges, and the smaller scale preferred by this reduced model remains unresolved.

The model does not fully account for the specific heat above $0.6$~K: between $0.6$ and $1.5$~K the ratio of the measured to the model $C_p/T$ is $1.9$ to $5.2$ for $J_{xy}=0.30$~K and $0.8$ to $2.0$ for $0.49$~K. A lattice term $\beta T^2$ with $\beta\approx6\times10^{-3}$~J\,K$^{-4}$\,mol$_{\rm Cu}^{-1}$, set by the rise above $2$~K, adds only $0.013$~J\,K$^{-2}$\,mol$_{\rm Cu}^{-1}$ at $1.5$~K. The data contain $0.29$~J\,K$^{-1}$\,mol$_{\rm Cu}^{-1}$ of entropy between $0.5$ and $2$~K, about $15\%$ of the doublet entropy $(R\ln2)/3=1.92$. This integral alone does not locate the remaining entropy below $0.5$~K, because the contribution above $2$~K has not been bounded. The effective model predicts that most of its doublet entropy lies below $0.5$~K, releasing only $0.09$ to $0.25$~J\,K$^{-1}$\,mol$_{\rm Cu}^{-1}$ above that temperature. Nor does it explain the in-plane magnetization. With the DFT-set DM sign an in-plane field does not couple to the doublet at first order. The pseudospin $g$ tensor then has entries $|G_{\perp\perp}|=0$, $|G_{zz}|=2.29$, and $|G_{z\perp}|=0.11$, so a field along $c$ acquires a small in-plane pseudospin component. These entries assume local Cu $g$ tensors that are axial about the CuO$_4$ plaquette normal, with $g=2.40$ along the normal and $2.22$ in the plaquette, chosen to reproduce the crystal $g$ values estimated from Curie--Weiss fits ($2.24$ along $c$, $2.30$ in the plane); for $g=2.40$ and $2.06$, $|G_{zz}|=2.15$ and $|G_{z\perp}|=0.21$; for an isotropic $g$, $G_{z\perp}=0$. The bulk Curie--Weiss estimates do not uniquely determine the local tensors. This local-tensor estimate is illustrative; the thermodynamic curves use the effective $g=2.22$, fixed independently by the plateau magnetization. The isolated-triangle Van Vleck response then gives $M_\perp=0.008$--$0.065\,\mu_B$/Cu at $0.13$~K between $0.5$ and $4$~T (fit set), only $1/6$ to $1/2$ of the measured $0.047$--$0.146\,\mu_B$/Cu.

\section{Remarks on Published Values}
\label{app:remarks}

We collect here remarks on published results that we used or compared with. Each states what we find with our conventions; none of them is needed for the arguments of the main text.

\paragraph{Exact solution for general couplings.} For couplings $J_x,J_y,J_z$ on the triangle edges and $Z$ on the inter-triangle bonds, Kells \emph{et al.}\ obtained the critical sphere $Z^2=J_x^2+J_y^2+J_z^2$ from a gap closing at $\bm{k}=0$, with the Abelian phase for $Z>|\bm{J}|$ and the non-Abelian phase for $Z<|\bm{J}|$ away from the planes $J_a=0$~\cite{Kells2010}. Their Eq.~(A8) also has solutions at other momenta, whose position depends on the gauge. We find that these bound an additional region with $\nu=0$, $J_a^2>J_b^2+J_c^2+Z^2$ with $J_a$ the largest triangle coupling, inside the part of the phase diagram assigned to the non-Abelian phase in Ref.~\cite{Kells2010}. For $(J_x,J_y,J_z)=(0.3,0.3,0.9)$, for example, the gap closes at $Z/|\bm{J}|=0.7977$, with $\nu=0$ below this value, $|\nu|=1$ between it and $Z=|\bm{J}|$, and $\nu=0$ for $Z>|\bm{J}|$. The region has a simple interpretation for $Z\ll|\bm{J}|$. The Majorana zero mode of a triangle has an amplitude on each vertex proportional to the coupling of the opposite edge, which has the color $a$ of the inter-triangle bond at that vertex, so the inter-triangle bonds couple the zero modes into a honeycomb Kitaev model with couplings $ZJ_a^2/|\bm{J}|^2$. For isotropic triangles this is the effective model derived by Dusuel \emph{et al.}~\cite{Dusuel2008}, at its gapless isotropic point; for anisotropic triangles its Abelian phase, $J_a^2>J_b^2+J_c^2$, is the limit $Z\to0$ of the additional region. The region does not intersect the family of isotropic triangles studied in the main text.

\paragraph{Majorana representation.} The identification given in Eqs.~(10) and~(A1) of Ref.~\cite{BurnellNayak2011} differs from Eq.~\eqref{eq:majorana-map} by an orthogonal transformation of determinant $-1$ of the four Majoranas, so that $b^x_ib^y_ib^z_ic_i=-1$ on the physical states. With it, $\ii b^a_ic_i$ represents on the physical states $-U\sigma^a_iU^\dagger$ with $U=\exp(-\ii\frac{\pi}{2}\sum_jS^z_j)$, as in Eq.~(A5) there; the overall sign drops out of every bond term, so the projected states are exact eigenstates of $UH_KU^\dagger$, which is $H_K$ with $\sigma^x\sigma^x$ and $\sigma^y\sigma^y$ interchanged on every bond. Because $H_K$ is quadratic in the spins, the same Hamiltonian is obtained with the global $\pi$ rotation $C_2^{[1\bar10]}$, and on the star lattice also with the mirror $S$ acting on the sites alone, which exchanges the $x$ and $y$ colors; it has the spectrum of $H_K$. When the lattice operations do not rotate the spins, $U$ commutes with all of them and with time reversal, and the identification of Ref.~\cite{BurnellNayak2011} can be used together with this rotation, as for the square-hexagon-dodecagon lattice~\cite{RajIqbalSHD}. On the star lattice the lattice operations carry spin rotations that do not commute with $U$, and we use Eq.~\eqref{eq:majorana-map}, with which the projected states are eigenstates of $H_K$ itself.

\paragraph{Honeycomb PSGs.} You, Kimchi, and Vishwanath list $144$ algebraic forms in Appendix~A of Ref.~\cite{YouKimchiVishwanath2012}. We find that all of them satisfy the relations of the group and that they represent $112$ gauge-inequivalent PSGs: in the classes (I)(B), (II)(B1), and (II)(B2), forms that differ in the sign of $\theta_1$, of $\theta_1-\theta_2$, or of $\theta_1+\theta_2$ are related by a global gauge transformation. Their Eq.~(13) for the Kitaev state is gauge equivalent to the (I)(B) form with $(\eta_1,\eta_9,\eta_{10},\eta_{14})=(+,+,+,-)$ and $\theta_1=\pm2\pi/3$ in the parametrization of their Eq.~(A33); the parameter set $(+,-,-,-)$ given in their Appendix~A describes a different (I)(B) PSG, which differs in the relation value of $\sigma^2$ ($\eta_{10}$). With $\theta_1\in\{0,\pm2\pi/3\}$, Eq.~(A33) gives $[G_{C_6}(B)G_{C_6}(A)]^3=-\eta_1\eta_9$, whereas Eq.~(A29) there has $+\eta_1\eta_9$; the sign of $\eta_9$ is therefore a convention of Eq.~(A33), in which the relation value of $C_6^6$ is $-\eta_9$. The identification of the Kitaev state in class (I)(B) is unaffected.

\paragraph{Phase boundaries of the Kitaev--Heisenberg model.} Every boundary of the exact phase diagram must map onto a boundary under Eq.~\eqref{eq:dualphi}. The boundaries of the Majorana mean-field phase diagram in Fig.~8 of Ref.~\cite{dOrnellasKnolle2024} are mutually dual to within $0.014\pi$, consistent with the $0.02\pi$ angular resolution of that calculation, whereas the values $0.83\pi$, $1.26\pi$, $1.63\pi$, and $1.83\pi$ quoted in the text there differ from dual pairs by up to $0.05\pi$ ($0.83\pi$ maps to $1.877\pi$ and $1.26\pi$ to $1.644\pi$). The iDMRG boundaries outside the spin-liquid windows, at $0.817\pi$ and $1.900\pi$, map onto $1.892\pi$ and $0.811\pi$, within $0.008\pi$. The iDMRG windows of the Kitaev spin liquid in Fig.~7 of Ref.~\cite{dOrnellasKnolle2024}, $[0.455,0.541]\pi$ (FM) and $[1.495,1.509]\pi$ (AFM), whose edges carry an uncertainty of $\pm0.005\pi$ from their extraction from that figure, are self-dual within this uncertainty: the FM window maps onto $[0.4530,0.5394]\pi$ and the AFM window onto $[1.4907,1.5049]\pi$, shifts of at most $0.004\pi$. Because the AFM window is narrow, this shift is about $30\%$ of its width: self-dual AFM windows consistent with the two iDMRG edges have widths between $0.0099\pi$ and $0.0183\pi$, so the ratio of the FM to the AFM width implied by these data lies between $4.6$ and $8.9$ (Table~\ref{tab:ratios}). If both edges of each self-dual window are allowed to deviate from these values by the extraction uncertainty of $\pm0.005\pi$, the range widens to about $4$ to $12$.

\paragraph{iPEPS energies of the Heisenberg model.} The bond energies printed in Ref.~\cite{JahromiOrus2018} include the coupling, $E_b=J_b\langle\bm{S}_i\cdot\bm{S}_j\rangle$: at large $\Je$ ($\Jt=0.05\Je$) the triangle value $-0.000475$ agrees with the second-order result $-3\Jt^2/(16\Je)=-0.00047$, and the bond energies reproduce the quoted energy per site, $-0.375238$ against $-0.375234$. At the isotropic point the bond energies, obtained with bond dimension $D=9$ (the dimension of the virtual tensor indices, which controls the accuracy), give $-0.4663$ per site, below the quoted $-0.464114$, below the value $-0.464299$ of Ref.~\cite{Ghosh2025}, and below the $42$-site ED energy $-0.4649$~\cite{Richter2004}. The quoted energy is an extrapolation to $D\to\infty$, but Fig.~5 there gives $\varepsilon_0\approx-0.464$ already at $D=9$, the bond dimension of the printed bond energies, so the difference is not due to the extrapolation, and its origin remains unclear. At large $\Jt$ ($\Je=0.05\Jt$) the discrepancy does not depend on such details: the printed bond energies sum to $-0.22556$ per site against the quoted $-0.255155$, and even if the triangle not shown is given its lowest energy, $-3\Jt/4$, and every $\Je$ bond is a perfect singlet, the energy per site is $-0.2518$, still above $-0.255155$. Reference~\cite{JahromiOrus2018} states that its $48$-site cell encompasses the $\sqrt3\times\sqrt3$ structure. The $12$- and $48$-site cells used there contain $2$ and $8$ unit cells of the star lattice, whereas a periodic state with the $18$-site $\sqrt3\times\sqrt3$ pattern requires a multiple of three unit cells, so by this counting neither cell can represent it.

\paragraph{DiMACuS.} The Supplemental Material of Ref.~\cite{Ishikawa2024} gives $D_\parallel/\Jt=0.493$ along $c$ and $D_\perp/\Jt=0.274$ [Eqs.~(S12)--(S15) there], while the main text prints the two labels exchanged. The doublet gap $\sqrt3D_\parallel\approx35$~K quoted in the main text, for $J=57$~K and $|D|/J=0.42$, requires the larger component along $c$: we find $36.6$~K, against $21.6$~K for the exchanged labels. Equations~(S4)--(S8), (S12)--(S14), and (S20) and Fig.~S3 of Ref.~\cite{Ishikawa2024} consistently use $A^{\alpha\beta}=\epsilon^{\alpha\beta\gamma}D^\gamma$ and $H_{\rm DM}=\bm D_{ij}\cdot(\bm S_i\times\bm S_j)$. The counterclockwise bonds then carry $D_z=-33.0$~K [Eq.~(S12) there, for the relativistic $J_T=66.9$~K], and the DFT set gives the case of Eq.~\eqref{eq:dimacus} with $\Gamma\neq0$ and the $13.8^\circ$ tilt of the easy axis. Since a global sign change of all DM vectors is equivalent, by a spin rotation about $c$, to reversing $D_\parallel$ alone, reversing all DM vectors would give the XXZ case; the published convention selects the first case. The compound of Ref.~\cite{Ji2022} is listed with DiMACuS among the spin-$1/2$ realizations of the star lattice in Ref.~\cite{dOrnellasKnolle2024}; its magnetic ions are high-spin Fe(III) with $S=5/2$~\cite{Ji2022}.

\bibliography{references}

\clearpage
\onecolumngrid
\makeatletter\let\@hangfrom@section\SM@hangfrom@section\let\@sectioncntformat\SM@sectioncntformat\@removefromreset{equation}{section}\makeatother
\setcounter{figure}{0}\setcounter{table}{0}\setcounter{equation}{0}\setcounter{section}{0}
\renewcommand{\thefigure}{S\arabic{figure}}\renewcommand{\theHfigure}{S\arabic{figure}}
\renewcommand{\thetable}{S\arabic{table}}\renewcommand{\theHtable}{S\arabic{table}}
\renewcommand{\theequation}{S\arabic{equation}}\renewcommand{\theHequation}{S\arabic{equation}}
\renewcommand{\thesection}{S\Roman{section}}\renewcommand{\theHsection}{S\Roman{section}}
\renewcommand{\thesubsection}{\Alph{subsection}}
\begin{center}
{\large\bfseries Supplemental Material for ``\papertitle''}\\[1ex]
Harsh Raj and Yasir Iqbal\\[0.5ex]
{\small\itshape Department of Physics, Indian Institute of Technology Madras, Chennai 600036, India}
\end{center}
\phantomsection\addcontentsline{toc}{section}{Supplemental Material}
\vspace{1ex}
This Supplemental Material lists the symmetry-allowed mean-field amplitudes and the realization ranges of all $\Z_2$ projective symmetry groups (PSGs)~\SMcite{SM:Wen2002} of the Kitaev--Heisenberg model on the star lattice, with the symmetry group $G_{\rm KH}\times\Z_2^\T$ of Sec.~\ref{sec:lattice} of the main text. Section~\ref{sec:sm-conv} summarizes the conventions, Sec.~\ref{sec:sm-trs} lists the $352$ time-reversal-symmetric PSGs, Sec.~\ref{sec:sm-chiral} the $80$ chiral PSGs~\SMcite{SM:BieriLhuillierMessio2016} in each of the four chiral classes, and Sec.~\ref{sec:sm-u1} the $U(1)$ PSGs of the Kitaev--Heisenberg and Heisenberg models. Equations, tables, and sections numbered with the prefix S belong to this Supplemental Material; all other numbers refer to the main text.

\section{Conventions}
\label{sec:sm-conv}

A $\Z_2$ PSG is labeled Z$n_{T_2}n_{\T S}n_{\T C_6}n_Sn_{SC_6}n_{C_6}t\,n_Rn_{TR}$ by the relation values of Eq.~(\ref{eq:labels}) of the main text (digit $1$ for $-1$), with a suffix -$a$ or -$b$ for the $160$ pairs of PSGs that share all nine relation values, and with the same suffix for the $32$ pairs of non-locked chiral PSGs that share all their relation values. The two PSGs of a pair differ in $\zeta=G_{R_x}(i_z)/G_{R_x}(i_x)=\pm1$, the ratio of the central gauge parts of $R_x$ on two sites of the same unit cell whose inter-triangle bonds are $z$ and $x$ bonds, for example $(0,0,1)$ and $(0,0,2)$; the suffix -$b$ denotes $\zeta=+1$ and -$a$ denotes $\zeta=-1$. Since $\zeta$ is a ratio of central matrices, it is invariant under gauge transformations and re-liftings (sign changes $G_g\to-G_g$ of gauge parts) and does not depend on the cell; a chiral $\Z_2$ PSG is labeled ZC$n_{T_2}n_Sn_{SC_6}n_{C_6}n_Rn_{TR}$. The PSGs are given in the canonical gauge of Appendix~\ref{app:forms} of the main text~\SMcite{SM:RajIqbalSHD}, and the type of a PSG refers to the gauge parts of the stabilizer of the site $(0,0,1)$ listed in Table~\ref{tab:onsite-sm}.

\begin{table}[h]
\caption{On-site gauge parts of $(S,R_x,R_y,\T)$ at the site $(0,0,1)$ in the canonical gauge (Table~\ref{tab:onsite} of the main text). For the chiral PSGs only $(S,R_x,R_y)$ apply.}
\label{tab:onsite-sm}
\begin{ruledtabular}
\begin{tabular}{ccccc}
type & $G_S$ & $G_{R_x}$ & $G_{R_y}$ & $G_\T$\\
\hline
I & $\tau^0$ & $\tau^0$ & $\tau^0$ & $\tau^0$\\
II & $\tau^0$ & $\tau^0$ & $\tau^0$ & $\ii\tau^1$\\
III & $\ii\tau^1$ & $\tau^0$ & $\tau^0$ & $\tau^0$\\
IV & $\ii\tau^1$ & $\tau^0$ & $\tau^0$ & $\ii\tau^1$\\
V & $\ii\tau^1$ & $\tau^0$ & $\tau^0$ & $\ii\tau^2$\\
VI & $\ii\tau^1$ & $\ii(\tau^1+\tau^2)/\sqrt2$ & $\ii(\tau^1-\tau^2)/\sqrt2$ & $\tau^0$
\end{tabular}
\end{ruledtabular}
\end{table}

For every PSG and every bond orbit of Table~\ref{tab:bonds} of the main text, the allowed space is the real solution space of the symmetry conditions on the reference bond, including the stabilizer of the bond and, for bonds that are reversed by a symmetry, the reversal. Its dimension is given as singlet/triplet, the dimensions of the projections of the allowed space onto the singlet coefficients $(\chi^0,\chi^3,\Delta^1,\Delta^2)$ and onto the twelve triplet coefficients of Eq.~(\ref{eq:links}) of the main text~\SMcite{SM:LiuHalaszBalents2021}. The realization columns give the invariant gauge group (IGG) of a generic ansatz with random amplitudes in all allowed directions of the listed orbits and without on-site multipliers, computed as the commutant of the bond amplitudes on a connected patch with an absolute singular-value threshold; an ansatz whose bonds do not connect the lattice is marked as disconnected.

\section{Time-reversal-symmetric PSGs}
\label{sec:sm-trs}

{\footnotesize
\setlength{\tabcolsep}{4pt}
\begin{longtable}{rlcccccccccccc}
\caption{All $352$ time-reversal-symmetric $\Z_2$ PSGs of $G_{\rm KH}\times\Z_2^\T$. Type: on-site gauge parts of $(S,R_x,R_y,\T)$ at the site $(0,0,1)$ in the canonical gauge (Table~\ref{tab:onsite-sm}). $J_t$ to $J_4$: real dimensions singlet/triplet of the allowed amplitudes on the reference bond ($0$: forbidden); $a_i$: dimension of the allowed on-site multiplier. $R_1$ to $R_4$: IGG of a generic ansatz with bonds up to $R_1=\{J_t,J_e\}$, $R_2=R_1\cup\{J_2\}$, $R_3=R_2\cup\{J_{3a},J_{3b}\}$, $R_4=R_3\cup\{J_4\}$, with singlet and triplet amplitudes / with singlet amplitudes only (Z: $\Z_2$, U: $U(1)$, S: $SU(2)$, --: disconnected).}
\label{tab:soc-sm}\\
\hline\hline
No. & PSG & type & $J_t$ & $J_e$ & $J_2$ & $J_{3a}$ & $J_{3b}$ & $J_4$ & $a_i$ & $R_1$ & $R_2$ & $R_3$ & $R_4$\\
\hline
\endfirsthead
\hline\hline
No. & PSG & type & $J_t$ & $J_e$ & $J_2$ & $J_{3a}$ & $J_{3b}$ & $J_4$ & $a_i$ & $R_1$ & $R_2$ & $R_3$ & $R_4$\\
\hline
\endhead
\hline\hline
\endfoot
1 & Z111111101-$a$ & V & 1/0 & 0 & 0/2 & 0 & 0/1 & 0/1 & 1 & -- & Z/-- & Z/-- & Z/--\\
2 & Z111111101-$b$ & V & 0/1 & 0 & 0/2 & 0 & 1/0 & 0/1 & 1 & -- & Z/-- & Z/-- & Z/--\\
3 & Z111111100-$a$ & V & 0/1 & 0 & 0/2 & 0 & 0/1 & 1/0 & 1 & -- & Z/-- & Z/-- & Z/--\\
4 & Z111111100-$b$ & V & 1/0 & 0 & 2/0 & 0 & 1/0 & 1/0 & 1 & -- & Z/Z & Z/Z & Z/Z\\
5 & Z111110101-$a$ & V & 1/0 & 0 & 0/2 & 0 & 0/1 & 0/1 & 1 & -- & Z/-- & Z/-- & Z/--\\
6 & Z111110101-$b$ & V & 0/1 & 0 & 0/2 & 0 & 1/0 & 0/1 & 1 & -- & Z/-- & Z/-- & Z/--\\
7 & Z111110100-$a$ & V & 0/1 & 0 & 0/2 & 0 & 0/1 & 1/0 & 1 & -- & Z/-- & Z/-- & Z/--\\
8 & Z111110100-$b$ & V & 1/0 & 0 & 2/0 & 0 & 1/0 & 1/0 & 1 & -- & Z/Z & Z/Z & Z/Z\\
\end{longtable}
}

\section{Chiral PSGs}
\label{sec:sm-chiral}

{\footnotesize
\setlength{\tabcolsep}{4pt}
\begin{longtable}{rlccccccccccccc}
\caption{The $80$ chiral $\Z_2$ PSGs of $G_{\rm KH}$ in each of the chiral classes A--D (Sec.~\ref{sec:chiral} of the main text), in the notation of Table~\ref{tab:soc-sm}; the type refers to the on-site gauge parts of $(S,R_x,R_y)$.}
\label{tab:chiral-sm}\\
\hline\hline
No. & PSG & class & type & $J_t$ & $J_e$ & $J_2$ & $J_{3a}$ & $J_{3b}$ & $J_4$ & $a_i$ & $R_1$ & $R_2$ & $R_3$ & $R_4$\\
\hline
\endfirsthead
\hline\hline
No. & PSG & class & type & $J_t$ & $J_e$ & $J_2$ & $J_{3a}$ & $J_{3b}$ & $J_4$ & $a_i$ & $R_1$ & $R_2$ & $R_3$ & $R_4$\\
\hline
\endhead
\hline\hline
\endfoot
1 & ZC111111 & A & VI & 0/1 & 0 & 1/3 & 0/1 & 0/1 & 1/1 & 0 & -- & Z/S & Z/S & Z/S\\
2 & ZC111110 & A & VI & 0/1 & 0 & 1/3 & 0/1 & 0/1 & 0 & 0 & -- & Z/S & Z/S & Z/S\\
3 & ZC111101-$a$ & A & III & 1/0 & 0 & 0/4 & 0/1 & 0/1 & 0/1 & 1 & -- & Z/-- & Z/-- & Z/--\\
4 & ZC111101-$b$ & A & III & 0/1 & 0 & 0/4 & 1/0 & 1/0 & 0/1 & 1 & -- & Z/-- & Z/-- & Z/--\\
5 & ZC111100-$a$ & A & III & 0/1 & 1/0 & 0/4 & 0/1 & 0/1 & 3/0 & 1 & U/-- & Z/-- & Z/-- & Z/--\\
6 & ZC111100-$b$ & A & III & 1/0 & 1/0 & 4/0 & 1/0 & 1/0 & 3/0 & 1 & U/U & Z/Z & Z/Z & Z/Z\\
7 & ZC111011 & A & VI & 0/1 & 0/1 & 1/3 & 0/1 & 0/1 & 0/2 & 0 & Z/-- & Z/S & Z/S & Z/S\\
8 & ZC111010 & A & VI & 0/1 & 0/1 & 1/3 & 0/1 & 0/1 & 1/3 & 0 & Z/-- & Z/S & Z/S & Z/S\\
\end{longtable}
}

\section{\texorpdfstring{$U(1)$}{U(1)} PSGs}
\label{sec:sm-u1}

For the IGG $U(1)=\{e^{\ii\alpha\tau^3}\}$~\SMcite{SM:Wen2002} every gauge part has the form $G_g(i)=e^{2\pi\ii\phi_g(i)\tau^3}(\ii\tau^1)^{w_g}$, where $w$ is a character of the symmetry group with values in $\Z_2$ that records whether $g$ preserves ($w_g=0$) or reverses ($w_g=1$) the $U(1)$ axis. The relations of $G_{\rm KH}$ force $w_{R_x}=w_{R_y}=0$, and $w$ is fixed by $(w_{C_6},w_S,w_\T)$, which defines eight sectors. In each sector we count the $U(1)$ PSGs by exact integer elimination of all phases $\phi_g(i)$ on a patch of the star lattice, including the on-site conditions at the site $(0,0,1)$; the Smith normal form of the remaining constraints gives the discrete PSGs and the one-parameter families. As a second method, every solution of the canonical-gauge search of Appendix~\ref{app:forms} in the dicyclic group of order $48$ whose generators are uniformly diagonal or uniformly off-diagonal is an explicit $U(1)$ PSG; we reduce these modulo gauge transformations and re-liftings. Table~\ref{tab:u1-sm} lists the results. In the KH columns each class of relation values contains two PSGs, which differ in the sign $\zeta=\pm1$ of Sec.~\ref{sec:sm-conv}, so the number of PSGs is twice the number of classes. The two methods give the same counts in every sector; in the sector $w=010$ the explicit search, being restricted to a finite subgroup of $SU(2)$, finds only the members of the families at flux $0$ and $\pi$. For the Heisenberg model a $U(1)$ PSG is called realizable if a singlet amplitude is allowed on some bond up to two lattice constants.

\begin{table}[h]
\caption{$U(1)$ PSGs of the star lattice by sector $w=(w_{C_6},w_S,w_\T)$. KH: symmetry group $G_{\rm KH}\times\Z_2^\T$; relation-value classes and PSGs from the integer elimination, and PSGs found by the explicit canonical-gauge search. Heisenberg: symmetry group $p6mm\times\Z_2^\T$; classes of the cohomology before the on-site condition, PSGs after it, and the number of realizable PSGs. $n\times U(1)$: $n$ one-parameter families of PSGs with a continuous flux.}
\label{tab:u1-sm}
\begin{ruledtabular}
\begin{tabular}{ccccccc}
 & \multicolumn{3}{c}{KH} & \multicolumn{3}{c}{Heisenberg}\\
$w$ & relation values & PSGs & explicit search & cohomology & PSGs & realizable\\
\hline
$000$ & $16$ & $32$ & $32$ & $16$ & $8$ & $4$\\
$001$ & $16$ & $32$ & $32$ & $32$ & $8$ & $8$\\
$010$ & $8\times U(1)$ & $16\times U(1)$ & $32$ points & $16\times U(1)$ & $4\times U(1)$ & $2\times U(1)$\\
$011$ & $16$ & $32$ & $32$ & $32$ & $8$ & $8$\\
$100$ & $32$ & $64$ & $64$ & $32$ & $16$ & $8$\\
$101$ & $16$ & $32$ & $32$ & $32$ & $8$ & $8$\\
$110$ & $16$ & $32$ & $32$ & $32$ & $8$ & $4$\\
$111$ & $16$ & $32$ & $32$ & $32$ & $8$ & $8$\\
\hline
total & $128+8\times U(1)$ & $256+16\times U(1)$ & $256+32$ points & $208+16\times U(1)$ & $64+4\times U(1)$ & $48+2\times U(1)$
\end{tabular}
\end{ruledtabular}
\end{table}

\begingroup
\makeatletter
\let\SM@label\label\let\SM@ref\ref\def\SM@last{LastBibItem}%
\renewcommand\label[1]{\def\SM@arg{#1}\ifx\SM@arg\SM@last\SM@label{SM:LastBibItem}\else\SM@label{#1}\fi}%
\renewcommand\ref[1]{\def\SM@arg{#1}\ifx\SM@arg\SM@last\SM@ref{SM:LastBibItem}\else\SM@ref{#1}\fi}%
\makeatother
\endgroup

\end{document}